\documentclass[%
 reprint,
 superscriptaddress,
preprintnumbers,
nofootinbib,
 amsmath,
 amssymb,
 aps,
 prd,
 longbibliography,
floatfix,
]{revtex4-2}

\usepackage{graphicx}
\usepackage{dcolumn}
\usepackage{bm}
\usepackage{xcolor}
\usepackage{lineno}
\usepackage{booktabs}
\usepackage{footmisc}
\usepackage{enumitem}
\usepackage{multirow}

\usepackage{subfigure}
\usepackage[normalem]{ulem}
\usepackage{hyperref}
\hypersetup{
    colorlinks=true,
    linkcolor=blue,      
    citecolor=blue,      
    filecolor=blue,      
    urlcolor=blue,       
}
\usepackage{makecell}

\newcommand{\tagn}{$\log_{10} T_{\mathrm{AGN}}$}

\newcommand{\journal}[1]{{\color{orange}#1}}
\renewcommand{\journal}[1]{#1}                     

\begin{document}

\preprint{DES-2025-0937}
\preprint{FERMILAB-PUB-26-0035-PPD}

\title{Dark Energy Survey Year 6 Results: Weak Lensing and Galaxy Clustering Cosmological Analysis Framework}

%


\author{D.~Sanchez-Cid}\email{david.sanchezcid@physik.uzh.ch}
\affiliation{Physik-Institut, University of Zürich, Winterthurerstrasse 190, CH-8057 Zürich, Switzerland}
\affiliation{Centro de Investigaciones Energ\'eticas, Medioambientales y Tecnol\'ogicas (CIEMAT), Madrid, Spain}
\author{A.~Fert\'e}\email{ferte@slac.stanford.edu}
\affiliation{SLAC National Accelerator Laboratory, Menlo Park, CA 94025, USA}
\author{J.~Blazek}
\affiliation{Department of Physics, Northeastern University, Boston, MA 02115, USA}
\author{S.~Samuroff}
\affiliation{Department of Physics, Northeastern University, Boston, MA 02115, USA}
\affiliation{Institut de F\'{\i}sica d'Altes Energies (IFAE), The Barcelona Institute of Science and Technology, Campus UAB, 08193 Bellaterra (Barcelona) Spain}
\author{A.~Amon}
\affiliation{Department of Astrophysical Sciences, Princeton University, Peyton Hall, Princeton, NJ 08544, USA}
\author{F.~Andrade-Oliveira}
\affiliation{Physik-Institut, University of Zürich, Winterthurerstrasse 190, CH-8057 Zürich, Switzerland}
\author{J.~M.~Coloma-Nadal}
\affiliation{Institute of Space Sciences (ICE, CSIC),  Campus UAB, Carrer de Can Magrans, s/n,  08193 Barcelona, Spain}
\author{J.~Muir}
\affiliation{Department of Physics, University of Cincinnati, Cincinnati, Ohio 45221, USA}
\affiliation{Perimeter Institute for Theoretical Physics, 31 Caroline St. North, Waterloo, ON N2L 2Y5, Canada}
\author{A.~Porredon}
\affiliation{Centro de Investigaciones Energ\'eticas, Medioambientales y Tecnol\'ogicas (CIEMAT), Madrid, Spain}
\affiliation{Ruhr University Bochum, Faculty of Physics and Astronomy, Astronomical Institute, German Centre for Cosmological Lensing, 44780 Bochum, Germany}
\author{J.~Prat}
\affiliation{Nordita, KTH Royal Institute of Technology and Stockholm University, Hannes Alfv\'ens v\"ag 12, SE-10691 Stockholm, Sweden}
% ...

\collaboration{DES Collaboration}

\date{\today}

\begin{abstract}
We present the methodology for the weak lensing and galaxy clustering analyses of the Dark Energy Survey (DES) Year 6 data set. In this work, we design and validate the analysis pipeline for the cosmic shear, galaxy clustering plus galaxy–galaxy lensing ($2 \times 2$pt), and the joint analysis in the $3 \times 2$pt. Our framework accounts for key theoretical uncertainties, such as baryonic feedback and galaxy bias, incorporating both linear and non-linear models. We apply scale cuts in regimes where theoretical modeling becomes unreliable. The robustness of the pipeline is validated using mock data and simulations, confirming unbiased cosmological constraints and highlighting the importance of posterior projection effects in the validation process. As a result, we deliver robust and validated analysis pipelines for cosmic shear, $2 \times 2$pt, and $3 \times 2$pt in $\Lambda$CDM and $w$CDM scenarios, including a well-defined set of scales suitable for real data analysis, a robust prescription for theoretical systematics, and the theoretical covariance of the signal. This comprehensive methodology also lays the groundwork for future galaxy surveys such as the Vera C. Rubin Observatory Legacy Survey of Space and Time.
\end{abstract}

\maketitle

\section{\label{sec:intro}Introduction}

The dynamics and composition of the late-time Universe, governed by dark energy and dark matter remain among the most profound open questions in modern science. Observational cosmology seeks to shed light on them through complementary approaches that include geometric probes and studies of the growth of cosmic structures. A particularly powerful method is the joint analysis of weak gravitational lensing~\cite{Kaiser_1992, Refregier_2003, Mandelbaum_2018} and galaxy clustering~\cite{peebles1973statistical, davis1983survey, desjacques2018large}, sliced in tomographic redshift bins to trace their evolution over cosmic time.

Weak gravitational lensing arises when the light from distant source galaxies is deflected by the gravitational field of the large-scale structure (LSS), leading to coherent distortions in their observed shapes. This effect not only enables mapping of the matter distribution in the Universe, but also probes the competition between gravitational attraction from dark matter and the accelerated expansion driven by dark energy. Similarly, galaxy clustering traces the growth of structure by measuring how galaxies cluster under the influence of gravity, providing complementary information on the interplay between dark matter and dark energy over cosmic time.

The cosmological information encoded in the weak lensing and galaxy density fields can be captured by two-point correlation functions (2PCFs). These functions involve the shapes of background (source) galaxies and the positions of foreground (lens) galaxies, with both lens and source galaxy samples further divided into tomographic redshift bins. The correlation of source galaxy shapes is known as \textit{cosmic shear}. The cross-correlation between distorted source galaxy shapes and lens galaxy positions is termed \textit{galaxy-galaxy lensing}. Finally, the correlation of lens galaxy positions defines the \textit{galaxy clustering} correlation function. The combination of these three correlation functions constitutes the so-called $3 \times 2$pt analysis. In contrast, the joint analysis of galaxy-galaxy lensing and galaxy clustering is typically referred to as $2 \times 2$pt.

The $3 \times 2$pt analysis was first developed and implemented by Stage-III imaging surveys \cite{albrecht2006reportdarkenergytask}, including the Dark Energy Survey (DES) \cite{Abbott_2018, Abbott_2022}, the Kilo-Degree Survey (KiDS) \cite{Heymans_2021}, and the Hyper Suprime-Cam (HSC) survey \cite{Sugiyama_2023, miyatake2023hypersuprimecamyear3}. Over time, both the methodology and the datasets have matured, enabling precision constraints on the total matter abundance, $\Omega_{\rm m}$, and the amplitude of matter density fluctuations, $\sigma_8$, in the late Universe. A joint analysis of weak lensing and galaxy clustering enhances the constraining power of cosmological models by breaking parameter degeneracies and is more resilient to systematics.

In this work, we present the methodology required to perform the weak lensing and galaxy clustering analyses with the final six-year dataset of DES \cite{bechtol2025darkenergysurveyyear}. To this end, we develop a framework robust against modeling uncertainties, considering the  latest advancements in theory, while accounting for the statistical properties of the data set, including deeper exposures and a higher number density of galaxies. With the increased statistical power in this final DES analysis it becomes essential to control theoretical uncertainties at a comparable level. This balance between statistical and theoretical precision will be even more critical in upcoming analyses from photometric Stage-IV surveys, such as the \textit{Vera C. Rubin Observatory} Legacy Survey of Space and Time (LSST) \cite{lsstdarkenergysciencecollaboration2012largesynopticsurveytelescope}, the \textit{Euclid mission} \cite{euclid2025}, and the \textit{Nancy Grace Roman Space Telescope} \cite{spergel2015}. We expect that the prescriptions presented in this work will serve as a foundational step for the analysis pipelines required by these next-generation surveys, \journal{aligned with the existing community effort toward accurate and robust 3×2pt modeling~\citep{Truttero:2026bec, Euclid:2026klv, Zhang:2026ttm}}.

In this work we carefully address the main theoretical systematics affecting weak lensing—namely baryonic feedback \cite{Chisari_2019} and intrinsic alignment \cite{Lamman_2024}—as well as those affecting galaxy clustering, which is primarily limited by galaxy bias \cite{nicola2023galaxybiaseralsst}. Additionally, we define the physical scales where this modeling remains accurate, ensuring that mis-modeled scales are excluded from the analysis via \textit{scale cuts}, thereby avoiding biased cosmological constraints. We present optimized prescriptions for the cosmic shear analysis~\cite{y6-1x2pt}, the $2 \times 2$pt~\cite{y6-2x2pt}, and the $3 \times 2$pt analysis~\cite{y6-3x2pt}.

This work is structured as follows. In Section~\ref{sec:theory}, we describe the 2PCF formalism for the joint analysis of weak lensing and galaxy clustering. In Section~\ref{sec:pipeline}, we outline the modeling choices made to address theoretical systematics and describe the elements that comprise the analysis pipeline. The Bayesian framework for parameter inference and the covariance matrix are described in Section \ref{sec:inference}. Section~\ref{sec:data} provides a detailed account of the properties of the synthetic data used throughout the analysis. Section~\ref{sec:cuts} describes scale cuts definition and validation with synthetic data and simulations. Section~\ref{sec:validation} presents the results and expected cosmological constraints. In Section ~\ref{sec:robustness-tests}, we present additional validation tests, demonstrating that the pipeline delivers unbiased and robust results and remains stable against variations in data and analysis configurations. Finally, we summarize and discuss the results in Section~\ref{sec:conclusions}.
\section{\label{sec:theory}Theory}

The central component of the \(3 \times 2\)pt analysis in configuration space is the set of 2PCFs, which summarizes the statistical properties of the projected galaxy overdensity field, \(\delta_{\rm obs}\), and the shear field, \(\gamma_\alpha\). The shear field is a spin-2 field, typically decomposed into E- and B-modes. The observed galaxy overdensity field have contributions from the underlying galaxy field $\delta_g$, and the lens magnification, $\delta_\mu$. The 2PCFs are real-space projections of the angular power spectra, $C_\ell$. 

In this framework, galaxy clustering, denoted as \(w(\theta)\), represents the auto-correlation of the lens galaxy positions within each tomographic redshift bin \(i\) for an angular separation $\theta$. In the \textit{curved-sky} projection:
\begin{equation}
    w^{i}(\theta) = \sum_{\ell} \frac{2\ell+1}{4\pi} P_\ell(\cos\theta) C^{ii}_{\delta_{\rm obs} \delta_{\rm obs}}(\ell),
\end{equation}
where \( P_\ell \) are the Legendre polynomials of order \(\ell\).

The galaxy-galaxy lensing signal, which captures the cross-correlation between the galaxy overdensity field in tomographic bin \(i\) and the E-mode component of the shear field in bin \(j\), is given by:
\begin{equation}
    \gamma_t^{ij}(\theta) = \sum_\ell \frac{2\ell+1}{4\pi} \frac{P_\ell^2(\cos\theta)}{\ell(\ell+1)} C^{ij}_{\delta_{\rm obs} E}(\ell).
\end{equation}

Finally, cosmic shear, which describes the correlation of the observed shapes of source galaxies, is expressed as:
\begin{equation}
    \begin{aligned}
        \xi^{ij}_{\pm}(\theta) =& \sum_{\ell} \frac{2\ell + 1}{4\pi} \frac{2(G^+_{\ell,2}(x) \pm G^-_{\ell,2}(x))}{\ell^2(\ell+1)^2} \times \\
        & \times [C^{ij}_{EE}(\ell) \pm C^{ij}_{BB}(\ell)],
    \end{aligned}
\end{equation}
Here, \(i\) and \(j\) denote the redshift tomographic bins being correlated, $G^{+/-}_{\ell, 2}(x)$ are analytic functions given in Appendix A of ~\cite{Friedrich_2021} , and $x = \cos(\theta)$.

The angular power spectra \(C^{ij}(\ell)\) appearing in the expressions above include contributions from various astrophysical and observational effects, such as intrinsic alignments (IA), lens magnification, and redshift-space distortions (RSD).

\begin{figure}[t]
    \centering
    \includegraphics[width=\columnwidth]{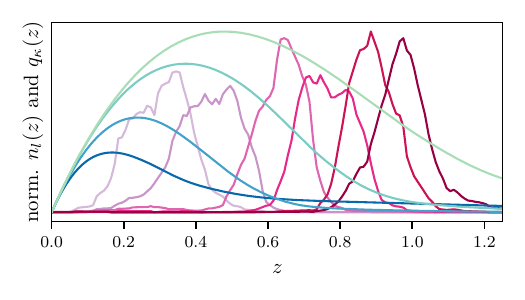}
    \caption{\label{fig:nz_kernels} Estimated redshift distributions for the lens or clustering catalog, divided into six redshift bins (pinks), presented in \protect\cite{giannini2025darkenergysurveyyear} and the lensing efficiency kernel for each of the four source redshift bins (blues), presented in \protect\cite{yin2025}, demonstrating that the DES Year 6 data is most sensitive between $z=0.1 - 1.0$.}
\end{figure}

We assume the Limber approximation for cosmic shear and galaxy-galaxy lensing, whose angular power spectra then take the form:
\begin{equation}
    C^{ij}_{\mathcal{AB}}(\ell) = \int d\chi \frac{q^i_\mathcal{A}(\chi)q^j_\mathcal{B}(\chi)}{\chi^2} P_m\left( k=\frac{\ell + 1/2}{\chi}, z\right),
\end{equation}
where \(\mathcal{A}\) and \(\mathcal{B}\) represent the fields being correlated, which in our case correspond to the galaxy overdensity field \(\delta_g\) and the convergence field \(\kappa\), derived from the shear field. The function \(P_m(k,z)\) represents the three-dimensional matter power spectrum, which we evaluate at a given redshift \(z\) and wavenumber \(k\) as described in Section \ref{subsec:spectrum}.  For the galaxy clustering angular power spectrum calculation,  we do not assume the Limber approximation and its full expression can be found in \cite{Fang_2020}. \journal{The Limber approximation is retained for cosmic shear and galaxy--galaxy lensing because their projection kernels are dominated by the broad lensing efficiency $q_\kappa(\chi)$, for which the approximation degrades only as $\mathcal{O}(\ell^{-2})$ at low $\ell$~\cite{LoVerde:2008re}; with the prescription $k=(\ell+1/2)/\chi$ adopted here, the shear power spectrum is sub-percent accurate for $\ell \gtrsim 10$~\cite{Kilbinger:2017lvu}, well below the characteristic $\ell \approx 43$ of our largest angular scale. Galaxy clustering, by contrast, depends on the narrow lens kernels $n^i(z)$ and requires the full non-Limber treatment~\cite{Fang_2020}, consistent with DES~Y3~\cite{krause2021darkenergysurveyyear}.} The kernel or radial weight function (displayed in Fig.~\ref{fig:nz_kernels} for the Year 6 setup) for the galaxy overdensity field:
\begin{equation}
    q^i_{\delta_g}(\chi) = b^i \, n^i_l(z(\chi)) \, \frac{dz}{d\chi} \, ,
\end{equation}
and for the convergence field, the lensing efficiency is
\begin{equation}
    q^i_\kappa(\chi) = \frac{3H_0^2 \Omega_{\rm m} \chi}{2a(\chi)} \int^{\chi_H}_{\chi} d\chi' \left( \frac{\chi' - \chi}{\chi'} \right) n^i_s(z(\chi'))\, \frac{dz}{d\chi'} \, ,
\end{equation}
where \(H_0\) is the Hubble constant, \(a(\chi)\) is the scale factor for a given comoving distance \(\chi\), \(n_{l/s}\) is the normalized redshift distribution of the lens/source galaxies, and \(b^i\) is the linear galaxy bias coefficient linking the galaxy distribution and the matter distribution \(\delta^i_g = b^i \delta^i_m\). We assign a separate linear bias coefficient to each redshift bin, as it is usually done in tomographic galaxy clustering analyses.  We assume that the evolution of the bias value with redshift within each bin is negligible, a valid assumption based on N-body simulations \cite{Porredon_2022}. 

Our model includes contributions from intrinsic alignment and lens magnification. The additional terms to each  angular power spectra are as follows:
\begin{equation}
    \begin{aligned}
        C^{ij}_{EE}(\ell) &= C^{ij}_{\kappa \kappa}(\ell) + C^{ij}_{\kappa I_E}(\ell) + C^{ji}_{\kappa I_E}(\ell) + C^{ij}_{I_E I_E}(\ell), \\
        C^{ij}_{BB}(\ell) &= C^{ij}_{I_B I_B}(\ell), \\
        C^{ij}_{\delta_{\rm obs} E}(\ell) &= C^{ij}_{\delta_g \kappa}(\ell) + C^{ij}_{\delta_g I_E}(\ell) + C^{ij}_{\delta_\mu \kappa}(\ell) + C^{ij}_{\delta_\mu I_E}(\ell), \\
        C^{ii}_{\delta_{\rm obs} \delta_{\rm obs}}(\ell) &= C^{ii}_{\delta_g \delta_g}(\ell) + C^{ii}_{\delta_g \delta_\mu}(\ell) + C^{ii}_{\delta_\mu \delta_\mu}(\ell),
    \end{aligned}
\end{equation}
where \(I_{E/B}\) refers to the E/B modes of the intrinsic alignment (IA) contribution, and \(\delta_\mu\) is the lens magnification contribution.

Magnification \cite{Joachimi_2010, Garcia_Fernandez_2018} affects the observed galaxy overdensity, which is particularly important for our lens sample and is modeled as:
\begin{equation}
    \delta^{i}_{\mu} = 2\big[\alpha^i(m_{\rm cut}) - 1\big] \kappa^{i},
\end{equation}
where $m_{\rm cut}$ is the magnitude limit applied in the lens galaxy sample selection, and the coefficient $\alpha^i(m)$ is defined by
\begin{equation}
    \alpha^i(m) = 2.5 \frac{d}{dm} \big[\log N^{i}_\mu(m)\big],
    \label{eqn:lens-mag}
\end{equation}
with $N_\mu(m)$ representing the cumulative number of lensed galaxies brighter than magnitude $m$ in bin $i$.

Intrinsic alignment (IA) refers to the effect whereby galaxies align with their local tidal field, mimicking the cosmic shear signal due to gravitational deflection ~\cite{Lamman_2024}. This phenomenon is one of the main systematics in cosmic shear analyses. Galaxy shape correlations are determined by the large-scale cosmological tidal field. The dominant effect at large scales and for central galaxies is the ``tidal alignment'' of galaxy shapes \cite{2004PhRvD..70f3526H}. More complex processes, like ``tidal torquing'', are relevant for the angular momentum of spiral galaxies. These require perturbative treatments, which we refer to as Tidal Alignment and Tidal Torquing (TATT) \cite{Blazek_2019}. In this extended model, the intrinsic galaxy shape field at the source galaxy's location is described by nonlinear cosmological perturbation theory \cite{Blazek_2015}, expanding in the matter density field \( \delta_m \) and the tidal field \( s_{ij} \):
\begin{equation}
    \gamma^{\rm IA}_{ij} = A_1 s_{ij} + A_{1\delta} \delta s_{ij} + A_2 \sum_k s_{ik}s_{kj}+... \;.
    \label{eqn:ia-expansion}
\end{equation}
We adopt different approaches for the IA modeling for the cosmic shear analysis and for $2 \times 2$pt and $3 \times 2$pt analyses, differing in the number of parameters involved in the analysis. Details on this are provided in Section \ref{subsec:model-ia}. IA includes both gravitational lensing-intrinsic (GI) and intrinsic-intrinsic (II) contributions, described by the following power spectra:

\begin{equation}
    C^{ij}_{\rm II}(\ell) = \int^{\chi_{\rm H}}_0 d\chi \frac{n^i(\chi)n^j(\chi)}{\chi^2}P_{\rm II} \left( k=\frac{\ell + 1/2}{\chi}, \chi \right), 
\end{equation}

\begin{equation}
    C^{ij}_{\rm GI}(\ell) = \int^{\chi_{\rm H}}_0 d\chi \frac{q^i_\kappa(\chi)n^j(\chi)}{\chi^2}P_{\rm GI} \left( k=\frac{\ell + 1/2}{\chi}, \chi \right),
\end{equation}

where

\begin{equation}
    \begin{aligned}
    P_{\rm GI}(k) &= A_1 P_{\rm m}(k) + A_{1 \delta} P_{0|0E}(k) + A_2 P_{0|E2}(k) , \\
    P_{\rm II, \,EE}(k) &= A_1^2 P_{\rm m}(k) + 2A_1 A_{1 \delta} P_{0|0E}(k) + A^2_{1\delta} P_{0E|0E}(k)\\
        &\quad + A_2^2 P_{E2|E2}(k) + 2A_1 A_2 P_{0|E2}(k) \\
        &\quad + 2A_{1 \delta} A_2 P_{0E|E2}(k) ,\\
    P_{\rm II, \,BB}(k) &= A^2_{1 \delta} P_{0B|0B}(k) + A_2^2 P_{B2|B2}(k) \\
        &\quad+ 2A_{1 \delta} A_2 P_{0B|B2}(k) .    
    \end{aligned}
\end{equation}

In this work, we evaluate $k$-dependent terms with \textsc{Fast-pt} \cite{Fang_2017, McEwen_2016} implementation in \textsc{CosmoSis} \cite{Zuntz_2015}. We refer the reader to \cite{Blazek_2019} for a detailed description of the expressions and modeling of these power spectra. 
\section{\label{sec:pipeline}Modeling Ingredients}

\begin{table*}
    \centering
    \resizebox{0.75\textwidth}{!}{%
    \begin{tabular}{llrc}
        \midrule \midrule
        Model ingredient & Baseline choice & Model misspecification & Section \\
        \midrule \midrule
        Matter power spectrum, $P_{\rm m}(k)$ & \textsc{HMCode 2020} & \textsc{EuclidEmulator2} & \ref{subsec:spectrum} \\
        Baryonic feedback & \tagn = 7.7 & [Gravity-only, \textsc{Bahamas 8.0}] & \ref{subsec:spectrum} \\
        Galaxy bias & Linear & Non-linear (\textsc{Fast-Pt}) & \ref{subsec:th-galaxy-bias} \\
        & Non-linear (\textsc{Fast-Pt}) & \makecell[r]{\textsc{Aemulus-HEFT}, \\ \textsc{Cardinal} simulation} & \ref{subsec:th-galaxy-bias} \\
        Intrinsic alignment (cosmic shear) & NLA \& TATT-4 & - & \ref{subsec:model-ia} \\
        Intrinsic alignment ($2$ and $3 \times 2$pt) & TATT-4 & - & \ref{subsec:model-ia} \\
        Lens magnification & Balrog Gaussian prior & - & \ref{subsec:model-magnification} \\
        \hline \hline
    \end{tabular}%
    }
    \caption{\label{tab:likelihood-pipeline}Summary of the modeling choices for our baseline analysis and the alternative models used to introduce contamination in the mock signal for robustness tests.}
\end{table*}

In this section, we describe the baseline modeling pipeline developed for the cosmic shear, $2\times2$pt, and $3\times2$pt analyses. The pipeline, summarized in Table~\ref{tab:likelihood-pipeline}, is implemented in \textsc{CosmoSIS} \cite{Zuntz_2015}.

\subsection{\label{subsec:spectrum}Matter Power Spectrum}

Accurate modeling of the matter power spectrum is essential for extracting cosmological information from the DES Year 6 data set. Our weak lensing data are sensitive to scales above $k \sim 0.1\,h\,{\rm Mpc}^{-1}$, extending into the non-linear regime \cite{Mandelbaum_2018, Preston2024, prat2025weakgravitationallensing}. Our analysis is therefore sensitive to a wide range of scales, requiring accurate predictions for both the linear and non-linear power spectrum. We compute the linear power spectrum with \textsc{Camb} \cite{Lewis_2000, Lewis_2002}. 

\begin{figure*}[t]
    \centering
    \includegraphics[width=0.9\linewidth]{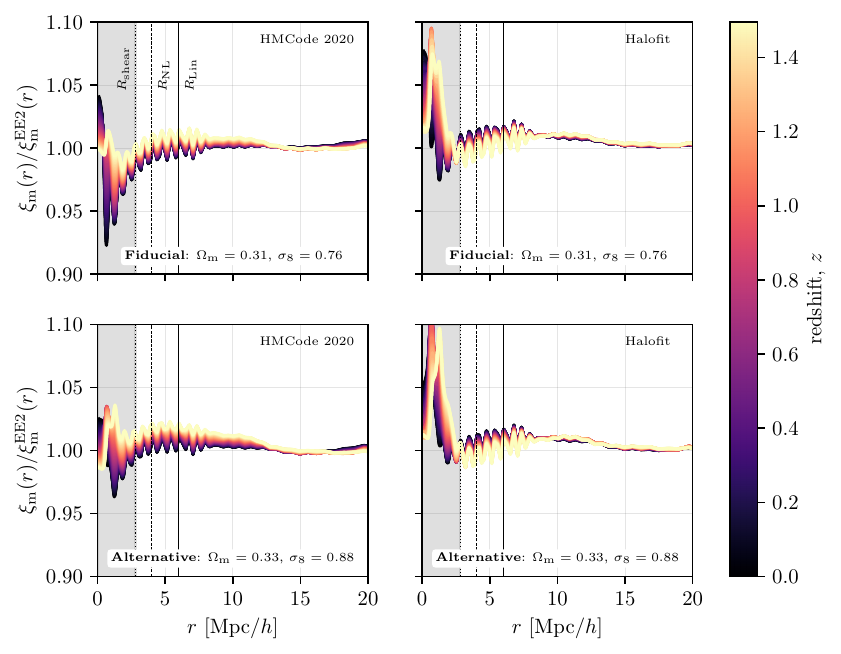}
    \caption{\label{fig:pk-ratio}Comparison of the accuracy of \textsc{Hm20} and \textsc{Hf} dark-matter-only matter correlation function predictions against the \textsc{Eemu} estimates over the redshift range \( 0 < z < 1.5 \). The comparison is shown for two different cosmological models: the fiducial cosmology adopted in this work (\( \Omega_{\rm m} = 0.31 \), \( \sigma_8 = 0.76 \)) and an alternative cosmology within the emulator range (\( \Omega_{\rm m} = 0.33 \), \( \sigma_8 = 0.88 \)). Smallest scale reached with the linear galaxy bias analysis is $R_{\rm Lin} = 6$ Mpc/$h$, while for the non-linear galaxy bias analysis $R_{\rm NL} = 4$ Mpc/$h$. For the cosmic shear analysis, the smallest scale considered is $R_{\rm shear} = 2.8$ Mpc/$h$.}
\end{figure*}

The non-linear power spectrum is shaped by two main effects: (i) the non-linear gravitational collapse of dark matter, and (ii) baryonic physics, the most dominant being baryonic feedback mechanisms, which describes energetic outflows from active galactic nuclei (AGN) and stellar feedback \cite{Chisari_2019, prat2025weakgravitationallensing}. For the dark-matter-only non-linear spectrum we compare three models: (1) the \textsc{Euclid Emulator v2} (\textsc{Eemu}, \cite{Euclid_2021}), a fast and precise emulator calibrated on suites of N-body simulations; (2) \textsc{HMCode 2020} (\textsc{Hm20}, \cite{Mead_2021}), an analytic halo-model framework tuned to hydrodynamical results; and (3) \textsc{Halofit} (\textsc{Hf}, \cite{Takahashi_2012}), a fitting formula historically adopted in DES Year 1 and Year 3 analyses. While \textsc{Eemu} achieves sub-percent accuracy, its calibration is restricted to a narrower parameter space than required for DES Year 6. \textsc{Hf}, though convenient, deviates by $5$--$10\%$ on non-linear scales and lacks baryonic extensions. By contrast, \textsc{Hm20} provides flexible modeling of baryonic and neutrino effects and agrees with \textsc{Eemu} to percent-level precision across DES scales. \textsc{Hm20} introduces the effect of baryons via the sub-grid heating parameter \tagn, indicating the strength of the AGN. Fig.~\ref{fig:pk-ratio} shows these comparisons across two cosmologies, highlighting the strong agreement of \textsc{Hm20} with the \textsc{Eemu} and the breakdown of \textsc{Hf} at small scales. We therefore adopt \textsc{Hm20} as our baseline model, and present in Appendix~\ref{app:pk-emu} the investigations on the use of an emulator which finally was not implemented in the analysis. In validating our modeling choices as described in Section~\ref{sec:cuts}, we use \textsc{Eemu} as the prescription for the dark-matter-only non-linear power spectrum in the contaminated data vector, on top of the baryon feedback as explained below.

On the scales that our data is most sensitive to, baryon feedback suppresses the matter power spectrum compared to a dark matter-only prediction. The extent and amplitude of the power suppression is highly uncertain.  
Cosmological hydrodynamical simulations admit a range of scenarios  with the amplitude of the suppression ranging from $\sim 10$--$30\%$ at $k \sim 0.1$--$10\,h\,{\rm Mpc}^{-1}$  \cite{McCarthy_2016, Schaye2023, SIMBA, FABLE, MilTNG, Siegel:2025ivd}. More recently, there have been constraints on the suppression of the matter power spectrum due to baryon feedback from analyzing gas observables, including constraints pointing to even stronger matter power spectrum suppression than fiducial simulations predict \cite{bigwood2024, Dalal2025, Pandey25, Kovac25, 2025arXiv250717742R}, which could directly bias the cosmological inference \cite{Amon2022,Preston23}.  

To account for this uncertainty, our primary strategy is to discard small-scale data where feedback has the largest impact. To minimize the loss of information, we include a moderate baryon feedback prescription in our fiducial model: \textsc{Hm20} with \tagn=7.7. \journal{Since the suppression is asymmetric, including a moderate feedback signal rather than a dark-matter-only model lets us retain slightly smaller scales. We fix this amplitude rather than marginalizing over it because, once the most feedback-sensitive scales are removed by the cuts, the data no longer meaningfully constrain \tagn; sampling it would chiefly add a poorly-constrained parameter and enhance projection effects, while the choice of prior would feed back into the scale-cut definition.} For validation, we bracket this with two alternative scenarios: a gravity-only case (minimal feedback) and the strong-feedback \textsc{Bahamas} model with \tagn=8.0 \cite{McCarthy_2016}, on top of an \textsc{Eemu} matter power spectrum in both cases. These edge-cases are used in Section~\ref{sec:cuts} to validate the robustness of our scale cuts. Baryonic modifications are applied via
\begin{equation}
    P_{\rm m}(k, z) = \frac{P_{\rm hydro}(k, z)}{P_{\rm DM}(k, z)} P_{\rm m}^{\textsc{Eemu}}(k, z),
\end{equation}
where $P_{\rm hydro}/P_{\rm DM}$ tabulates suppression across redshift and scale for each simulation.  
Thus, our fiducial model is validated against both minimal- and high-feedback scenarios, ensuring robust predictions across the full DES Year 6 parameter space. \journal{The strong-feedback bracket adopted here, $\log_{10}T_{\rm AGN}=8.0$, is moreover consistent with the suppression preferred by recent joint analyses of weak lensing and gas observables~\cite{bigwood2024, Siegel:2025ivd}, which favour feedback comparable to the high-AGN \textsc{Bahamas} model; we verify on cosmic shear analyses that more extreme scenarios induce shifts in $S_8$ of at most $\sim1.5\sigma$ under our fiducial cuts, further reduced when marginalizing over \tagn. We note that the companion cosmic shear analysis~\cite{y6-1x2pt} explicitly tests the robustness of this choice, finding negligible shifts in $S_8$ even when \tagn $\,$ is varied up to $9.0$, well beyond the range favored by current feedback constraints.}

An alternative approach to constrain cosmology by reconstructing the matter power spectrum from observations is presented in Appendix ~\ref{app:pk-reconstruction}, and ~\cite{y6-1x2pt,y6-3x2pt} will present results following this method.

\subsection{\label{subsec:th-galaxy-bias}Galaxy Bias}

\begin{table*}
     \centering
     \begin{tabular}{lcccccc|cccccc} \midrule \midrule
         \multicolumn{13}{c}{\textbf{Eulerian Perturbation theory}} \\ \midrule \midrule
         \multicolumn{1}{c}{} & \multicolumn{6}{c}{\textsc{MagLim} Year 3-like (baseline)} & \multicolumn{6}{c}{\textsc{Redmagic}-like} \\ \midrule
         Bias coefficient & $z$-bin 1 & $z$-bin 2 & $z$-bin 3 & $z$-bin 4 & $z$-bin 5 & $z$-bin 6 & $z$-bin 1 & $z$-bin 2 & $z$-bin 3 & $z$-bin 4 & $z$-bin 5 & $z$-bin 6 \\ \midrule
         Linear, $b_1$ & 1.54 & 1.81 & 1.85 & 1.76 & 1.93 & 1.9 & 1.66 & 1.82 & 1.97 & 2.11 & 2.22 & 2.32 \\
         Local quadratic, $b_2$ & 0.28 & 0.5 & 0.53 & 0.46 & 0.59 & 0.57 & 0.37 & 0.5 & 0.63 & 0.75 & 0.85 & 0.93 \\
         Quadratic tidal, $b_{s^2}$ & -0.31 & -0.46 & -0.49 & -0.43 & -0.53 & -0.51 & -0.38 & -0.47 & -0.55 & -0.63 & -0.7 & -0.75 \\
         Third-order non local, $b_{3nl}$ & 0.54 & 0.81 & 0.85 & 0.76 & 0.93 & 0.9 & 0.66 & 0.82 & 0.97 & 1.11 & 1.22 & 1.32 \\
         \midrule \midrule \multicolumn{13}{c}{\textbf{Aemulus-HEFT} (\cite{DeRose_2023}, Lagrangian)} \\ \midrule \midrule
         \multicolumn{1}{c}{} & \multicolumn{6}{c}{\textsc{MagLim} Year 3-like (baseline)} & \multicolumn{6}{c}{\textsc{RedMagic}-like} \\ \midrule \midrule
         Bias coefficient & $z$-bin 1 & $z$-bin 2 & $z$-bin 3 & $z$-bin 4 & $z$-bin 5 & $z$-bin 6 & $z$-bin 1 & $z$-bin 2 & $z$-bin 3 & $z$-bin 4 & $z$-bin 5 & $z$-bin 6 \\ \midrule
         Linear, $b_1$ & 0.54 & 0.81 & 0.85 & 0.76 & 0.93 & 0.9 & 0.66 & 0.82 & 0.97 & 1.11 & 1.22 & 1.32 \\
         Local quadratic, $b_2$ & 0.14 & 0.38 & 0.42 & 0.34 & 0.48 & 0.46 & 0.24 & 0.38 & 0.52 & 0.66 & 0.76 & 0.86 \\
         Quadratic tidal, $b_{s^2}$ & 0.02 & 0.05 & 0.05 & 0.05 & 0.05 & 0.05 & 0.04 & 0.05 & 0.05 & 0.03 & 0.0 & -0.04 \\
         Higher-derivative, $b_{\nabla^2 \delta}$ & -0.31 & -0.77 & -0.86 & -0.67 & -1.06 & -0.98 & -0.49 & -0.8 & -1.16 & -1.56 & -1.9 & -2.25 \\ 
     \end{tabular}
     \caption{\label{tab:bias-coeff}Fiducial values of the linear and higher-order galaxy bias coefficients for two galaxy samples: one representing the \textsc{MagLim} Year 3 sample~\cite{Porredon_2022}, and the other resembling the \textsc{RedMagic} sample~\cite{Pandey_2022}. Galaxy bias coefficients are derived using the Eulerian perturbation theory approach, which is the fiducial method in our analysis, as well as the Lagrangian perturbation theory treatment of galaxy bias, implemented within the Aemulus framework and its hybrid effective field theory approach.}
\end{table*} 

The galaxy density field is a biased tracer of the underlying dark matter distribution, a relationship characterized by the galaxy bias \cite{desjacques2018large}. Due to the incomplete understanding of this connection, galaxy bias remains one of the main sources of uncertainty in galaxy clustering analyses \cite{nicola2023galaxybiaseralsst}. In this work, we explore two distinct modeling strategies to describe galaxy bias: one that restricts the analysis to \textit{linear} scales, and another extending to \textit{non-linear} scales, incorporating a more complex bias model.

On large scales, the galaxy and dark matter density fields can be related through a linear bias model, characterized by a single parameter per tomographic lens bin, $b^i_1$ \cite{Barreira_2021}. On smaller scales, however, the bias becomes non-local and non-linear \cite{Somerville_2001}, requiring a more sophisticated description. We adopt an Eulerian Perturbation Theory (EPT) approach \cite{Pandey_2020}, as implemented in \textsc{Fast-pt} \cite{McEwen_2016, Fang_2017}. Perturbative bias was developed to describe quasi-linear scales. EPT enables a controlled expansion of the galaxy overdensity field in terms of the dark matter overdensity, where higher-order non-linear corrections become increasingly relevant on smaller scales. Within this setup, the perturbed galaxy-matter power spectrum is computed as

\begin{equation}
    \begin{aligned}
        P_{\rm gm}(k,z) &=  b_1 P_{\rm mm}(k,z) +  \frac{1}{2}b_2 P_{b_1 b_2}(k,z) \\
        & + \frac{1}{2}b_{s^2}P_{b_1 s^2}(k,z) + \frac{1}{2} b_{\rm 3nl} P_{b_1 b_{\rm 3nl}}(k,z) \\
        & + b_{\rm k} k^2 P_{\rm mm}(k,z),
    \end{aligned}
\end{equation}
and the galaxy power spectrum is given by
\begin{equation}
    \begin{aligned}
        P_{\rm gg}(k,z) &= b_1^2 P_{\rm mm}(k,z) + b_1 b_2 P_{b_1 b_2}(k,z) \\
        & + b_1 b_{s^2} P_{b_1 s^2} (k,z) + b_1 b_{\rm 3nl} P_{b_1 b_{\rm 3nl}}(k,z) \\
        & + \frac{1}{4} b^2_2 P_{b_2 b_2}(k,z) + \frac{1}{2}b_2 b_{s^2} P_{b_2 s^2}(k,z) \\
        & + \frac{1}{4} b^2_{s^2} P_{s^2 s^2} (k,z) + 2 b_1 b_{\rm k} k^2 P_{\rm mm}(k,z).
    \end{aligned}
    \label{eqn:galaxy-galaxy-pk}
\end{equation}
The precise range of validity of this method is subtle and depends on both the galaxy population under study and the statistical power of the survey -- it was demonstrated in \cite{Pandey_2020} to have sufficient accuracy for DES on scales above $\sim 4$ Mpc$/h$, when using the fully nonlinear matter power spectrum $P_{mm}$ for the linear bias contribution. Although not fully consistent within the EPT framework, this treatment improves the accuracy of the most significant term and captures a significant part of nonlinearity on these scales \cite{nicola2023galaxybiaseralsst}. 

Perturbation theory introduces several parameters: the local quadratic bias $b_2$, quadratic tidal bias $b_{s^2}$, third-order non-local bias $b_{3\mathrm{nl}}$, and higher derivative bias $b_k$. The kernels that appear in the power spectrum calculations, such as $P_{b_1 b_2}$, $P_{b_1 s^2}$, $P_{b_1 b_{3\mathrm{nl}}}$, $P_{b_2 s^2}$, and $P_{s^2 s^2}$, are described in \cite{Saito_2014}. This perturbation theory approach adds four degrees of freedom per lens tomographic redshift bin. For the constraining power and galaxy sample considered here~\cite{Pandey_2022}, the parameters $b_{s^2}$ and $b_{3\mathrm{nl}}$ can be fixed to their co-evolution values \cite{McDonald_2009, Saito_2014}, given by
\begin{equation}
    b_{s^2} = -\frac{4}{7}(b_1 - 1),
    \label{eqn:coevol1}
\end{equation}
and 
\begin{equation}
    b_{3\mathrm{nl}} = b_{1} - 1,
    \label{eqn:coevol2}
\end{equation}
while fixing $b_k=0$~\cite{Pandey_2020}. We take advantage of these relations for our analysis, thus adding just two free galaxy bias coefficients, $b^i_1$ and $b^i_2$, per lens redshift bin.

We also consider an alternative biasing approach based on Lagrangian Perturbation Theory (LPT) to generate mock signals in order to validate the EPT modeling. Specifically, we use the implementation of the Hybrid Effective Field Theory (HEFT) galaxy bias described in \textsc{Aemulus} \cite{DeRose_2023}, which we refer to hereafter as Aemulus-\textsc{HEFT}.

In order to obtain consistent galaxy bias injections with both approaches, and to be able to translate from one picture to the other during the validation process and generation of the contaminated signal, it is essential to establish the relationship between coefficients in EPT and LPT frameworks. Given the linear galaxy bias coefficients in the Eulerian picture (denoted with subscript \( E \)), we can translate them into the Lagrangian coefficients (denoted by subscript \( L \)) following the equations prescribed in Table~3 of \cite{Zennaro_2022}:

\begin{equation}
    b_1^L = b_1^E - 1.
\end{equation}
Higher-order coefficients are then obtained as
\begin{equation}
    b^L_2 (b^L_1) = 0.01677(b^L_1)^3 - 0.005116(b^L_1)^2 + 0.4279 b^L_1 - 0.1635,
\end{equation}
\begin{equation}
    b^L_{s^2} (b^L_1) = -0.3605\,(b^L_1)^3 + 0.5649\,(b^L_1)^2 - 0.1412\,b^L_1 - 0.01318,
\end{equation}
and 
\begin{equation}
    b^L_{\nabla^2\delta} (b^L_1) = 0.2298\,(b^L_1)^3 - 2.096\,(b^L_1)^2 + 0.7816\,b^L_1 - 0.1545.
\end{equation}

These relations in the form of polynomial fitting functions are obtained for galaxies in the \textsc{Bacco} dark matter simulations \cite{Angulo_2021}.
Table~\ref{tab:bias-coeff} shows the resulting bias parameters used in EPT and Aemulus-\textsc{HEFT} to test the validity of our modeling approaches.

\subsection{Point-mass marginalization}

The tangential shear is inherently non-local: measurements at any angular separation $\theta$ receive contributions from all smaller scales within that radius. Consequently, mismodeling the halo-matter correlation function on small scales propagates to large scales. To mitigate this, we marginalize over a \textit{point-mass} parameter \cite{MacCrann2020}, modeling the impact of unmodeled small-scale mass as equivalent to a point mass equal to the enclosed mass within the smallest scale used. While alternative methods exist for handling this non-locality, comparative studies demonstrate nearly identical performance \cite{Prat2023}. We adopt point-mass marginalization for its computational efficiency and well-established validation in previous DES analyses. This enables significantly smaller scale cuts while maintaining unbiased cosmological constraints. For example, DES Year 3 galaxy-galaxy lensing measurements were reliably modeled down to 6~Mpc$/h$, compared to the 12~Mpc$/h$ limit required in DES Year 1, despite increased precision.

The point-mass marginalization methodology models small-scale contributions as a point-mass term with characteristic $1/R^2$ (or $1/\theta^2$ in angular coordinates) scaling. This is incorporated into the tangential shear model as $\gamma_t^{ij}(\theta) = \gamma_{t,\mathrm{model}}^{ij}(\theta) + A^{ij}/\theta^2$. Rather than explicitly sampling the unknown amplitude $A^{ij}$ during parameter estimation, the method analytically marginalizes over this contamination by modifying the inverse covariance matrix to account for uncertainty in small-scale contributions. Because we assume an effectively infinitely wide prior on the point-mass amplitude, the modification can only be applied to the inverse covariance matrix, effectively removing sensitivity to all scales below a chosen threshold.

We adopt the \textit{thin lens} approximation, assuming the point-mass amplitude evolves slowly across the redshift width of each lens bin. This allows marginalization over one parameter per lens bin rather than one per lens-source pair. Under this approximation, the enclosed unmodeled mass for a given lens bin is a single value, and its varying contributions to $\gamma_t$ for different source bins within the same lens bin have relative scales that are purely geometric and can be computed analytically (see  Section~2.3.1 of \cite{Prat2023}). The thin lens approximation was validated in the DES Year 3 analysis: Fig.~4 of \cite{Prat2023} demonstrates minimal differences between the full and geometric approaches for a $2\times2$pt analysis, with even smaller differences expected for $3\times2$pt. We apply this methodology for both our linear and nonlinear bias analyses.

\subsection{\label{subsec:model-ia}Intrinsic alignment}

Intrinsic alignment of galaxy shapes with the large-scale tidal field can masquerade as a shear signal if unmodeled and bias cosmological constraints. Direct measurements of IA in spectroscopic and imaging surveys \cite[e.g.][]{Singh2015, Fortuna2021, Samuroff_2023, navarro2025,Siegel2025} are well described by the widely used Nonlinear Linear Alignment (NLA) model \cite{Bridle_2007} and the more flexible Tidal Alignment and Tidal Torquing (TATT) model \cite{Blazek_2019}. Furthermore, previous weak lensing and clustering analyses consistently find small but non-zero amplitudes of IA model parameters \cite{Abbott_2022,deskids, y3-cosmicshear, y3-shear}, with trends depending on galaxy type \cite{Samuroff2019, McCullough_2024,Wright2025}. Given the increased statistical precision of DES Year 6 data, as well as the added depth giving more high-redshift galaxies, we utilize both the TATT and NLA models, as described below, with TATT as the fiducial model for our full $3 \times 2$pt analysis.

The TATT parameterization \cite{Blazek_2019}, from the expansion in Eq.~\ref{eqn:ia-expansion}, includes contributions from \textit{linear alignment}:

\begin{equation}
    A_1(z) = -A_1 \bar{C}_1 \frac{\rho_{\rm crit}\Omega_{\rm m}}{D(z)} \left( \frac{1+z}{1+z_0}\right)^{\eta_1},
\end{equation}

\noindent as well as \textit{tidal torquing} and the impact of \textit{source density weighting}, given by:

\begin{equation}
    A_2(z) = 5A_2\bar{C}_1 \frac{\rho_{\rm crit}\Omega_{\rm m}}{D(z)^2} \left( \frac{1+z}{1+z_0}\right)^{\eta_2},
\end{equation}

\begin{equation}
    A_{1\delta}(z) = b_{ta} A_1(z),
\end{equation}

\noindent where $\bar{C}_1$ is a normalization constant fixed to $\bar{C}_1 = 5 \times 10^{-14} \, h^{-2} \textrm{Mpc}^2$ \cite{Brown_2002}, and the pivot redshift is set to $z_0 = 0.3$ (note that this differs from the value used in the DES Year 3 analysis $z_0 = 0.62$ \cite{Troxel_2018}, see Appendix~\ref{app:ia}). In the limit where $A_2, \; b_{ta} \to 0$, we recover the widely used NLA model.

We adopt different parameterizations of the IA of galaxies for the cosmic shear study and the $2 \times 2$pt and $3 \times 2$pt analyses, optimizing the constraining power of each probe and the complexity of the respective models. For the cosmic shear, we present two parallel analyses: one using the NLA model, and another employing a simplified version of TATT in which $b_{ta}=1$ (hereafter referred to as TATT-4). In Section~\ref{sec:robustness-tests}, we demonstrate on simulated data that fixing $b_{ta}$ is expected to have a minor impact on cosmological inference. For each IA model, we design optimized scale cuts (see Section~\ref{subsec:cutsprocess}). Similarly, in the $2 \times 2$pt and $3 \times 2$pt analyses, we adopt the TATT-4 parameterization.

\subsection{\label{subsec:model-magnification}Lens magnification}

We marginalize over the lens magnification coefficients per redshift bin assuming Gaussian priors (see Table~\ref{tab:priors}), derived in \cite{legnani2025} and informed by the Balrog image simulations specifically designed for DES Year~6 \cite{Anbajagane_2025}. In our analysis, we sample over the coefficient $\alpha$, defined as the slope of the cumulative number counts in Eq.~\eqref{eqn:lens-mag}. This parameter is directly related to the effective magnification bias coefficient, $C_{\rm sample}$, following the standard definition \cite{Joachimi_2010, Garcia_Fernandez_2018, Elvin_Poole_2023}:  
\begin{equation}
    C^i_{\rm sample} = C^i_{\rm flux} - 2 = 2 \left(\alpha^i - 1\right),
\end{equation}
where $C_{\rm flux}$ represents the contribution from flux magnification, while the ``$-2$'' term accounts for the dilution of number density due to the increase in solid angle.

\subsection{\label{subsec:nz-mode}Redshift calibration and multiplicative shear bias}

Redshift distributions of both the lens and source galaxy samples are calibrated with \textit{mode projection}, a novel method introduced in~\cite{DES:2025szw}. In short, the model for the $n(z)$ function of a given sample $i$ is
\begin{equation}
 n_i(z) = \bar{n}_i(z) + \sum_{j=1}^{M_i} u_{ij}\, U_{ij}(z) \;,
\end{equation}
where $\bar{n}(z)$ is the mean redshift distribution,  $M$ is the number of redshift modes $U(z)$, and $u$ are the scalar coefficients distributed according to standard Gaussians, which are being marginalized over during the inference. 
We adopt redshift modes as our fiducial parametrization because, as demonstrated in~\cite{DES:2025szw}, they provide a mathematically well-defined and complete description of redshift distribution uncertainties that is guaranteed to propagate correctly into the cosmological observable variance. In contrast, commonly used shift-and-stretch parameterizations do not have such guarantees and can fail to capture the full space of relevant redshift uncertainties, leading to biased or underestimated cosmological errors.
Source galaxy redshift distribution is calibrated with 7 modes for the 4 redshift bins. Meanwhile, the lens galaxy redshift distribution is calibrated with 3 modes per tomographic redshift bin, amounting up to 18 parameters in total. These modes and the mean-distributions are constructed from samples of($\sim 10^{3}$) $n(z)$ candidates, generated with photometric  \cite{giannini2025darkenergysurveyyear, yin2025} and clustering redshift \cite{dassignies2025} information.

The uncertainty in the shape galaxy measurement and its response to galaxy blending is calibrated with the multiplicative shear bias \cite{MacCrann_2021}. We follow the parameterization proposed in \cite{Heymans_2006, Huterer_2006} relating the measured ellipticity $e_{j}$ to the true shear $\gamma_j$ in each bin by
\begin{equation}
    e^i_j = \left(1 + m^i\right) \gamma^i_j,
\end{equation}
where $m$ is the shear bias, and assumed to be redshift and scale independent within each bin. This effect is included in the analysis by marginalizing over Gaussian priors (see Table \ref{tab:priors}) which are informed by the Balrog image simulations.

\subsection{Higher-order lensing}\label{subsec:model-higher-lensing}

The weak lensing formalism assumed in our analysis is limited to leading-order effects. As a result, known higher-order corrections—such as reduced shear, source magnification and source clustering—are not included in our modeling. Recent studies, including those from KiDS and Euclid \cite{linke2024euclidkids1000quantifyingimpact, euclidcollaboration2023euclidpreparationxxviiimodelling}, have shown that these effects are subdominant given the current observational precision and can therefore be safely neglected. Similarly, previous analyses using DES Year 3 data found consistent results, as discussed in Section V.B.5 of \cite{krause2021darkenergysurveyyear} and in Section V.C of \cite{Prat_2022}. Moreover, \cite{legnani2025} has measured the magnification parameters for the DES Year 6 \textsc{Metacalibration} sample and found they have comparable values to the DES Year 3 sample \cite{Elvin_Poole_2023}. 
\section{\label{sec:inference}Inference framework}
In this section, we present the calculation of the covariance matrix of the signal and how we estimate error bars for the 2PCFs. Then, we describe the parameter inference framework based on the Bayesian statistics formalism.

\subsection{\label{subsec:covmat}Covariance matrix}

\begin{figure*}[t]
    \centering
    \includegraphics[width=\linewidth]{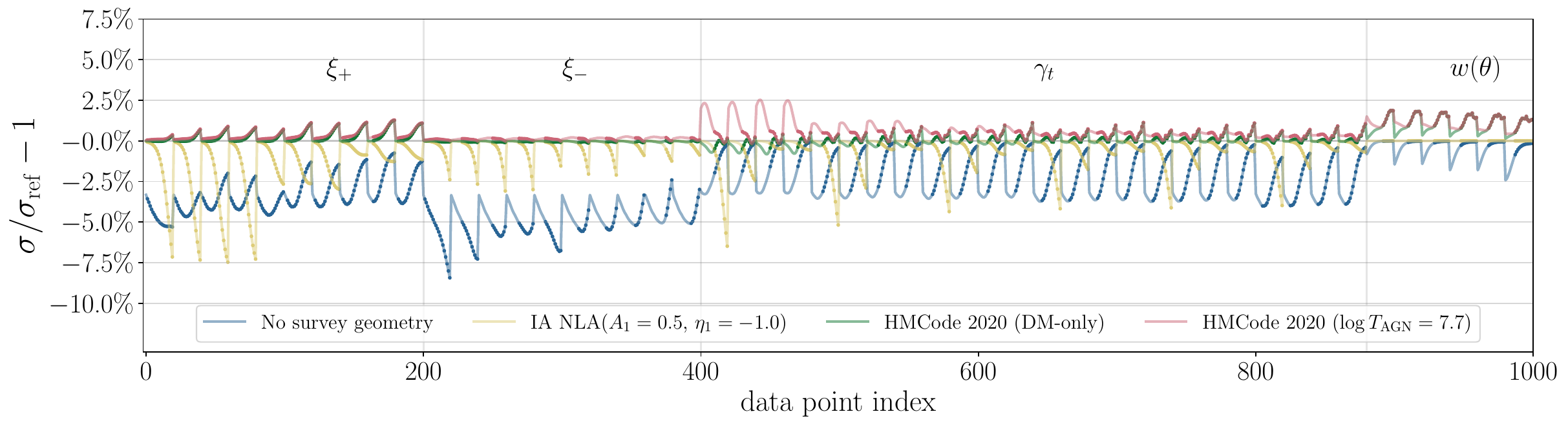}
    \caption{\label{fig:covmat-mask}Impact of different modeling assumptions on the covariance matrix, shown as the ratio of the standard deviation extracted from the modified covariance to that from the reference covariance. The reference covariance is computed using a \textsc{Halofit} dark matter-only matter power spectrum, neglecting intrinsic alignments, and including survey geometry effects. The ratio is plotted as a function of the data point index, from left to right: cosmic shear, galaxy--galaxy lensing, and galaxy clustering. Solid translucent lines display errorbars for the 1000 data points in the data vector, while circles denote the 660 points entering in the linear galaxy bias analysis after applying scale cuts.}
\end{figure*}

The covariance of the $3 \times 2$pt signal is computed analytically using \textsc{CosmoCov} \cite{Krause_2017, Fang_2020_fft}, which incorporates both Gaussian (G) and non-Gaussian contributions. The Gaussian term captures correlations on large scales, while the non-Gaussian terms are evaluated using the halo model. These include the connected non-Gaussian (cNG) term, which models the nonlinear evolution of small-scale structures, and the super-sample covariance (SSC) term \cite{2015JCAP...08..042W, 2017JCAP...06..053B, 2017JCAP...11..051B, 2018JCAP...06..015B, 2018MNRAS.479..162S}, which accounts for modes larger than the survey footprint. The SSC term reflects the impact of observing a finite region of the sky embedded in a larger-scale over- or underdensity.

\textsc{CosmoCov} employs the Limber approximation for the cosmic shear and galaxy--galaxy lensing probes, but goes beyond this approximation for modeling angular galaxy clustering at large angular separations \cite{Fang_2020}. The computation also includes corrections to shape noise and shot noise arising from the survey geometry \cite{Troxel2018}, using the angular power spectrum of the survey mask. Additionally, the covariance accounts for uncertainties introduced by marginalizing over different methods used to derive observational systematics weights, which correct for spurious clustering signals induced by varying observing conditions—such as air mass and seeing—that modulate the observed galaxy density \cite{y6-maglim,  Rodr_guez_Monroy_2022}.

The covariance estimation code was thoroughly validated for the DES Year 3 $3 \times 2$pt analysis \cite{Friedrich_2021}. In this work, we verify that the slight updates in the Year 6 assumptions have a well-understood and controlled impact on the error bar estimates.

Specifically, for the Year 6 analysis, we adopt non-zero central values for the multiplicative shear bias (see Table~\ref{tab:priors}). As a result, we update the effective number density of source galaxies \cite{yamamoto2025darkenergysurveyyear} to reflect this, as:
\begin{equation}
n_{\rm source}^i \;\longrightarrow\; n_{\rm source}^i \,
\left(1 + m^{i}\right)^2.
\end{equation}
This modification leads to the updated values reported in Section~\ref{sec:data}, namely $n_{\rm source}^i = \{ 2.54, 2.28, 2.44, 1.70 \}$ galaxies per arcmin$^2$.

Secondly, we compute the matter power spectrum used in the analytical covariance with \textsc{Hf}, assuming a gravity–only scenario, while the mock signal and baseline modeling adopt \textsc{Hm20} with baryonic feedback defined by  \tagn = 7.7. 

We quantify the impact of this mismatch by comparing the standard deviation of the data vector derived from covariances with and without these effects included. We find deviations of up to 2.5\% due to the power spectrum choice and baryonic feedback (Fig. \ref{fig:covmat-mask}).

Additionally, we assume zero contribution from IA in the covariance calculation, although the mock signal used in the validation includes a non-zero IA contribution (see fiducial IA values in Table~\ref{tab:priors}). 

To assess the impact of this choice, we activate the available IA modeling via the NLA parameterization and evaluate it at values drawn from Year 3 analysis posteriors ($A_1 = 0.5$, $\eta_1 = -1$). We find a stable and consistent impact, with the cosmic shear error bars reduced by up to $10\%$. We thus conclude that the assumption of excluding IA contributions in the covariance is conservative for the purposes of scale cut definition and validation (Fig. \ref{fig:covmat-mask}). Therefore, we can safely use the covariance both to obtain the results in this work and to obtain cosmology constraints in the Year 6 data analysis~\cite{y6-1x2pt,y6-2x2pt,y6-3x2pt}.

Finally, we assess how survey geometry affects covariance estimation. Following the DES Year 3 analysis methodology, we correct for this effect by computing the power spectra of the survey mask itself. We expect larger geometric effects in DES Year 6 because its mask contains significantly more small-scale structure than the Year 3 mask~\cite{y6-mask}. Our analysis confirms this expectation: while the ratio of mask $C_\ell$ spectra agrees well up to multipole $\ell = 100$, differences reach up to 50\% at smaller angular scales. Neglecting survey geometry in covariance calculations leads to an underestimation of cosmic shear error bars by up to 10\% (Fig.~\ref{fig:covmat-mask}). Therefore, we incorporate this correction into our fiducial covariance matrix. 

\subsection{\label{sec:likelihood} Statistical inference}
We validate and test the robustness of our analysis pipeline using simulated likelihood analyses based on mock data vector produced as described in Section \ref{sec:data} and evaluated at fiducial values in Table~\ref{tab:priors}. This mock signal is compared to theoretical predictions evaluated at model parameters $\mathbf{p}$, defined as $\mathbf{M}(\mathbf{p}) \equiv \left\{ \xi^{ij}_{\pm}(\mathbf{p}), \gamma^{ij}_t(\mathbf{p}), w^i(\mathbf{p}) \right\}$.

We assume a Gaussian likelihood of the form,

\begin{align}
    \mathcal{L}(\mathbf{D}|\mathbf{p}) \propto \exp\left(-\frac{1}{2} \left[\mathbf{D} - \mathbf{M}(\mathbf{p})\right]^T  \mathbf{Cov}^{-1} \left[\mathbf{D} - \mathbf{M}(\mathbf{p})\right]\right),
\end{align}
where $\mathbf{Cov}$ is the analytic covariance matrix of the $3 \times 2$pt signal described in Section \ref{subsec:covmat}. 

Once the likelihood is defined, we run Markov Chain Monte Carlo (MCMC) with a nested sampling algorithm to obtain the evidence and sample the posterior distributions of the model parameters. To obtain the results presented in this work, we use \textsc{Nautilus} \cite{lange2023}, a nested sampling algorithm designed for robustness and efficiency in high-dimensional parameter spaces powered by neural networks. The \textsc{Nautilus} configuration used in our analysis includes: \texttt{num\_live\_points} = 6000, \texttt{n\_batch} = 1200, and \texttt{n\_networks} = 16. See Appendix \ref{app:mcmc_samplers} for a performance comparison of the \textsc{Nautilus} sampler with \textsc{Polychord} \cite{Handley_2015}, the baseline code in the Year 3 analyses, in different scenarios and configurations.

\subsection{\label{subsec:parameters-and-priors}Parameters and priors}
In this analysis, we sample over both cosmological and nuisance parameters considering the priors in Table~\ref{tab:priors}. On the cosmological parameters front, we sample over the total matter abundance $\Omega_{\rm m}$, the reduced Hubble parameter $h$, the baryon abundance $\Omega_{\rm b}$, the scalar spectral index $n_s$, and the amplitude of the primordial power spectrum, parameterized as $A_s \times 10^9$ (see Appendix \ref{app:parametrizing-amplitude} for the justification of this choice). Additionally, we sample over the sum of neutrino masses, $m_\nu$. In the $w$CDM extension, we also vary the dark energy equation of state parameter $w$, using a flat prior within the range $[-2, -1/3]$. 

In Year 6, we revise intrinsic alignment priors relative to Year 3. We maintain flat priors on the IA amplitudes $A_1$ and $A_2$, but narrow their ranges to $[-1,3]$ and $[-3,3]$, respectively. The limit at 3 is motivated by sampling efficiency -- since existing results already constrain the quadratic amplitude in similar samples to within $A_2\sim \pm 3$ (see, for example, \cite{y3-cosmicshear} Fig. 8 and \cite{deskids}), and $A_1$ to well below this, it was deemed safe to reduce the range of the priors while remaining uninformative. Note that, in practice, these limits are well clear of the posteriors for all Year 6 probes (see the discussion in \cite{y6-1x2pt}). We additionally impose $A_1 > -1$: while blue galaxies may exhibit small negative alignments, empirical limits \cite{Samuroff_2023,Siegel2025} indicate that large negative values are implausible. For the tidal alignment–density parameter we fix $b_{\rm TA}=1$ (see Sec.~\ref{sec:robustness-tests} and \cite{y6-1x2pt} for tests of this on simulated and real data). For the redshift evolution parameters $\eta_1$ and $\eta_2$ we impose wide Gaussian priors, $\mathcal{N}(0,3)$, chosen for numerical stability. These downweight extreme values smoothly, while not excluding them completely. Together with the updated pivot redshift (Appendix~\ref{app:ia}), this mitigates projection effects in $\eta_1$ and $\eta_2$ noted in Year 3 \cite{y3-cosmicshear}. For NLA analyses we adopt the same priors on $A_1, \eta_1$.

For the galaxy bias, we sample over flat priors for the linear coefficients, in [0.8, 3]. When extending to the non-linear description, coefficients are sampled in the range [-3, 3].

We account for mild correlations among the calibration parameters of the source galaxy sample, namely the multiplicative shear bias $m$ and the redshift distribution modes $u$. These parameters are assigned Gaussian priors with non-zero means (see Table \ref{tab:priors}), with prior widths informed by the correlation matrix derived from sky image simulations \cite{Anbajagane_2025}. This correlation, introduced after the main results of this work were obtained, is further examined in Appendix~\ref{app:corrpriors}, where we show that its impact on the cosmological posteriors is negligible.

The choice of priors and parameterization, together with the marginalization over both cosmological and nuisance parameters, brings the approach of this work close to the conditions expected in the real data analysis, once the 2PCFs are unblinded. This makes the present study particularly valuable as a realistic validation step.
\section{\label{sec:data}Synthetic Data}

\begin{table*}
    \centering
        \begin{tabular}{lp{5cm}r}
            \midrule \midrule
            Parameter & Fiducial & Prior \\ 
            \midrule \midrule
            \multicolumn{3}{l}{\textbf{Cosmology}} \\
            $\Omega_{ \rm m}$ & 0.31 & [0.1, 0.6] \\ 
            $A_{\rm s} \times 10^9$ & 1.831 & [0.5, 5] \\ 
            $S_8$ & 0.77 & - \\
            $\sigma_8$ & 0.76 & - \\ 
            $h$ & 0.69 & [0.58, 0.8] \\ 
            $\Omega_{\rm b}$ & 0.051 & [0.03, 0.07] \\ 
            $n_{\rm s}$ & 0.965 & [0.93, 1.00] \\ 
            $w$ & -1 & [-2, -1/3] \\
            $m_\nu$ [eV] & 0.077 & [0.06, 0.6] \\ 
            \midrule
            \multicolumn{3}{l}{\textbf{Intrinsic alignment}} \\
            $z_{0}$ & 0.3 & - \\
            $A_1$ & 0.13 & [-1, 3]  \\
            $A_2$ & -0.2 & [-3, 3] \\
            $\eta_1$ & 2.0 & $\mathcal{N}(0.0,3.0) \in [-5,5]$ \\
            $\eta_2$ & 2.0 &  $\mathcal{N}(0.0,3.0) \in [-5,5]$ \\
            $b_{\rm TA}$ & 1.0 & fixed \\
            \midrule
            \multicolumn{3}{l}{\textbf{Lens galaxy bias}} \\
            $b^1_i (i \in [1, 6])$ & $\{1.54, 1.81, 1.85,$ $1.76, 1.93, 1.90\}$ & [0.8, 3]\\
            $b^2_i (i \in [1, 6])$ & $\{0.28, 0.50, 0.53,$ $0.46, 0.59, 0.57\}$ & [-3, 3]\\
            \midrule
            \multicolumn{3}{l}{\textbf{Baryonic feedback}} \\
            $\log_{10} T_{\rm AGN}$ & 7.7 & fixed \\
            \midrule
            \multicolumn{3}{l}{\textbf{Lens magnification}} \\
            $\alpha_1$ & 1.58 & $\mathcal{N}(1.58, 0.04)$\\
            $\alpha_2$ & 1.38 & $\mathcal{N}(1.38, 0.11)$\\
            $\alpha_3$ & 2.04 & $\mathcal{N}(2.04, 0.08)$\\
            $\alpha_4$ & 2.21 & $\mathcal{N}(2.21, 0.08)$\\
            $\alpha_5$ & 2.45 & $\mathcal{N}(2.45, 0.14)$\\
            $\alpha_6$ & 2.42 & $\mathcal{N}(2.42, 0.13)$\\
            \midrule
            \multicolumn{3}{l}{\textbf{Lens n($z$) modes}} \\
            $u^{l}_{i, m} \; (i \in [1, ..., 6], \; m \in [1,..., 3])$  & 0 & $\mathcal{N}(0, 1) \in [-3,3]$ \\
            \midrule
            \multicolumn{3}{l}{\textbf{Source n(z) modes}} \\
            $u^{s}_{m} \; (m \in [1,..., 7])$  & 
              0  &
              $\mathcal{N}(0, \, \sigma^2) \in [-3,3]$ \\
            \midrule
            \multicolumn{3}{l}{\textbf{Shear calibration}} \\
            $m_{i} \; (i \in [1,..., 4])$  & 
              $\{ -3.40, 6.46, 15.94, 1.70 \} \times 10^{-3}$ &
              $\mathcal{N}(\mu, \, \sigma^2) \in [-1,1]$ \\
            \midrule \midrule
        \end{tabular}
    \caption{\label{tab:priors} Fiducial values and priors for the cosmological, astrophysical, and calibration parameters. Parameters are sampled using either flat priors, denoted by [min, max], or Gaussian priors, following the notation $\mathcal{N}(\mu, \, \sigma^2)$ for a normal distribution with mean $\mu$ and variance $\sigma^2$. The priors on the calibration parameters of the source galaxy samples—namely the multiplicative shear bias $m$ and the source redshift distribution modes—are correlated. The width of their Gaussian priors is determined by the covariance matrix of the parameters (see Section~\ref{subsec:parameters-and-priors}).}
\end{table*}

\begin{figure*}[t]
    \centering
    \includegraphics[width=\linewidth]{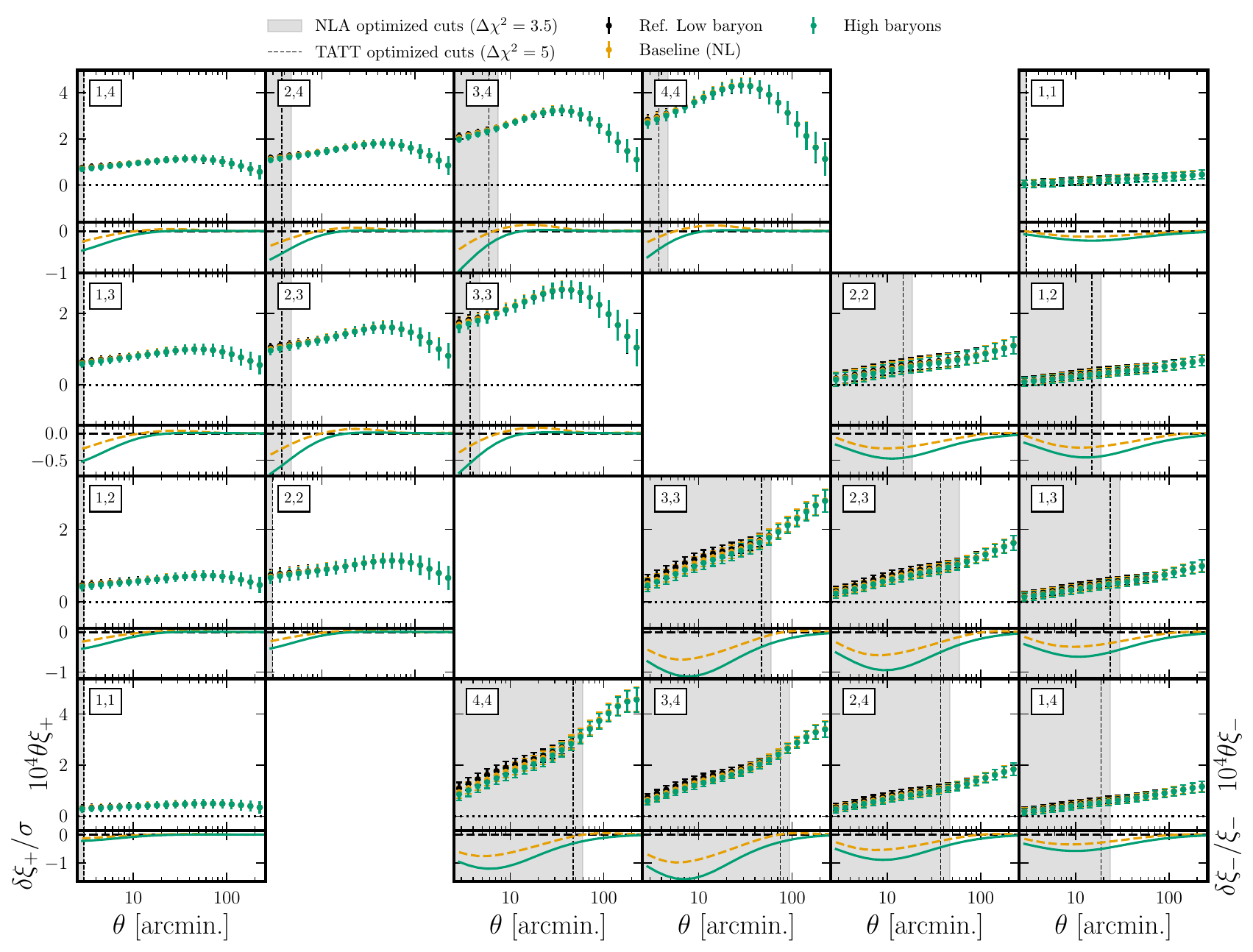}
    \caption{\label{fig:signal-shear}Cosmic shear 2PCFs measured in four redshift bins and 20 logarithmic angular bins over the range $2.5 < \theta < 250$ arcmin. The figure shows the noiseless mock signal for three scenarios: the low-baryon (dark matter only) case with \textsc{Eemu} matter power spectrum (black), the fiducial case obtained with \textsc{Hm20} and \tagn = 7.7 (orange), and the high-baryon scenario (\textsc{Bahamas 8.0}) with \textsc{Eemu} (green). The lower panels display the residuals of each signal with respect to the low-baryon case. The grey shaded region and vertical dashed lines indicate data points excluded from the analysis due to unmodeled baryonic feedback and nonlinearities in the matter power spectrum, corresponding to the optimized cosmic shear analysis under the NLA and TATT intrinsic alignment parameterizations, respectively. The pairs of indices on each panel denote the the correlated source redshift bins.}
\end{figure*}

\begin{figure*}[t]
    \centering
    \includegraphics[width=\linewidth]{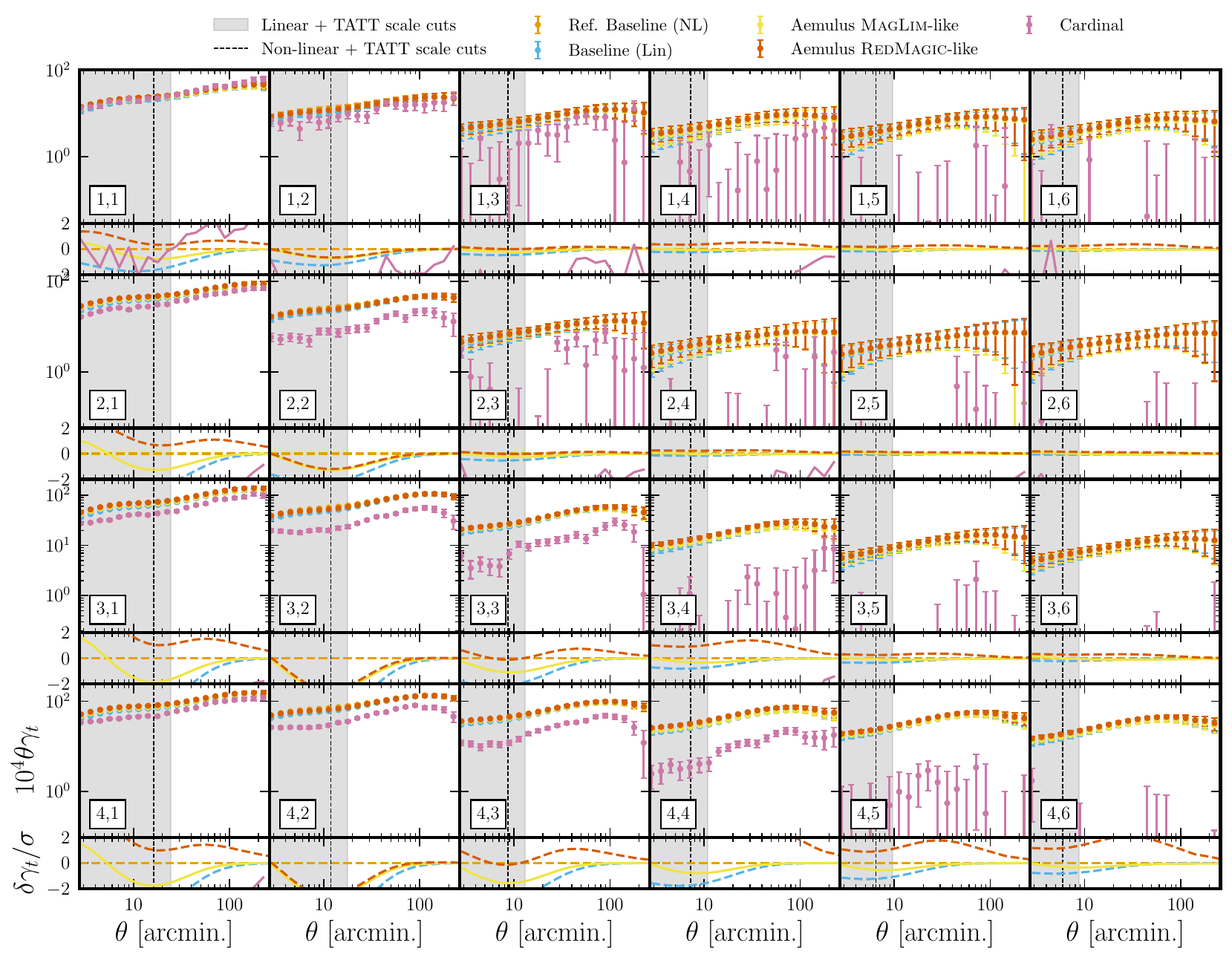}
    \caption{\label{fig:signal-gglensing}Galaxy-galaxy lensing 2PCFs estimates in four source redshift bins and six lens bins, measured in 20 logarithmic angular bins over the range $2.5 < \theta < 250$ arcmin. The figure shows the noiseless mock signals both with linear (blue) and non-linear galaxy bias (orange) contributions following the fiducial \textsc{Fast-pt} formalism. Additionally, alternative Aemulus-\textsc{Heft} galaxy bias signals mimicking the \textsc{MagLim} (yellow) and \textsc{Redmagic} (red) samples, are displayed, along with the Cardinal (magenta) measurement. The lower panels show the residuals of each signal with respect to the non-linear galaxy bias baseline case. The grey shaded region and vertical dashed line indicate data points excluded from the analysis due to unmodeled galaxy bias, baryonic feedback, and non-linearities in the matter power spectrum, corresponding to scales of 6 and 4 Mpc$/h$. The pairs of numbers within each panel denote the indices of the correlated source and lens redshift bins.}
\end{figure*}

\begin{figure*}[t]
    \centering
    \includegraphics[width=\linewidth]{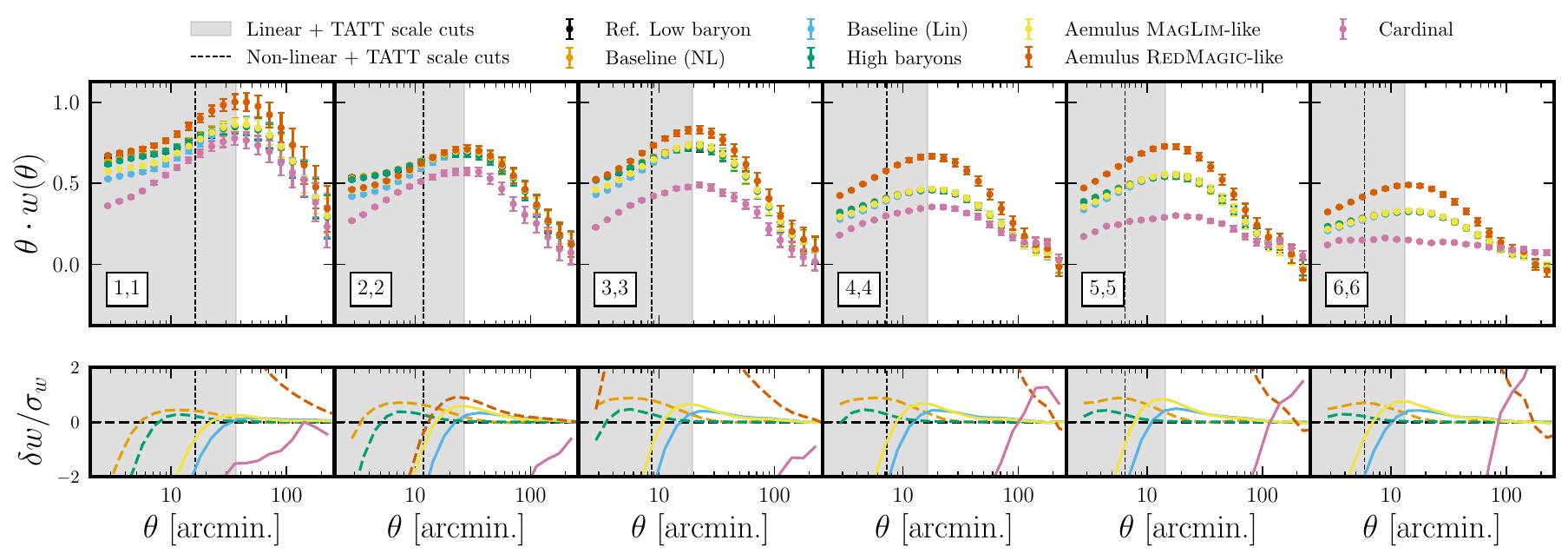}
    \caption{\label{fig:signal-wtheta}Galaxy clustering 2PCFs estimates in six redshift bins and 20 logarithmic angular bins over the range $2.5 < \theta < 250$ arcmin, displaying just the autocorrelations $w^{i}$. The figure presents the noiseless mock signal for three scenarios: the low-baryon (dark matter only) case using the \textsc{Eemu} matter power  (black), the fiducial case obtained with \textsc{Hm20} and \tagn = 7.7 (orange), and the high-baryon scenario (\textsc{Bahamas 8.0}) with \textsc{Eemu} (blue). All cases include non-linear galaxy bias contributions following the fiducial \textsc{Fast-pt} formalism. Additionally, alternative Aemulus-\textsc{heft} galaxy bias signals mimicking the \textsc{MagLim} (yellow) and \textsc{Redmagic} (red) samples, are shown alongside the Cardinal (magenta) measurement. The lower panels display the residuals of each signal with respect to the low-baryon case. The grey shaded region and vertical dashed line indicate data points excluded from the analysis due to unmodeled galaxy bias contributions, corresponding to scales of 9 and 4 Mpc$/h$. The pairs of numbers within each panel denote the indices of the correlated lens redshift bins.}
\end{figure*}

The investigations presented in this work are conducted in a blinded manner, following \cite{Muir_2020}. To prevent experimenter bias from directly examining the data, we generate synthetic noiseless 2PCFs mimicking the properties of the Year 6 source and lens galaxy samples.  The mock \textit{data vector} contains the 2PCFs, $\mathbf{D} \equiv \left\{ \xi^{ij}_{\pm}(\theta), \, \gamma^{ij}_t(\theta), \, w^i(\theta) \right\}$ generated with an angular binning of 20 log-spaced bins in the range $[2.5, 250]$ arcmin., the analytical covariance matrix of the signal, and the redshift distribution of both the source and lens galaxy samples. See Appendix~\ref{app:larger-scales} for an exploration of an alternative angular binning. This noiseless synthetic signal (Figures \ref{fig:signal-shear}, \ref{fig:signal-gglensing}, and \ref{fig:signal-wtheta}) is generated using \textsc{CosmoSis} \cite{Zuntz_2015}, and the covariance is estimated with \textsc{CosmoCov} \cite{Krause_2017, Fang_2020} (Section \ref{subsec:covmat}). Additionally, we test our framework with a noisy 2PCF measurement of the \textsc{Cardinal} mock galaxy catalog \cite{to2023buzzardcardinalimprovedmock} .

For the production of the noiseless mock signal we assume the following cosmology values and properties of the galaxy samples (similar to the ones considered in the real data analysis):

\begin{itemize}[label=-]
    \item A \textit{fiducial cosmology} that describes a realistic scenario with values drawn from DES Year 3 cosmology results \cite{Abbott_2022}, and encompassed in range of values used to train the \textsc{Euclid Emulator v2} (\textsc{Eemu}, \cite{Euclid_2021}) of the matter power spectrum. The set of chosen values are listed in Table \ref{tab:priors}.
    \item The \textsc{Metadetect} source galaxy sample \cite{yamamoto2025darkenergysurveyyear}, with a tomographic number density, $n_{\rm source}^i = \{ 2.56, 2.25, 2.37, 1.69 \}$ galaxies per arcmin$^2$, in four redshift bins spanning $0.1 < z < 2.5$, a shape noise of $\sigma_e = 0.30$, and redshift distribution obtained with \textsc{Sompz + wz} \cite{yin2025,  campos2024enhancingweaklensingredshift}. 
    \item The \textsc{MagLim} lens galaxy sample \cite{weaverdyck2025}, with a tomographic number density of lenses $n_{\rm lens}^i = \{ 0.127, 0.092, 0.097 , 0.123, 0.096 , 0.097 \}$ galaxies per arcmin$^2$, in six redshift bins spanning $0.2 < z < 1.05$, and the redshift distribution obtained with \textsc{Sompz + wz} \cite{giannini2025darkenergysurveyyear, campos2024enhancingweaklensingredshift}.
\end{itemize}

Although some redundancy exists in the analysis, as the DES provides two source and lens galaxy samples, we focus the design and validation of the theoretical framework on the \textsc{Metadetect} source sample \cite{yamamoto2025darkenergysurveyyear} and the \textsc{MagLim} lens sample \cite{Porredon_2022}. The analysis pipeline (Section \ref{sec:pipeline}) is also applicable to the Bayesian Fourier Domain source sample \cite{Bernstein_2014}, and the \textsc{RedMagic} lens sample \cite{Rodr_guez_Monroy_2022} with minimal modifications. 

\subsection{\label{subsec:data-cardinal}Cardinal}

The \textsc{Cardinal} mock galaxy catalog \cite{to2023buzzardcardinalimprovedmock}, based on N-body simulations, is an updated version of the \textsc{Buzzard} mock catalog suite \cite{derose2019buzzardflockdarkenergy}, developed to support both current and future cosmological surveys. This synthetic catalog is based on a one-quarter-sky simulation extending to redshift $z=2.35$, with an $r$-band limiting magnitude of $m_r = 27.0$. It is available only as a single realization. The underlying cosmology assumes the following parameters: $\Omega_{\rm m} = 0.286$, $\Omega_{\rm b} = 0.047$, $\sigma_8 = 0.82$, $n_{\rm s} = 0.96$, $h = 0.7$, and three massless neutrino species with an effective number of relativistic species $N_{\rm eff} = 3.046$. \textsc{Cardinal} improves upon its previous iteration by implementing an updated sub-halo abundance matching model for galaxy–halo assignment \cite{Vale_2004, conroy2006modeling}, as well as a novel method for assigning galaxy colors. As a result, it serves as a valuable validation tool, accurately reproducing the observed properties of DES galaxy samples.

\section{\label{sec:cuts}Scale cuts}

In this section we describe the procedure to define scale cuts that mitigate the impact of poor modeling due to an incomplete description of theoretical systematics at small scales, such as baryonic feedback, the non-linear matter power spectrum, and galaxy bias.

\subsection{\label{subsec:cutsprocess}Procedure}

\begin{figure*}
    \centering
    \includegraphics[width=0.9\linewidth]{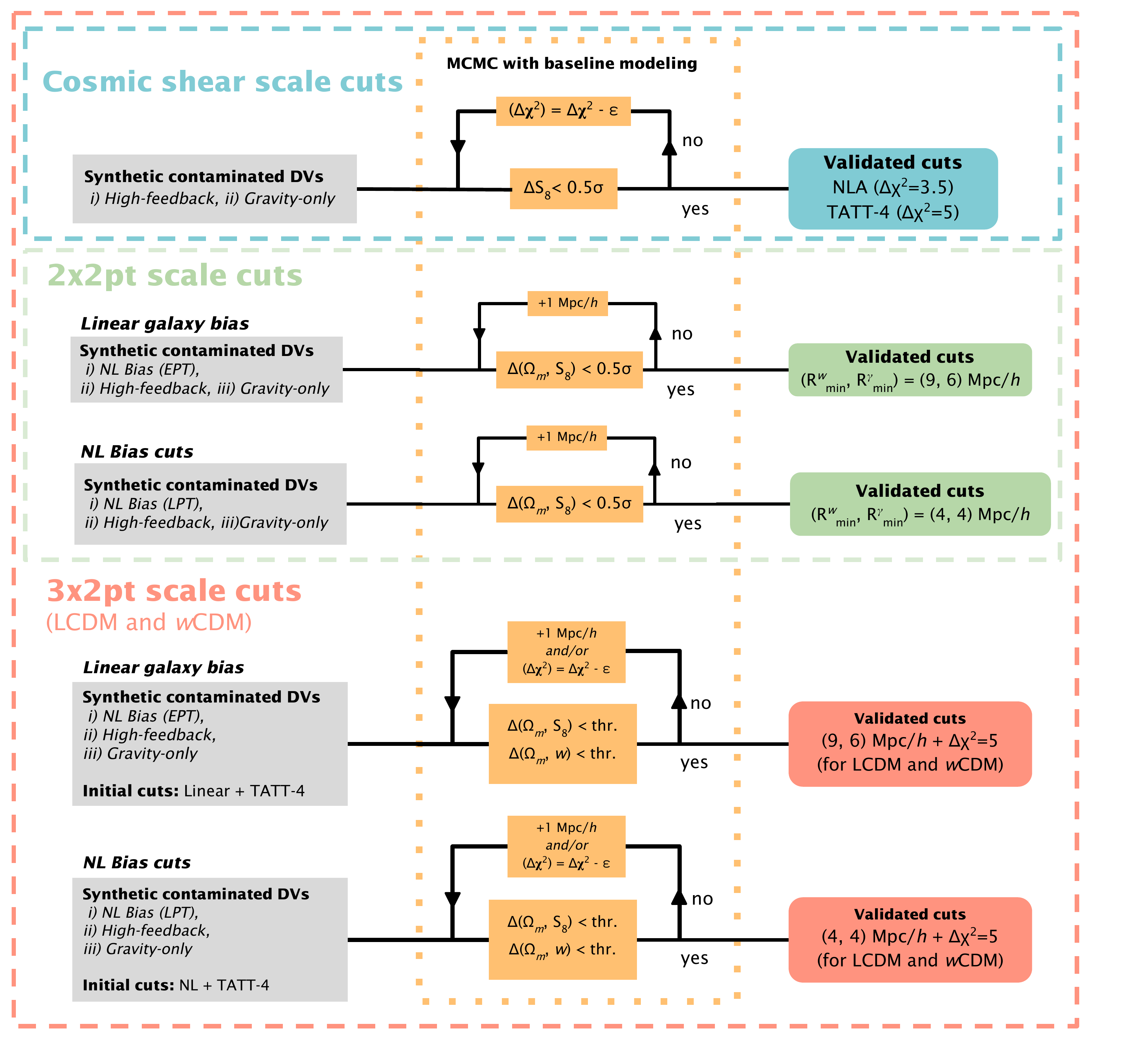}
    \caption{\label{fig:flow-shear-cuts}Scale cuts decision process. \textit{Cosmic shear cuts}: the procedure begins with an initial $\Delta\chi^2$ computed between the baseline signal and two contaminated cases: the high-feedback and gravity-only scenarios. Both contaminated scenarios incorporate \textsc{Eemu} matter power spectrum. The scale cut is set to the largest of the minimum scales obtained from these two cases. An MCMC chain is then run to evaluate the posterior shift in $S_8$ relative to the known input value. If $\Delta S_8 < 0.5\sigma$, the scale cut is accepted; otherwise, $\Delta\chi^2$ is incremented by $\epsilon$, and the process is repeated until convergence is achieved. We explore $\Delta\chi^2$ in the range [0.1, 5]. \textit{$2 \times 2$pt} cuts: we derive cuts for the linear and non-linear galaxy bias analyses, considering as contaminated data vectors those containing non-linear galaxy bias inject following EPT and LPT descriptions (Section~\ref{subsec:th-galaxy-bias}), and also high-feedback and gravity-only scenarios. Scale cuts are determined by the smallest physical scale for which the bias model remains valid, denoted as \((R_{\text{min},\, w(\theta)},\, R_{\text{min},\, \gamma_t})\). Validation criteria is defined as the shift induced in $\Delta(\Omega_m, S_8)$ due to unmitigated contamination. If the criteria is not met, we remove more scales in steps of $1$ Mpc$/h$. \textit{$3 \times 2$pt cuts}: we derive scale cuts for the linear and non-linear analyses in $\Lambda$CDM and $w$CDM. For the latter, we also add the condition on the shift in the plane of $\Omega_m -w$. The starting point is the union of cosmic shear and $2 \times 2$pt cuts. In the case cuts fail to mitigate the contamination, we remove cosmic shear data points with a more restrictive $\Delta \chi^2$ or points from $2 \time 2$pt in steps of $1$ Mpc$/h$. The MCMC runs required to apply this validation scheme are highly computationally demanding, totaling 1.75 million CPU-hours for validating cosmic shear, as well as the linear and non-linear cuts.}
\end{figure*}

To derive a set of scale cuts for our joint-probe analysis, we follow an iterative procedure that builds upon that of the Year~3 analysis \cite{Abbott_2022}. We first generate two mock $3\times2$pt data vectors designed to bracket the uncertainty in modeling the non-linear matter power spectrum, motivated in Sec.~\ref{subsec:spectrum}: (1) a \textit{gravity–only} scenario using \textsc{Eemu}, and (2) a \textit{high-feedback} scenario (\textsc{Bahamas-8.0}). These correspond to the cases plotted in Figs.~\ref{fig:signal-shear}--\ref{fig:signal-wtheta}.

We first obtain scale cuts for cosmic shear only and 2$\times$2pt analyses independently. Our shear scale cuts are defined using these two data vectors, by finding a minimum scale $\theta_{\rm min}$ in each bin pair $ij$ that maintains:
\begin{multline}
\textrm{max} \left[ \chi^2(\xi^{ij}_{\pm, \rm base}, \xi^{ij}_{\pm, \rm max}),
\chi^2(\xi^{ij}_{\pm, \rm base}, \xi^{ij}_{\pm, \rm min} )  \right] 
\\ < \Delta \chi^2, 
\end{multline}

\noindent
where we set $\Delta \chi^2$ based on the level of bias it  induces in the cosmological parameter estimates. We make an initial guess at $\Delta \chi^2$, run $1\times2$pt analyses of our two contaminated data vectors assuming those cuts, and evaluate the bias relative to our baseline data vector (described in Section~\ref{sec:data}, evaluated at fiducial values in Table~\ref{tab:priors}). Iteratively selecting $\Delta \chi^2$ values then allows us to find a set of scale cuts just below, but not exceeding, our predetermined error thresholds on $S_8$ and $S_8-\Omega_{\rm m}$. A visual representation of this algorithm can be seen in Fig. \ref{fig:flow-shear-cuts}.

We next need an initial guess at $2\times2$pt scale cuts. For this, we take the scale cuts determined in DES Year 3 \cite{Pandey_2020} for the linear and non-linear galaxy bias analyses. These are $(4,4)~\mathrm{Mpc}/h$ (for $w$ and $\gamma_t$, respectively) for non-linear bias, and $(8,6)~\mathrm{Mpc}/h$ for linear bias. These cuts are tested by running synthetic $2\times2$pt analyses on the same two baryon + emulator contaminated data vectors described above. Additionally, we run a third test analyzing data that includes galaxy bias contributions not captured by our baseline model. To validate the linear galaxy bias cuts we analyse a signal contaminated with non-linear galaxy bias injected with \textsc{Fast-pt}. For the validation of the non-linear galaxy bias cuts, we produce synthetic signals mimicking a \textsc{MagLim}-like and \textsc{Redmagic}-like samples following the Aemulus-\textsc{Heft} prescription (see Section~\ref{subsec:th-galaxy-bias}), with the galaxy bias coefficients in Table \ref{tab:bias-coeff}. If posteriors fail our validation thresholds, we progressively tighten scale cuts by 1~Mpc$/h$ until both linear and non-linear galaxy bias models pass validation. The choice of whether to cut $w$ or $\gamma_t$ data points at each step is guided by evaluating their respective impacts on the total $\chi^2$. \journal{Note that, unlike the cosmic shear cuts, which are tuned via the $\Delta\chi^2$ criterion until the validation metric is saturated, the 2\,$\times$\,2pt and 3\,$\times$\,2pt cuts cannot be pushed to saturate their thresholds, since this would require including scales below the validity of the galaxy bias model. The corresponding validation metrics therefore lie notably below the predefined thresholds (Table~\ref{tab:shifts}), and the constraining power of the non-linear bias parameterization is not fully exploited. Recovering additional small-scale data points would in any case not guarantee improved constraints, given the interplay with projection effects (Fig.~\ref{fig:shift-projeff}).}

We begin our $3\times2$pt analysis by taking the union of scale cuts identified separately for shear and $2\times2$pt analyses. This combined cut serves as our initial guess, which we then validate against contamination from the baryon plus non-linear matter power spectrum and galaxy bias. However, the additional constraining power of the full $3\times2$pt signal means that cuts passing individual probe tests are not guaranteed to remain valid in the combined analysis. If our initial cuts would fail these validation tests, we would follow the same procedure as in the $2\times2$pt case: selecting which probe to cut based on their contribution to the total $\chi^2$.

It should be noted that this entire iterative process is significantly model-dependent. Rather than limiting all probes to the most restrictive case, we derive separate cuts tailored to each probe's specific systematic uncertainties and theoretical limitations. The different cases we consider are:

\begin{itemize}
    \item Cosmic shear $\Lambda$CDM, for i) NLA, and ii) TATT-4
    \item Linear bias, for i) $2\times2$pt $\Lambda$CDM, and ii) $3\times2$pt $w/\Lambda$CDM
    \item Non-linear bias, for i) $2\times2$pt $\Lambda$CDM, and ii) $3\times2$pt $w/\Lambda$CDM
\end{itemize}

\noindent
Since we do not meaningfully constrain the $w$CDM model space with shear and $2\times2$pt separately, we do not derive separate cuts for those cases. By default, we use the TATT model for all $2\times2$pt and $3\times2$pt analyses, and so again we do not derive NLA cuts for anything but cosmic shear. 

\subsection{\label{subsec:validation-metric}Validation metrics}

\begin{figure}
    \centering
    \includegraphics[width=\linewidth]{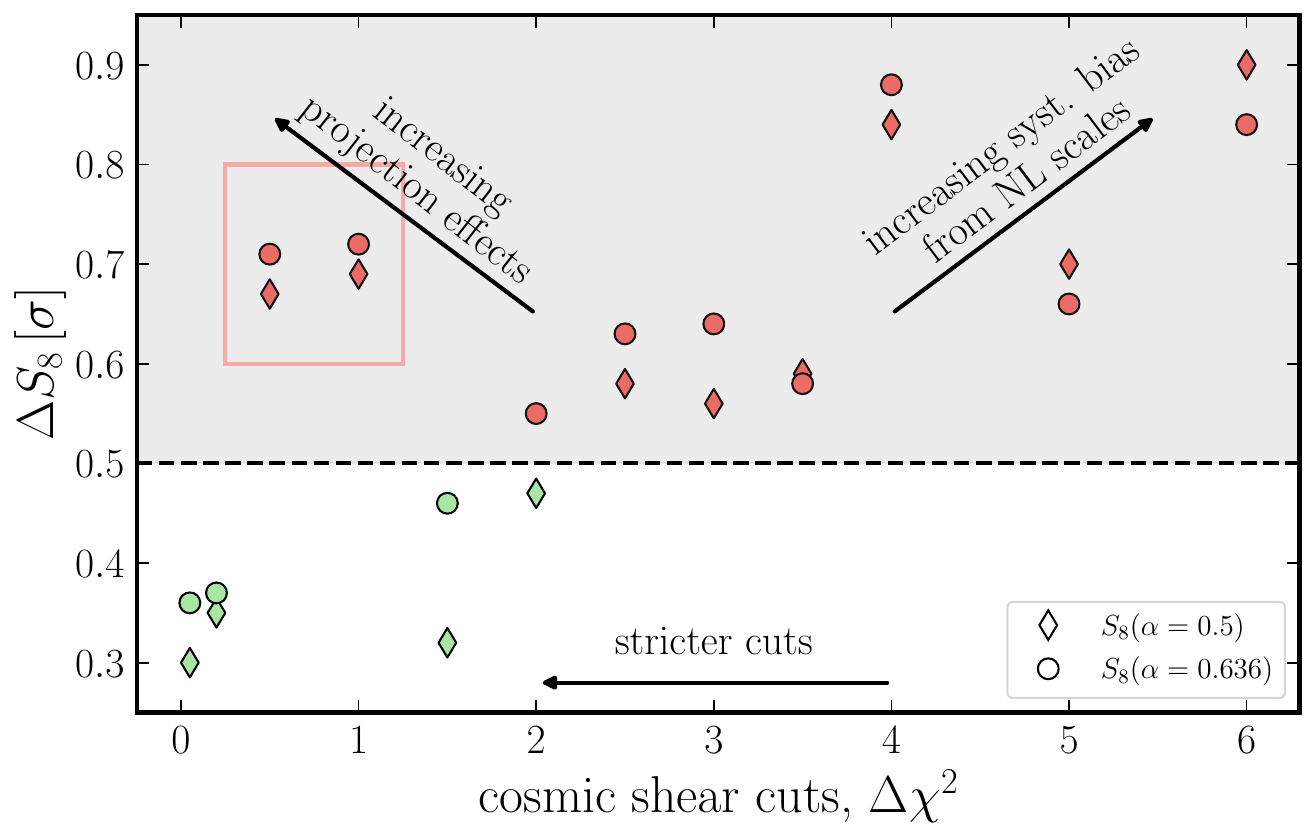}
    \caption{\label{fig:shift-projeff}Scale cuts and projection effects interplay in the $3 \times 2$pt $\Lambda$CDM analysis, with linear galaxy bias. The plot shows posterior shifts $\Delta S_8$ (in standard deviations) between baseline and baryonic feedback plus \textsc{Eemu}-contaminated data vectors, the difference between signals is given by the $\Delta \chi^2$. Scale cuts for galaxy clustering and galaxy-shear correlations are held fixed at $(9, 6)$ Mpc$/h$ while cosmic shear cuts are varied through $\Delta \chi^2 \in [0.2, 6]$. Diamonds show standard $S_8$, circles show the best-constrained direction $S_8(\alpha = 0.636)$ from Eq.~(\ref{eq:s8_alpha}) with color red and green displaying whether those cuts pass or not the validation metric. Because of projection/volume effects, stricter scale cuts (smaller $\Delta \chi^2$) do not systematically reduce parameter shifts even for the most-constrained direction, as can be seen in the cases highlighted inside the red rectangle.}
\end{figure}

\begin{figure}[t]
  \centering
  \includegraphics[width=\columnwidth]{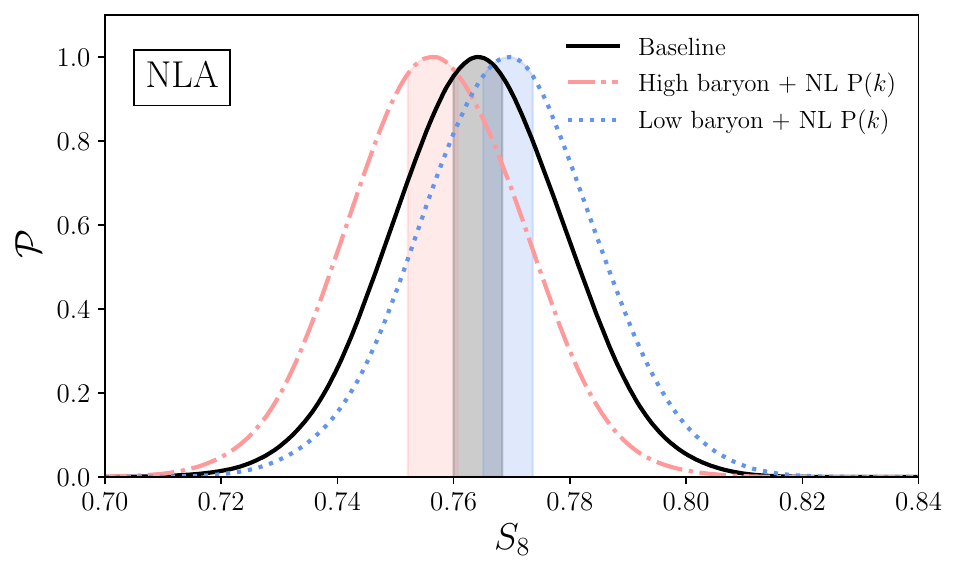}
  
  \vspace{0.5em} 

  \includegraphics[width=\columnwidth]{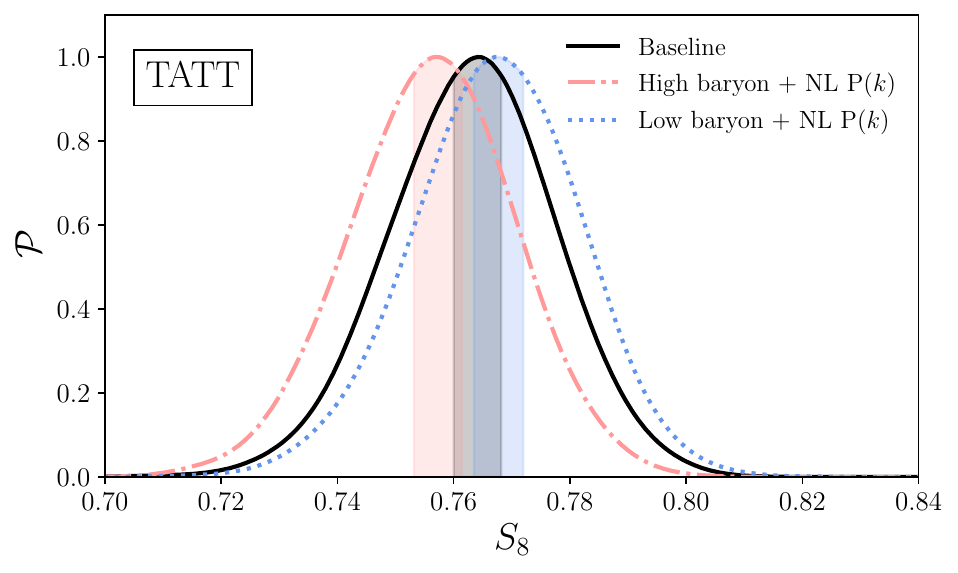}
  
  \caption{\label{fig:shear_cuts_validation}Posterior distributions of $S_8$ from the cosmic shear scale cuts validation, shown for the NLA (top) and TATT (bottom) intrinsic alignment models. The gravity-only case consists of a gravity-only power spectrum modeled with the \textsc{Eemu}, while the high-baryon scenario includes baryonic suppression following \textsc{Bahamas-8.0} and also includes the same modulations in the power spectrum. The shaded regions indicates the $0.5\sigma$ area.}
\end{figure}

We now present the criteria used to assess the efficiency of scale cuts in mitigating various sources of contamination. Validation thresholds are defined for the best-constrained parameters of each analysis and depend on whether a given systematic is classified as a major or minor contaminant, based on its impact on a specific probe. Different threshold values are assigned accordingly. The validation metrics and bias thresholds are motivated by the requirement that the combined effect of all systematics remains below $1\sigma$. For minor systematics, we define a $0.3\sigma$ threshold, such that the cumulative effect of ten independent systematics would produce at most a $1\sigma$ shift in the $(\Omega_{\rm m}, S_8)$ plane when added in quadrature. For major systematics, as well as for cases where systematic effects are expected to be coupled (e.g., the modeling of baryonic feedback and nonlinearities in the matter power spectrum), we adopt a looser threshold of $0.5\sigma$. With these considerations, we impose:

\begin{itemize}
    \item For the cosmic shear analysis, the validation metric is the bias in the marginal mean of $S_8$, as we find this to be the most stringent criteria. We observe that if the test passes based on the shift in the 1D posterior distribution, it will also pass in the 2D posterior $(\Omega_{\rm m}, S_8)$. To validate the scale cuts for mitigating baryonic feedback and the nonlinear evolution of the matter power spectrum, we set the criterion $\Delta(S_8) < 0.5\sigma$.
    
    \item For the $2 \times 2$pt analysis, the validation metric is the bias in the marginal mean of the 2D posterior $(\Omega_{\rm m}, S_8)$. We consider galaxy bias to be a major source of contamination in this analysis, whereas baryonic feedback and the non-linear evolution of the matter power spectrum are considered minor sources of contamination. Consequently, we set the thresholds for galaxy bias contamination mitigation tests to be $\Delta(\Omega_{\rm m}, S_8) < 0.5\sigma$. For baryonic feedback and matter power spectrum modeling, the criteria is $\Delta(\Omega_{\rm m}, S_8) < 0.3\sigma$. 

    \item For the $3 \times 2$pt $\Lambda$CDM analysis, both baryonic feedback plus nonlinear evolution of the matter power spectrum and galaxy bias are considered major sources of contamination. Thus, the validation criteria are $\Delta(\Omega_{\rm m}, S_8) < 0.5\sigma$ for both sources of contamination. For the $3 \times 2$pt $w$CDM analysis, we impose that the shift due to baryons plus nonlinear evolution of the matter power spectrum is $\Delta(\Omega_{\rm m}, S_8) < 0.3\sigma$, and $\Delta(\Omega_{\rm m}, w) < 0.3\sigma$, adding the requirement on the dark energy equation of state space. For the galaxy bias contamination, requirements are $\Delta(\Omega_{\rm m}, S_8) < 0.5\sigma$ and $\Delta(\Omega_{\rm m}, w) < 0.5\sigma$.
\end{itemize}

\begin{figure*}[t]
    \centering
    \includegraphics[width=\linewidth]{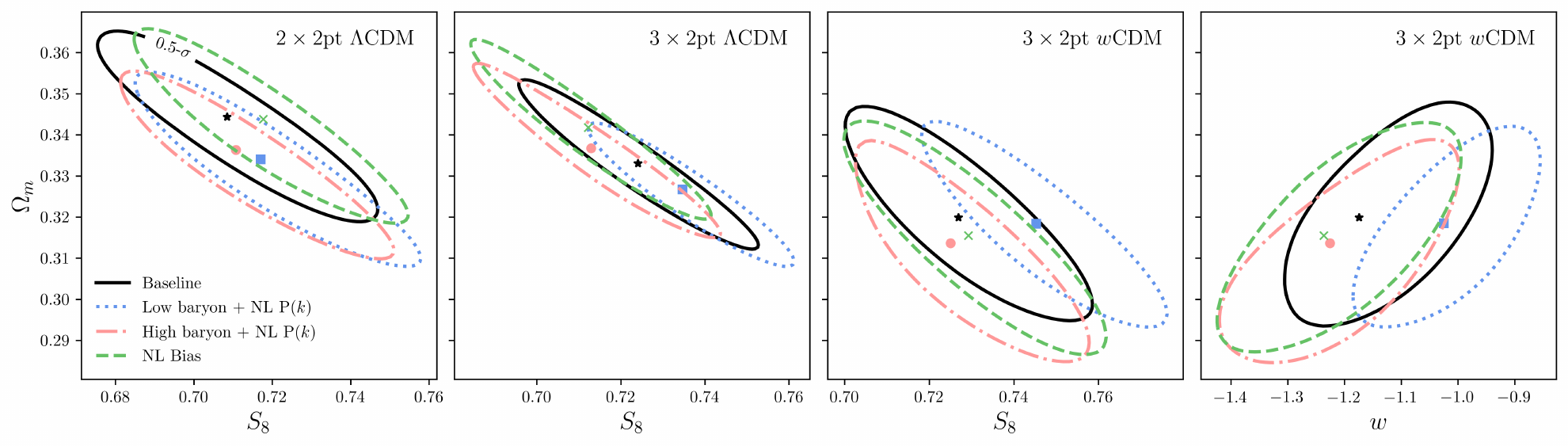}
    \vspace{1em} 
    \includegraphics[width=\linewidth]{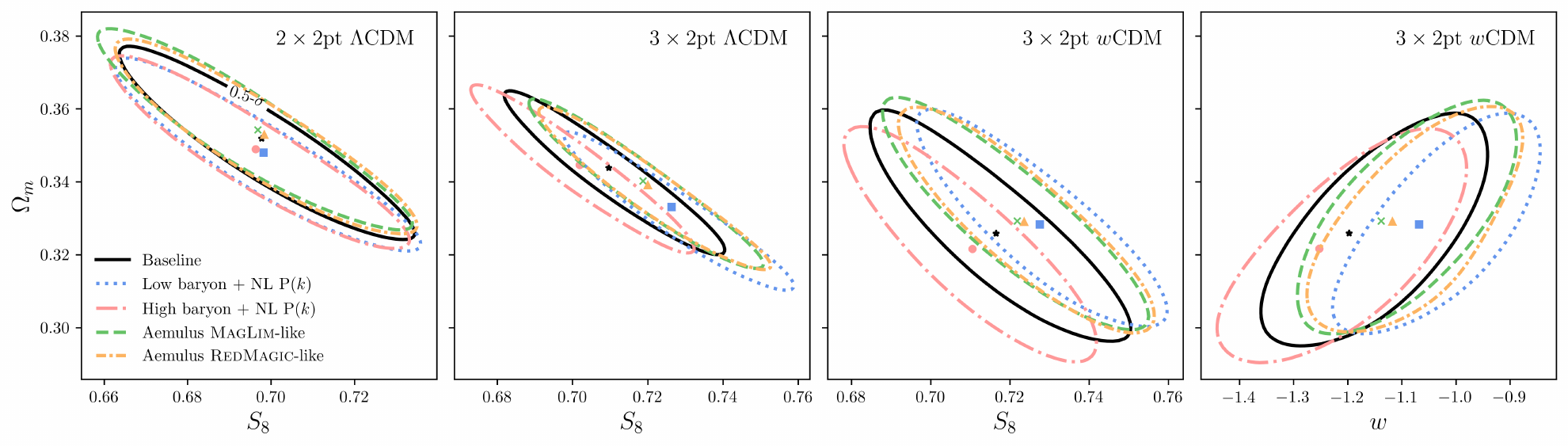}
    \caption{\label{fig:validation-cuts}
    Robustness against baryonic effects, non-linear power spectrum and galaxy bias modeling: $0.5\sigma$ confidence levels contours for the $2 \times 2$pt and $3 \times 2$pt $w/\Lambda$CDM analyses for different contaminated datavectors in comparison with baseline scale cuts we have determined in each case, presented in Table~\ref{tab:cuts-npoints}. All cases show shifts below our established 0.5$\sigma$ criterion. The top panel assumes linear galaxy bias, while the bottom panel  non-linear galaxy bias, and markers indicate posterior means.}
\end{figure*}

\journal{We note that baryonic feedback and the non-linear matter power spectrum are always injected jointly in our contaminated data vectors. A fully combined test additionally including galaxy bias is not pursued here, as it would require a self-consistent joint model of all three effects—most robustly obtained from hydrodynamical simulations—which we leave to future work; since baryonic feedback is a subdominant contaminant where galaxy bias dominates, we expect their residual coupling to be second-order.}

So-called projection or prior volume effects can result in marginal posteriors that are shifted away from the fixed truth (see~\ref{subsec:forecast}) in a potentially unintuitive way. This is not a bias in the inference, as variation over the full prior range would show the truth lying within an $X$ per cent credible interval $X$ per cent of the time~\cite{DES:2024xij, DES:2024jgw, DES:2025akz}. This effect tends to become more pronounced when the data is less constraining and, as a result, shifts can arise both when scale cuts are insufficiently strict, due to the assumed model not matching the true data generating process, or when scale cuts are too strict, due to increased projection effects from the loss of constraining power.

The magnitude of these effects depends on the specific signal features and contamination levels even in noiseless cases, creating a complex interplay between constraining power, systematic shifts, and projection/volume effects. While cosmic shear analysis is less affected by these projection effects due to its reduced parameter space, the multi-probe analyses present more complex constraints due to degeneracies with galaxy bias and $\sigma_8$ (see Sec~\ref{sec:error_budget} where we show that fixing galaxy bias is the most effective way to remove projection effects).

Due to these complications, we do not use $\Delta S_8$ as a validation metric for $2\times2$pt and $3\times2$pt analyses, as projection effects interfere with our scale cut decision-making process. The $2\times2$pt analysis constrains matter abundance more tightly than $S_8$, while $3\times2$pt analysis constrains both $\sigma_8$ and $\Omega_{\rm m}$ at similar levels, making the 2D $\Omega_{\rm m}$--$S_8$ plane more appropriate for validation. The counter-intuitive behavior where cutting more scales increases the shift can be seen in Fig.~\ref{fig:shift-projeff} for $S_8$ for the $3\times2$pt analysis,with the two problematic cases enclosed in the red rectangle. Consequently, we must carefully balance the trade-off between projection effects and systematic bias: including smaller scales increases constraining power and can reduce projection effects, but at the cost of higher sensitivity to theoretical systematics, while excessive scale cutting can make posteriors more susceptible to projection effects. 
In Fig.~\ref{fig:shift-projeff} we also show the shifts for the best-constrained direction in our fiducial case, an alternative definition of $S_8$ with exponent $\alpha = 0.636$, instead of the usual 0.5 (see Eq.~(\ref{eq:s8_alpha})). While this alternative parameterization generally shows smaller and more stable shifts compared to the standard $S_8$ (even while having less uncertainty), it still exhibits sensitivity to projection effects, with our fiducial configuration ($\Delta\chi^2 = 5$) producing shifts around $0.65$--$0.7\sigma$. This demonstrates that even the best-constrained parameter direction cannot fully eliminate projection artifacts in our inference setup, reaffirming our approach of using 2D validation metrics for multi-probe analyses.

Additionally, we note that we found the projection effects to be smaller in the real data cosmological analyses ~\cite{y6-1x2pt, y6-2x2pt, y6-3x2pt}, because of differences between the cosmology chosen in our simulated analysis and the real cosmology. We thus are confident our scale cuts choice remain robust for the level of projection effects of the real data case. 

\subsection{\label{subsec:cuts-results}Scale cuts results}

\begin{table*}
    \centering
    \begin{tabular}{lllcc}
        \midrule \midrule
        \multicolumn{5}{c}{\textbf{Cosmic shear}} \\
        \midrule \midrule
        Case & Scale cuts & Contamination & $\Delta(S_8)$ \\ 
        \midrule
        \multirow{4}{*}{$1 \times 2$pt} & NLA $\Delta \chi^2=3.5$ & High baryon + NL $P(k)$ & 0.50\\  
         & NLA $\Delta \chi^2=3.5$ & Low baryon + NL $P(k)$ & 0.33 \\
         & TATT-4 $\Delta \chi^2=5$ & High baryon + NL $P(k)$ & 0.49 \\  
         & TATT-4 $\Delta \chi^2=5$ & Low baryon + NL $P(k)$ & 0.27 \\
        \midrule \midrule
        \multicolumn{5}{c}{\textbf{Linear galaxy bias}} \\
        \midrule \midrule
        Case & Scale cuts & Contamination & $\Delta(\Omega_{\rm m}, S_8)$ & $\Delta(S_8^{\alpha=0.636})$\\
        \midrule
        \multirow{3}{*}{$2 \times 2$pt} & \multirow{3}{*}{(9, 6) Mpc/$h$} & NL Galaxy bias & 0.17 & 0.51\\ 
         &  & High baryon + NL $P(k)$ & 0.19 & 0.52 \\
         &  & Low baryon + NL $P(k)$ &  0.12 & 0.34\\
         \midrule
        \multirow{3}{*}{$3 \times 2$pt $\Lambda$CDM} & \multirow{3}{*}{(9, 6) Mpc/$h$ + $\Delta \chi^2=5$} & NL Galaxy bias & 0.13 & 0.03\\
         & & High baryon + NL $P(k)$ & 0.30 & 0.66 \\
         & & Low baryon + NL $P(k)$ & 0.10 & 0.22 \\
         \midrule
         &  &  & $\Delta(\Omega_{\rm m}, w)$ & $\Delta(\Omega_{\rm m}, S_8)$ \\
         \midrule
        \multirow{3}{*}{$3 \times 2$pt $w$CDM} & \multirow{3}{*}{(9, 6) Mpc/$h$ + $\Delta \chi^2=5$} & NL Galaxy bias & 0.37 & 0.10\\
         & & High baryon + NL $P(k)$ & 0.24 & 0.28\\
         & & Low baryon + NL $P(k)$ & 0.21 & 0.35\\        
        \midrule \midrule
        \multicolumn{5}{c}{\textbf{Non-linear galaxy bias}} \\ 
        \midrule \midrule
        &  &  & $\Delta(\Omega_{\rm m}, S_8)$ & $\Delta(S_8^{\alpha=0.636})$\\
        \midrule
        \multirow{4}{*}{$2 \times 2$pt} & \multirow{4}{*}{(4, 4) Mpc/$h$} & Aemulus \textsc{MagLim} & 0.015 & 0.15  \\
         & & Aemulus \textsc{RedMagic} & 0.001 & 0.01  \\
         & & High baryon + NL $P(k)$ & 0.14 & 0.46 \\
         & & Low baryon + NL $P(k)$ & 0.19 & 0.52 \\
        \midrule
        \multirow{4}{*}{$3 \times 2$pt $\Lambda$CDM} & \multirow{4}{*}{(4, 4) Mpc/$h$ + $\Delta \chi^2=5$} &  Aemulus \textsc{MagLim} & 0.14 & 0.46 \\
         & & Aemulus \textsc{RedMagic} & 0.16 & 0.46 \\
         & & High baryon + NL $P(k)$ & 0.31 & 0.72 \\
         & & Low baryon + NL $P(k)$ & 0.14 & 0.31 \\
         \midrule
        &  &  & $\Delta(\Omega_{\rm m}, w)$ & $\Delta(\Omega_{\rm m}, S_8)$ \\
        \midrule
        \multirow{4}{*}{$3 \times 2$pt $w$CDM} & \multirow{4}{*}{(4, 4) Mpc/$h$ + $\Delta \chi^2=5$} & Aemulus \textsc{MagLim} & 0.04 & 0.09 \\
         &  & Aemulus \textsc{RedMagic} & 0.02 & 0.14 \\
         & & High baryon + NL $P(k)$ & 0.24 & 0.31 \\
         & & Low baryon + NL $P(k)$ & 0.07 & 0.29 \\
        \midrule \midrule 
    \end{tabular}%
    \caption{\label{tab:shifts}Summary of the scale cuts. Each entry reports the difference between the posterior derived from the contaminated signal and that from the fiducial (baseline) signal, expressed in units of the posterior standard deviation. Contamination scenarios are described in Section~\ref{subsec:cutsprocess}, and validation criteria are defined in Section~\ref{subsec:validation-metric}. For $\Lambda$CDM, we additionally report the shift along the best-constrained parameter direction $S_8(\alpha)$, which is given by $\alpha = 0.636$. }
\end{table*}

After applying the  procedure described above, we arrive at the final cuts summarized in Table~\ref{tab:shifts}.

For cosmic shear, the scale cuts are determined separately for the NLA and TATT-4 intrinsic alignment models. In the NLA case, we find a cut corresponding to $\Delta \chi^2 = 3.5$, while for TATT-4, we find a more permissive threshold of $\Delta \chi^2 = 5$. These cuts result in $S_8$ shifts of $0.50\sigma$ and $0.33\sigma$ (NLA), and $0.49\sigma$ and $0.27\sigma$ (TATT-4), for the high- and low-baryon cases, respectively. The posterior distributions are shown in Fig.~\ref{fig:shear_cuts_validation}.

We determine the linear galaxy bias scale cuts for the $2 \times 2$pt analysis to be $(R_{\text{min},\; w(\theta)}, R_{\text{min},\; \gamma_t}) = (9, 6)$~Mpc$/h$. When extending to the $3 \times 2$pt analyses in both $\Lambda$CDM and $w$CDM, we actually find that the union of the 2$\times$2pt and cosmic shear cuts passes our criteria.

For the non-linear galaxy bias scenario, we find $(R_{\text{min},\; w(\theta)}, R_{\text{min},\; \gamma_t}) = (4, 4)$~Mpc$/h$ scale cuts for the $2 \times 2$pt analysis. These cuts also work for the $3 \times 2$pt analyses in both $\Lambda$CDM and $w$CDM together with cosmic shear scale cuts determined by the $\Delta \chi^2 = 5$ criterion. The results of this validation are summarized in Fig.~\ref{fig:validation-cuts}. We find galaxy bias to be the main limiting systematic in the $2 \times 2$pt analysis, with baryonic feedback playing a subdominant role. This justifies why posteriors derived from the analysis of different baryonic-contaminated signals are similar, as in the $2 \times 2$pt linear case. For the $3 \times 2$pt analyses, both baryons and galaxy bias appear as limiting systematics on equal footing.

The final set of cuts results from the iterative validation procedure outlined in Fig.~\ref{fig:flow-shear-cuts}, which is extremely computationally demanding. In total, approximately 30, 90, and 52 MCMC chains were run to validate the cosmic shear, linear, and non-linear cuts, respectively. Assuming each configuration used 300 CPUs, this corresponds to roughly 1.75 million CPU-hours. These figures highlight the scale of the challenge posed by this validation process and underscore the need to explore and develop technical tools to enable successful future analyses.
\journal{As discussed in App. \ref{app:pk-emu}, a non-linear power spectrum emulator would considerably reduce the runtime by a factor of around 3, but it is interesting to note that switching the sampler from \textsc{polychord} to \textsc{nautilus} reduced the runtime more significantly (see App. \ref{app:mcmc_samplers}).}

\subsection{\label{subsec:cuts-cardinal}Testing the scale cuts with the Cardinal simulation}

In this work, we use the $2 \times 2$pt signal (Fig.~\ref{fig:signal-gglensing}  and~\ref{fig:signal-wtheta}) measured in the \textsc{Cardinal} mock galaxy catalog ~\cite{to2023buzzardcardinalimprovedmock} with \textsc{TreeCorr} \cite{Jarvis_2004}  to validate the galaxy bias modelling. We analyse the signal using both linear, analysing data points down to $(R_{\mathrm{min}, w(\theta)}, R_{\mathrm{min}, \gamma_t}) = (8,6)$~Mpc$/h$\footnote{The test performed with \textsc{Cardinal} was carried out prior to the validation of the scale cuts using the noiseless mock signal described in Section~\ref{subsec:cutsprocess}. Thus, it employed $(R_{\mathrm{min}, w(\theta)}, R_{\mathrm{min}, \gamma_t}) = (8,6)$~Mpc$/h$ instead of the subsequently validated cuts of $(9,6)$~Mpc$/h$. Since the outcome with the less conservative cuts was satisfactory, a repetition of the validation was unnecessary.}, and non-linear galaxy bias, down to  $(4,4)$~Mpc$/h$, and applying scale cuts based on the DES Year 3 prescription \cite{Pandey_2020}. The goal is to assess whether the estimated values of $S_8$ and $\Omega_{\rm m}$ from the $\Lambda$ and $w$CDM analyses are consistent with the input cosmology of the simulation. We do not impose a strict requirement on the bias in cosmological parameters in this test as we are analysing the 2PCFs measured in one single realization of the simulation, and thus the inferred cosmological parameters are sensitive to cosmic variance and noise. Therefore, this test provides a useful complementary validation approach to check robustness under noisy conditions. We perform the analysis and check that the recovered parameters are reasonably close to the input values, typically within $2\sigma$. In light of recent DES Clusters multi-probe pipeline validation results obtained with \textsc{Cardinal}~\cite{to2025darkenergysurveymodeling}, observing a cosmological shift exceeding that expected from statistical noise would warrant a more detailed examination of the data cuts and the analysis pipeline to search for possible issues.

First, we determine the linear galaxy bias coefficients $b_1$ that best describe the signal, while fixing the higher-order coefficients $b_2$, $b_{s^2}$, and $b_{3\mathrm{nl}}$ to zero and setting the cosmological parameters to their known input values. The goal of this first step is to update the covariance matrix with the actual lens sample galaxy bias. We run a $2 \times 2$pt $\Lambda$CDM chain in which only the linear galaxy bias coefficients are sampled over the fiducial priors (Table~\ref{tab:priors}), applying the linear scale cuts. As a result, we obtain the following bias coefficients as the mean of the posteriors:
\begin{align*}
    b^1_1 &= 1.35 \pm 0.04, & b^2_1 &= 1.46 \pm 0.03, \\
    b^3_1 &= 1.43 \pm 0.02, & b^4_1 &= 1.39 \pm 0.02, \\
    b^5_1 &= 1.40 \pm 0.02, & b^6_1 &= 1.13 \pm 0.03.
\end{align*}

After estimating the linear galaxy bias coefficients for the \textsc{MagLim}-like sample we update the theoretical covariance matrix using the newly inferred bias values. In the second step, we run both a $2 \times 2$pt $\Lambda$CDM chain and a $2 \times 2$pt $w$CDM chain, sampling over cosmological parameters and marginalizing over nuisance parameters, while modeling galaxy bias with a linear description, applying the corresponding scale cuts, and considering the updated covariance in the likelihood calculation.

Finally, in the third stage, we develop $2 \times 2$pt $\Lambda$CDM and $2 \times 2$pt $w$CDM analyses with non-linear galaxy bias modeling and the respective scale cuts. In this case, we sample over cosmological and nuisance parameters while fixing the galaxy bias coefficients beyond the local quadratic term $b_2$, to their co-evolution values.

\begin{figure}[t]
    \centering
    \includegraphics[width=\linewidth]{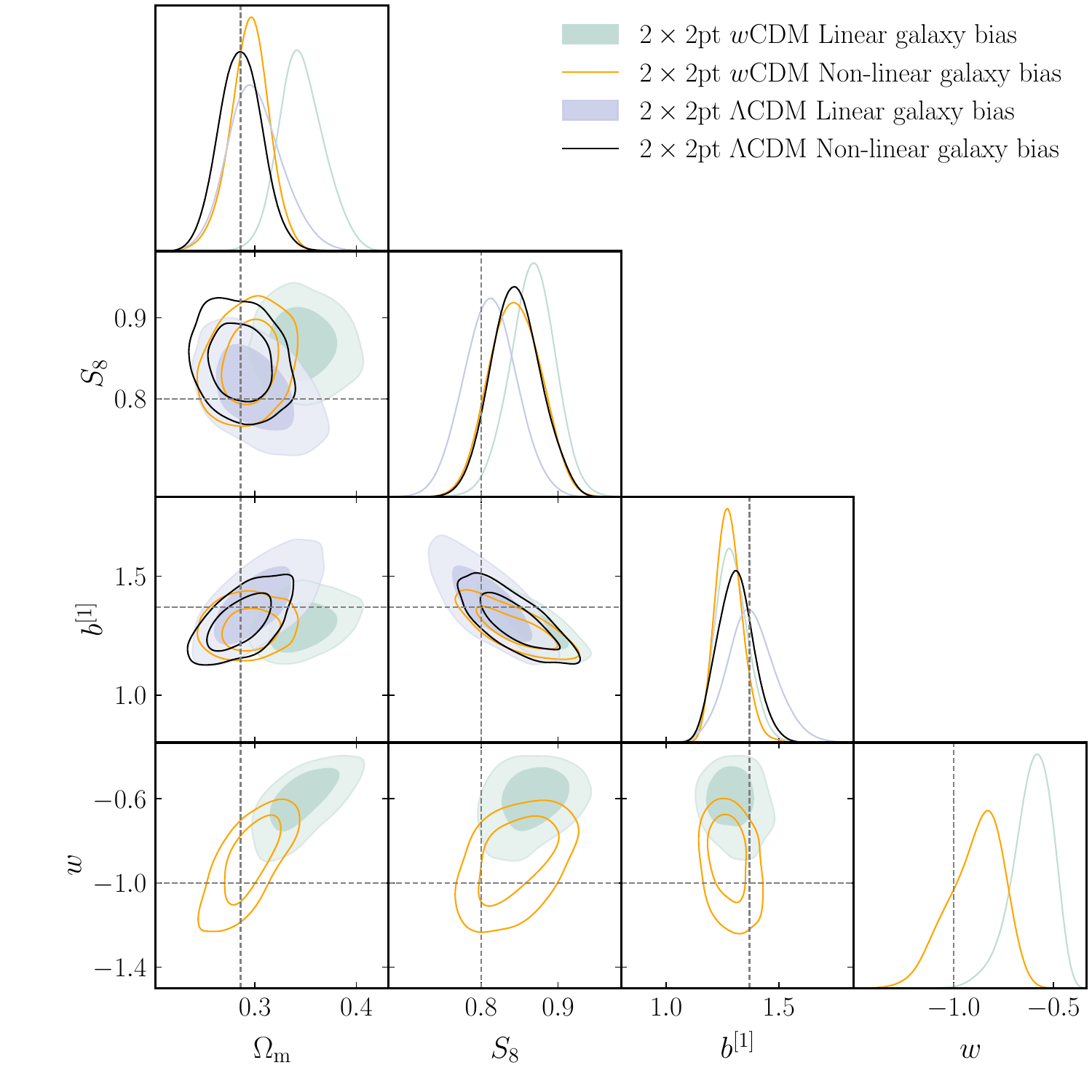}
    \caption{\label{fig:cardinal-results}Scale cuts validation with the Cardinal simulation. Posterior distributions of cosmological parameters and linear galaxy bias coefficients from the Cardinal $2 \times 2$pt $w/\Lambda$CDM analysis, comparing linear and non-linear galaxy bias models with scale cuts of (8, 6) Mpc/$h$ and (4, 4) Mpc/$h$, respectively. The figure highlights the most constrained parameters: $\Omega_{\rm m}$ and $S_8$, along with the linear galaxy bias coefficient for the first redshift bin and the dark energy equation of state parameter, $w$. Dashed lines indicate the input values.}
\end{figure}

The results are shown in Fig. \ref{fig:cardinal-results}. In the $\Lambda$CDM scenario, the posterior distributions obtained from both linear and non-linear bias models encompass the known input values within the 1-$\sigma$ region. This result gives us further confidence that the applied scale cuts effectively mitigate unmodeled galaxy bias contributions in an analysis of noisy data, serving as a complement to the validation on noiseless signal presented in the previous section.

Similarly, in the $w$CDM case, we find that the posterior distribution derived from the non-linear galaxy bias model remains consistent with the input values at approximately the 1-$\sigma$ level. However, the results obtained using the linear galaxy bias model exhibit larger shifts.

\section{\label{sec:validation} Expected cosmological constraints from Year 6}

In this section, we first characterize the DES Year 6 signal after applying the validated scale cuts, discuss volume effects, and analyze the contributions to the constraining power from different redshift ranges and scales. For this, we use the synthetic signal introduced in Sect.~\ref{sec:data}. We  then assess the robustness of our baseline setup against further modifications to the baseline analysis choices, besides the ones already used to define the scale cuts in the previous section. 

\subsection{\label{subsec:forecast}Baseline setup characterization}

\begin{figure}
    \centering
    \includegraphics[width=\linewidth]{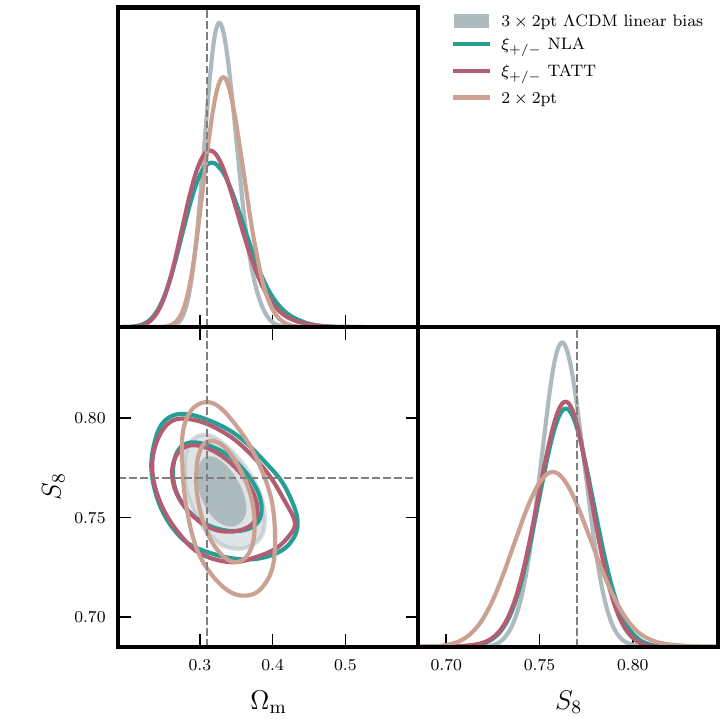}
    \caption{\label{fig:fore-lin}Posterior distributions of the abundance of matter and the amplitude of matter density fluctuations derived from the analysis of mock noiseless Year 6 data in cosmic shear analyses with NLA and TATT parameterizations, and with $2 \times 2$pt and $3 \times 2$pt linear galaxy bias analyses.}
\end{figure}

\begin{figure}
    \centering
    \includegraphics[width=\linewidth]{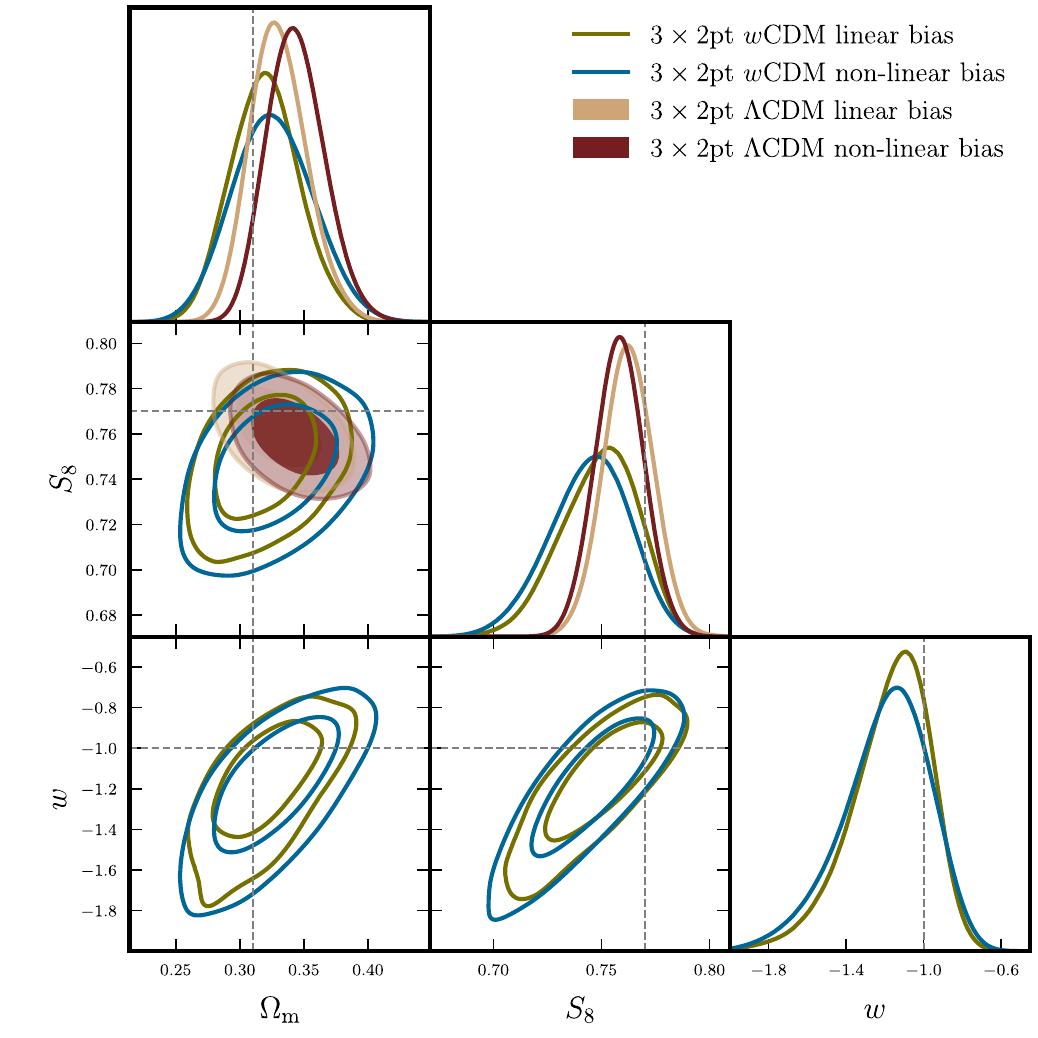}
    \caption{\label{fig:fore-nl}Posterior distributions of the abundance of matter, amplitude of matter density fluctuations, and the dark energy equation of state obtained from the analysis of mock noiseless Year 6 data in linear and non-linear galaxy bias $3 \times 2$pt analyses assuming $\Lambda$CDM and $w$CDM models.}
\end{figure}

\begin{figure}
    \centering
    \includegraphics[width=\linewidth]{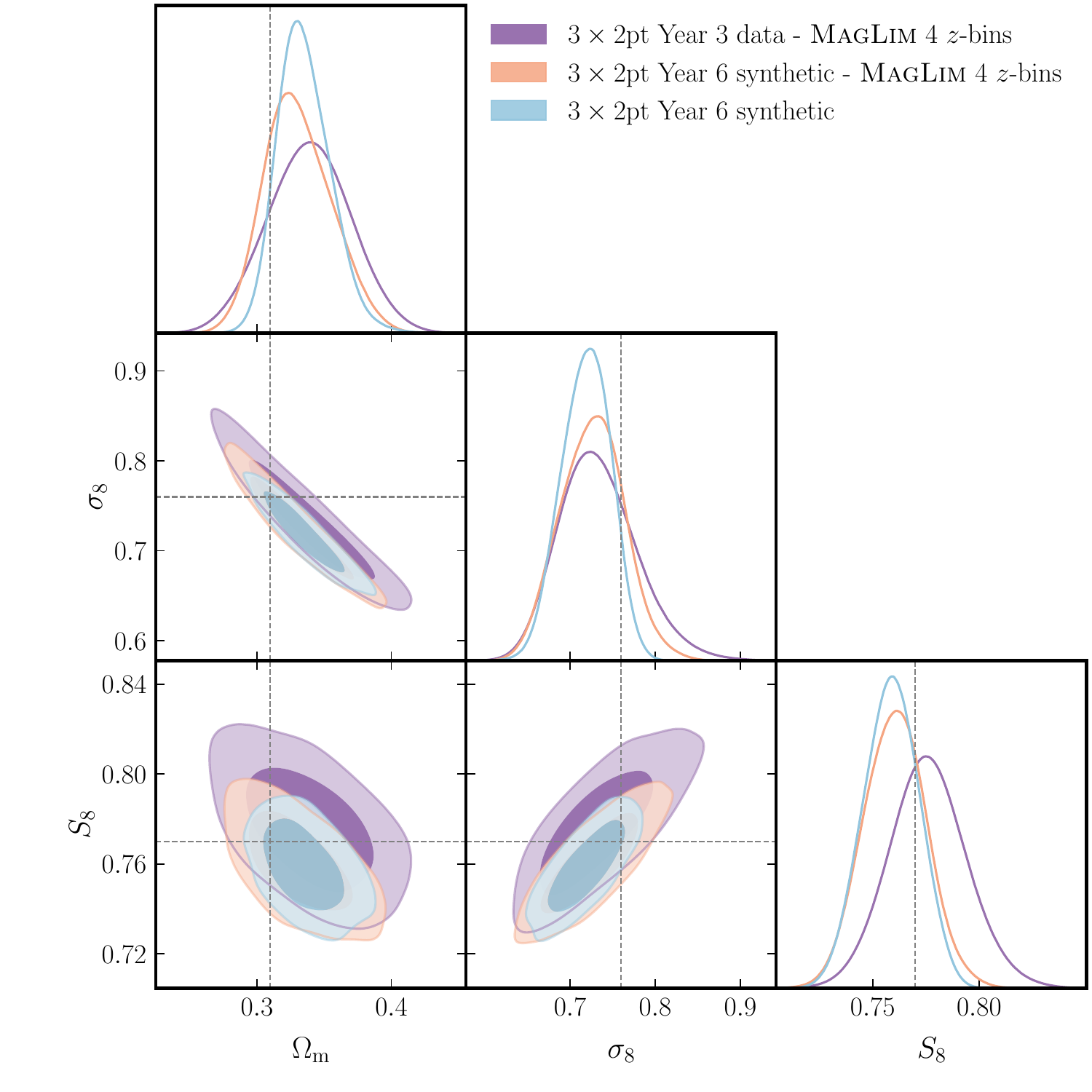}
    \caption{\label{fig:3x2pt-maglim4} Year 3 and Year 6 with 4 lens redshift bins. The figure shows the contours of the best-constrained parameters from the DES Year 3 \journal{real data} analysis, which considered the four lowest $z$-bins, included shear ratios,  and applied scale cuts of (8, 6) Mpc$/h$ (purple). For comparison, we present results from the Year 6 synthetic signal analysis, which uses similar scale cuts but does not include shear ratios, both for a similar $z$-bin configuration of the first four bins (orange) and for the fiducial setup accounting for all six lens bins (blue). Dashed lines indicate the input values. \journal{Note that the posteriors are shown without recentering although we are comparing real data Year 3 and mock signal Year 6 results and perfect alignment is not expected.}}
\end{figure}

\begin{table*}
    \centering
    \begin{tabular}{llcccccc} 
    \midrule\midrule 
    Analysis &  Scale cuts & $\xi_+$ & $\xi_-$ & $\gamma_t$ & $w(\theta)$ & Total & SNR \\
    \midrule\midrule 
    Cosmic shear NLA & $\Delta \chi^2 = 3.5$ & 178 & 89 & - & - & 267 & 37.52 \\
    Cosmic shear TATT-4 & $\Delta \chi^2 = 5$ & 183 & 99 & - & - & 282 & 39.76 \\
    \midrule
    $2 \times 2$pt $\Lambda$CDM Linear & (9, 6) Mpc/$h$ & - & - & 312 & 66 & 378 & 91.47 \\
    $3 \times 2$pt $w / \Lambda$CDM Linear & (9, 6) Mpc/$h$ + $\Delta \chi^2=5$ & 183 & 99 & 312 & 66 & 660 & 95.26 \\
    \midrule 
    $2 \times 2$pt $\Lambda$CDM Non-linear & (4, 4) Mpc/$h$ & - & - & 348 & 87 & 435 & 151.31 \\
    $3 \times 2$pt $w/\Lambda$CDM Non-linear & (4, 4) Mpc/$h$ + $\Delta\chi^2=5$ & 183 & 99 & 348 & 87 & 717 & 153.31 \\
    \midrule 
    \end{tabular}
    \caption{\label{tab:cuts-npoints} Number of data points included in each analysis, based on the applied scale cuts. The scale cuts are defined by the minimum physical scales at which galaxy bias can be accurately and robustly modeled ($R_{\text{min}, \,w(\theta)}$, $R_{\text{min}, \, \gamma_t}$), and by the cosmic shear cuts $\Delta \chi^2$. The corresponding SNR for each configuration is also shown.}
\end{table*}

\subsubsection{Signal-to-noise ratio}
Having established the validity and efficiency of the scale cuts in mitigating theoretical systematics, we now summarize the properties of the 2PCFs after removing the affected data points.  

Table~\ref{tab:cuts-npoints} collects the number of data points and the signal-to-noise ratio (SNR) for each analysis, considering the expression:
\begin{equation}
    {\rm SNR} = \sqrt{\mathbf{D}^T \mathbf{Cov}^{-1}\mathbf{D}}.
\end{equation}

For the cosmic shear NLA analysis, the final data vector contains 267 points, corresponding to an SNR of 37.5. The TATT analysis includes 282 data points, with a slightly higher SNR of 39.8, as shown in Table~\ref{tab:cuts-npoints}. The $2 \times 2$pt linear galaxy-bias analysis consists of 378 data points from $w(\theta)$ and $\gamma_t$, with an SNR of 91.5. In the non-linear bias case, the number of data points increases to 435, yielding a SNR of 151.3. Finally, the $3 \times 2$pt analyses include 660 data points for the linear-bias $\Lambda$CDM and $w$CDM cases, and 717 for the corresponding non-linear bias analyses, with SNRs of 95.3 and 153.3, respectively. 

Fig.~\ref{fig:fore-lin} displays the posteriors on the abundance of matter and $S_8$ forecasted for the main analysis in $\Lambda$CDM, from cosmic shear with NLA and TATT, the $2 \times 2$pt, and the combination into the $3 \times 2$pt linear analysis result. The figure evidences the different degeneracies in the $\Omega_{\rm m}-S_8$ plane from the different probes and how constraining power is enhanced in the multi-probe analysis. It is also expected that  both cosmic shear cases yield comparable constraining power despite differences in the number of points---this is by design of the scale cuts procedure.

\subsubsection{Parameter space: projection effects and best constrained parameters}

In our fiducial setup we observe some projection effects in $\Omega_{\rm m}$ and $S_8$. Projection, or volume effects refer to systematic shifts in marginalized parameter posteriors away from the peak of the full-dimensional posterior. This is a model-dependent effect that arises when changes in the parameter space volume at a given likelihood dominate over the changes in the likelihood itself, and so can be particularly present in models where the curvature of the likelihood depends strongly on the parameters, as is often the case with highly non-linear models. 

When the data is very constraining, the likelihood falls away quickly from the maximimum-likelihood point and volume effects become negligible; conversely, they can be exacerbated for models with poorly-constrained parameters.
Reparameterization of the likelihood, or the adoption of informative priors (e.g. a Jeffreys prior to make the likelihood parameterization-invariant), are two approaches to reducing the impact of prior volume effects (see e.g. \cite{Hadzhiyska_2023}).

In 3$\times$2pt $\Lambda$CDM linear galaxy bias we find $S_8$ to be 0.65$\sigma$ lower than the true value (0.60$\sigma$ for 2$\times$2pt). 

We find $\Omega_{\rm m}$ to be higher than the true, up to 1.50$\sigma$ for the 3$\times$2pt $\Lambda$CDM analysis. As expected, these shifts are worst in the more complex $w$CDM case, with $S_8$ being as high as 1.37$\sigma$ away form the truth for 3$\times$2pt. Estimates of the shifts in the posterior due to projection effects in $\Lambda$ and $w$CDM are summarized in Table~\ref{tab:proj_effects_lcdm} and~\ref{tab:proj_effects_wcdm}, respectively.

Such shifts indicate that there might be a better constrained direction in parameter space that is less sensitive to projection effects.
Using a noiseless data vector, we find the best-constrained direction in the $\Omega_{\rm m}$–$\sigma_8$ parameter space:
\begin{equation}\label{eq:s8_alpha}
    S_8(\alpha) = \sigma_8 \left( \frac{\Omega_{\rm m}}{0.3} \right)^{\alpha},
\end{equation}
with $\alpha = 0.636$, which we find it to be less prone to projection effects than $S_8(\alpha=0.5)$. 
We also use this parameter in Section~\ref{subsec:validation-metric} to understand the interplay of our scale cuts procedure and projection effects.  Accordingly, we report the shifts in this derived parameter in Table~\ref{tab:shifts} (in~\cite{y6-3x2pt} we derive an updated $\alpha$ using simulated data from the best-fit cosmology and final covariance matrix since this can also change due to noise.) 

This best-constrained direction is similar as to what was found in DES Year 3 (see Fig.~\ref{fig:3x2pt-maglim4}). In the $S_8-\Omega_{\rm m}$ panel it can be appreciated that for both analysis, the degeneracy is not completely broken, while we find the contours to be closer to a circle in the case of $\alpha = 0.636$. See \cite{Jain1997} for a discussion on how redshift and angular binning could affect the best-constrained degeneracy direction. 

\subsubsection{Reporting constraints on cosmological parameters}

We choose to quote the cosmological results on the real data analyses ~\cite{y6-3x2pt, y6-2x2pt, y6-1x2pt} as the mean of the 1D marginalized posterior of the parameter of interest with its associated standard deviation. 
However, because of projection effects, it is particularly useful to also report the maximum \textit{a posteriori} (MAP) values, \textit{i.e.} the parameter value at the highest posterior point.

We therefore report 1D marginalized constraints as:
\begin{equation}
    \text{mean} \;\pm\; \text{standard deviation} \; (\text{MAP}).
\end{equation}

We note that, as previously explained, we make use of the 1D mean of the marginalized posterior when estimating the shift in one dimension for the scale cuts validation rather than the MAP. 
In fact, the MAP point from an MCMC chain is a poor estimate of the global MAP of a given posterior. We therefore estimate the MAP as follows: We select the $N_{\rm guess}$ samples from a chain that have the highest posterior values and use each of these points as an initial guess for an optimization algorithm.  We then select the highest posterior output from these $N_{\rm guess}$ posterior maximization searches. In practice we choose $N_{\rm guess}=20$, and optimize using the Powell algorithm in {\tt scipy.optimize} with a tolerance of 0.05. Tests of this procedure can be found in Appendix~\ref{app:mapsearches}.

\subsubsection{Resulting constraints from DES Year 6 simulated data}

In Fig.~\ref{fig:fore-nl} we compare the $\Lambda$CDM contours with the $w$CDM ones, both for the linear and non-linear bias analyses. We find that the constraining power is similar for the linear and non-linear bias cases in $\Lambda$CDM, with a figure of merit (FoM) defined as 
\begin{equation}
    \text{FoM}_{\Omega_{\rm m}, \,\sigma_8} \equiv \left( \det{\text{Cov}(\Omega_{\rm m}, \sigma_8)} \right)^{-1/2},    
\end{equation}
of 4375 and 4562, respectively.

In $w$CDM, the constraining power in the case of using more scales, \textit{i.e.} assuming a non-linear galaxy bias model, is actually slightly worst than the more stringent scale cuts case, due to the additional parameters in the bias model varied in the analysis, washing out the constraining power. The FoM decreases from 2344 with the linear model to 1927 when assuming non-linear bias. By opening the parameter space we constrain non-linear coefficients at the expense of precision in $w$.

\begin{table}[h]
\centering
\caption{Projection effects assessment for $\Lambda$CDM quantified by the shift in the posterior mean relative to the true value in units of standard deviations.}
\label{tab:proj_effects_lcdm}
\begin{tabular}{lccc}
\hline
Parameter & $3\times2$pt & $2\times2$pt & $3\times2$pt NL \\
\hline
$S_8$ & $-$0.65 & $-$0.60 & $-$1.03 \\
$\Omega_m$ & $+$0.91 & $+$0.95 & $+$1.50 \\
$\sigma_8$ & $-$1.03 & $-$1.02 & $-$1.73 \\
\hline
\end{tabular}
\end{table}

\begin{table}[h]
\centering
\caption{Projection effects assessment for $w$CDM quantified by the shift in the posterior mean relative to the true value in units of standard deviations.}
\label{tab:proj_effects_wcdm}
\begin{tabular}{lcc}
\hline
Parameter & $3\times2$pt & $3\times2$pt NL \\
\hline
$S_8$ & $-$1.17 & $-$1.37 \\
$\Omega_m$ & $+$0.37 & $+$0.51 \\
$\sigma_8$ & $-$1.13 & $-$1.35 \\
$w$ & $-$0.83 & $-$0.86 \\
\hline
\end{tabular}
\end{table}

\subsection{Comparison to DES Year 3}

DES Year~3 analysis \cite{Abbott_2022} included 462 data points in the $3 \times 2$pt signal after applying linear galaxy-bias scale cuts, to be compared with the number of data points after linear cuts in Table~\ref{tab:cuts-npoints}. The majority of the added points in DES Year 6 come from the cosmic shear part, since actually  2$\times$2pt are similar among the analyses: (8, 6) Mpc/$h$ in DES Year 3 vs. (9, 6) Mpc/$h$ in DES Year 6, for $w$ and $\gamma_t$ respectively. 

We also compare the DES Year 6 $3\times2$pt constraining power in Fig.~\ref{fig:3x2pt-maglim4} with respect to the DES Year 3. To enable direct comparison, we also evaluate the constraining power when excluding the two highest-redshift lens bins of the \textsc{Maglim} sample, thereby matching the Year 3 tomographic configuration. This allows us to compare the DES Year 3 results from the linear galaxy-bias analysis (which include shear ratios in the likelihood and four lens tomographic bins) with our synthetic Year 6 analysis under two scenarios: the full six-bin configuration and a restricted four-bin setup. Note that the posteriors are shown without recentering; given that we are comparing contours from real data with simulated data vectors, perfect alignment is not expected. Even under the conservative scenario where the two highest-redshift lens bins are excluded in Year 6, we forecast substantial improvements in constraining power, with the FoM in the $\Omega_{\rm m}$--$\sigma_8$ plane increasing from 2068 to 3397.

\subsection{Understanding the Error Budget and Volume Effects}\label{sec:error_budget}

We now turn to a set of tests designed to assess how calibration and modeling choices impact the error budget in the final cosmological constraints, and how mildly constrained parameters in our analysis—such as the sum of neutrino masses—contribute to the observed projection effects. 

\begin{figure*}[htbp]
  \subfigure{
    \includegraphics[width=0.99\textwidth]{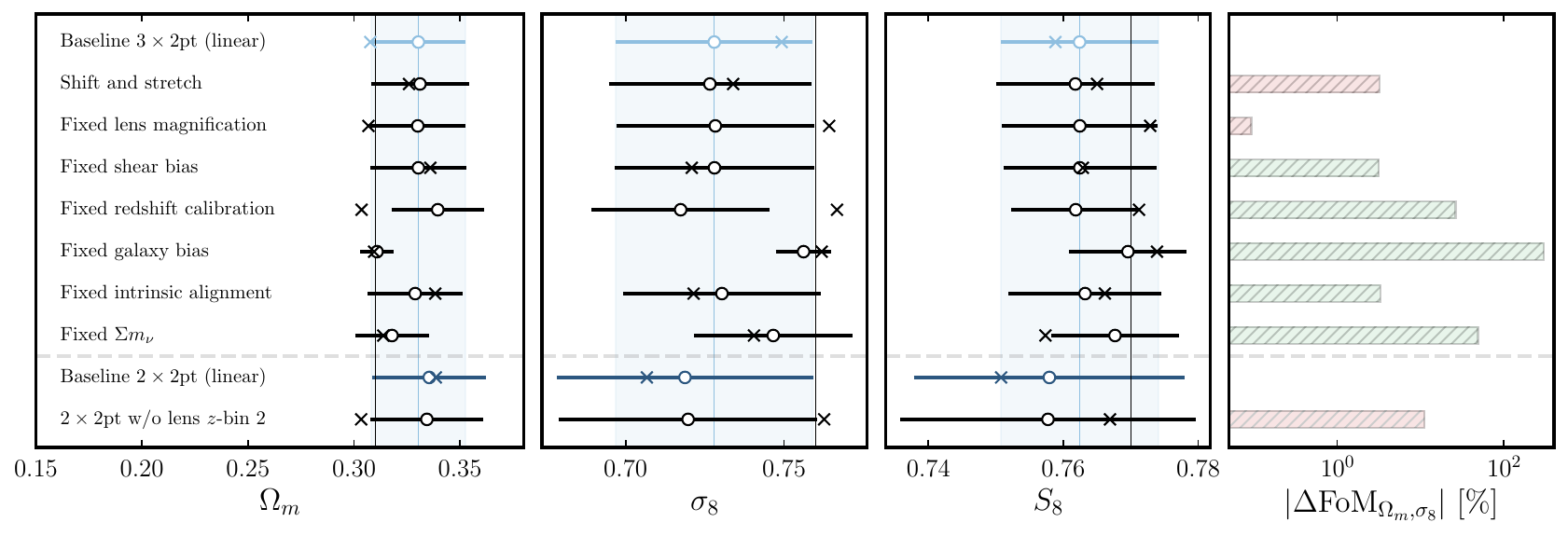}
  }
  
  \subfigure{
    \includegraphics[width=0.99\textwidth]{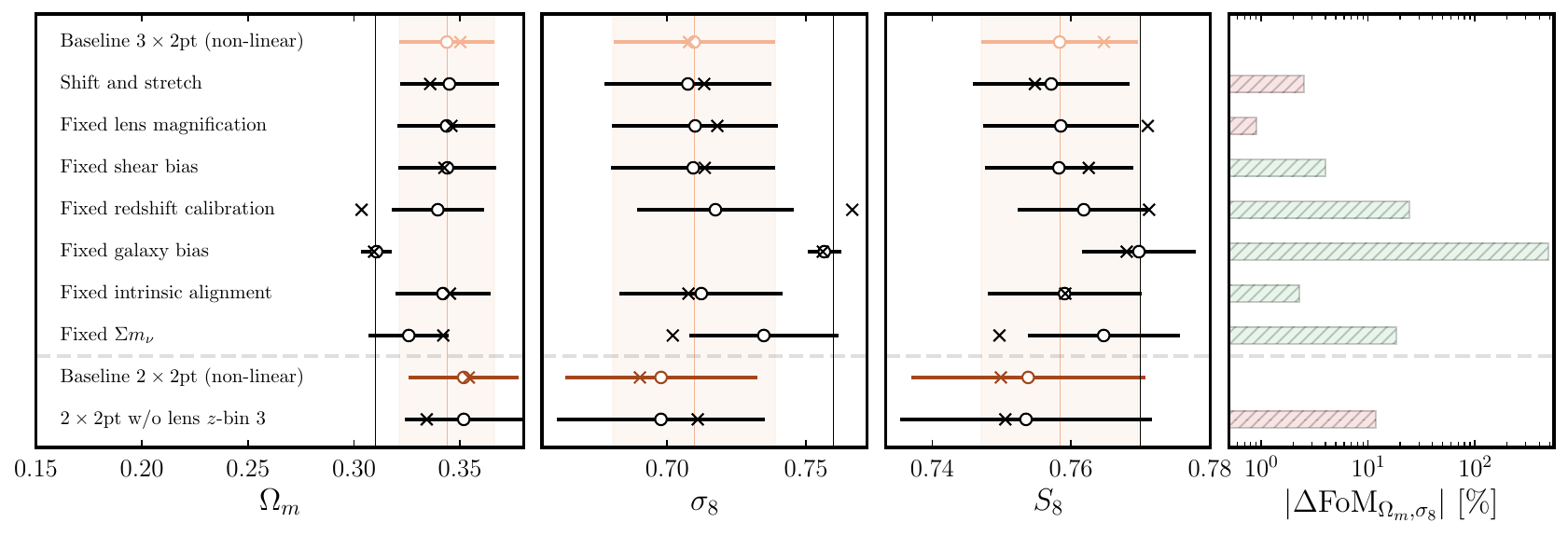}
  }
  
  \caption{\label{fig:robustness-tests} Summary of robustness test results for the: linear (top panel), and non-linear galaxy bias (bottom panel) analyses. The panels show the posteriors of $\Omega_{\rm m}$, $\sigma_8$, and $S_8$, where circles mark the posterior mean, crosses indicate the maximum a posteriori (MAP), and shaded regions denote the $1-\sigma$ credible interval. The vertical black line corresponds to the input value used to generate the mock signal, providing a reference for assessing projection effects relative to the posterior mean of the baseline analyses (blue for the linear and orange for the non-linear case). The right-most panel displays the absolute relative difference in constraining power with respect to the baseline, with bar colors indicating whether the figure-of-merit decreases (red) or increases (green).}
\end{figure*}

\paragraph*{Fixing parameters with Gaussian priors} We begin by fixing parameters with Gaussian priors, namely the lens magnification coefficients, the multiplicative shear bias, and the redshift calibration modes, setting them to their fiducial values. As shown in Fig.~\ref{fig:robustness-tests}, fixing the redshift parameters produces the largest increase in constraining power, with a change of $\sim 25\%$ in the FoM of the $\Omega_{\rm m}-S_8$ plane, fixing multiplicative shear biases and fixing lens magnification produce negligible changes in the FoM ($\sim 3\%$ and $\sim 1\%$, respectively). Projection effects are not significantly reduced by fixing these parameters, as indicated by the stability of the posterior means around the baseline results.  

\paragraph*{Fixing parameters with flat priors} In contrast, fixing parameters sampled with flat priors—specifically the galaxy bias and the sum of neutrino masses—produces a noticeable improvement in projection effects, with posterior means shifted closer to the input values used in the mock data generation. Still, we find that fixing intrinsic alignments (for which we have broad flat priors) does not help with reducing projection effects on $\Omega_{\rm m},\ \sigma_8$, or $S_8$.   

\paragraph*{Removing redshift bins} Additionally, we test the constraining power removing the highest-SNR lens tomographic redshift bin from the $2 \times 2$pt analysis. In the linear galaxy bias analysis, the highest-SNR bin is the second one with a SNR of 60.37. When we remove this bin we find a  negligible change in constraining power and no shift in the posterior compared to the baseline $2 \times 2$pt result. In the non-linear analysis, the highest-SNR bin is the third, with SNR of 84.28. We find that the mean value of the posterior is robust against removing this bin, with a reduction of $\sim 10$ points in the FoM (see Fig.~\ref{fig:robustness-tests}).

\begin{figure}
    \centering
    \includegraphics[width=\linewidth]{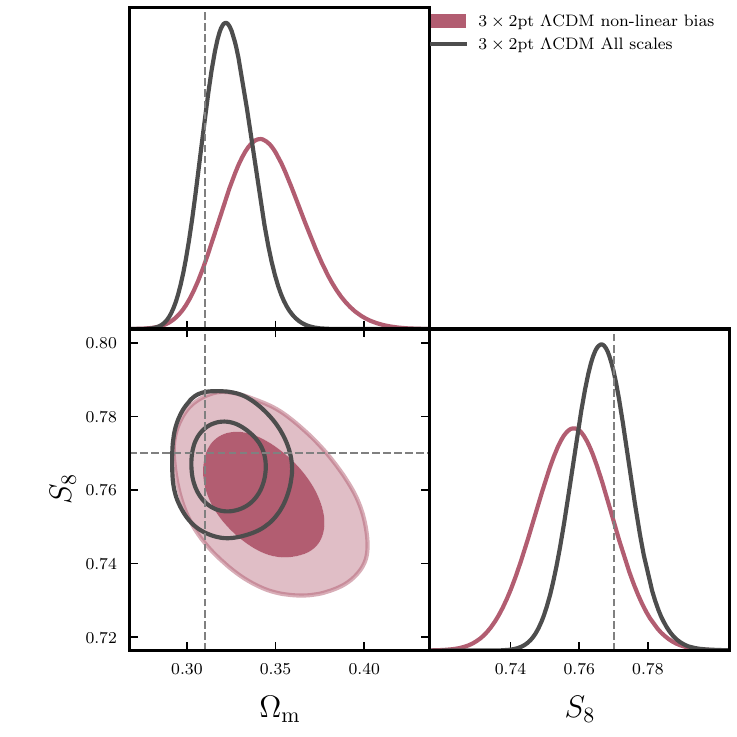}
    \caption{\label{fig:nocuts}Constraining power comparison between the fiducial $3 \times 2$pt non-linear galaxy bias analysis and a analysis including all scales, modeling with non-linear galaxy bias and sampling over \tagn $\in [7.3, 8.3]$.}
\end{figure}

\paragraph*{Using all scales.} We now explore the impact on constraining power from assuming extreme model choices.  To do so, we consider both the simulated linear and non-linear galaxy bias data vectors and use all angular scales $\theta$ between 2.475 and 250 arcmin., marginalizing over \( \log_{10} T_{\mathrm{AGN}} \) and assuming a non-linear galaxy bias model. These assumptions are of course extreme, the perturbative non-linear galaxy bias model not having been validated for scales below 4 Mpc/h. We here want to assess the potential constraining power contained in the DES Year 6 3x2pt data, assuming our model was correct. The results of this test are displayed in Fig.~\ref{fig:nocuts}. When analysing all scales, we obtain $\text{FoM}_{\Omega_{\rm m},\sigma_8}=9052$, which supposes a factor of 2 gain in constraining power compared to the fiducial non-linear galaxy bias analysis, while at the same time reducing the level of projection effects. In addition to enhanced constraining power, we also recover higher posterior accuracy in this test. However, this improved accuracy must be interpreted cautiously: in this idealized scenario, we have exact knowledge of the small-scale physics, an assumption that cannot be justified when analyzing real data.

\section{\label{sec:robustness-tests}Additional Robustness Tests}

\begin{figure*}
    \centering
    \includegraphics[width=0.9\linewidth]{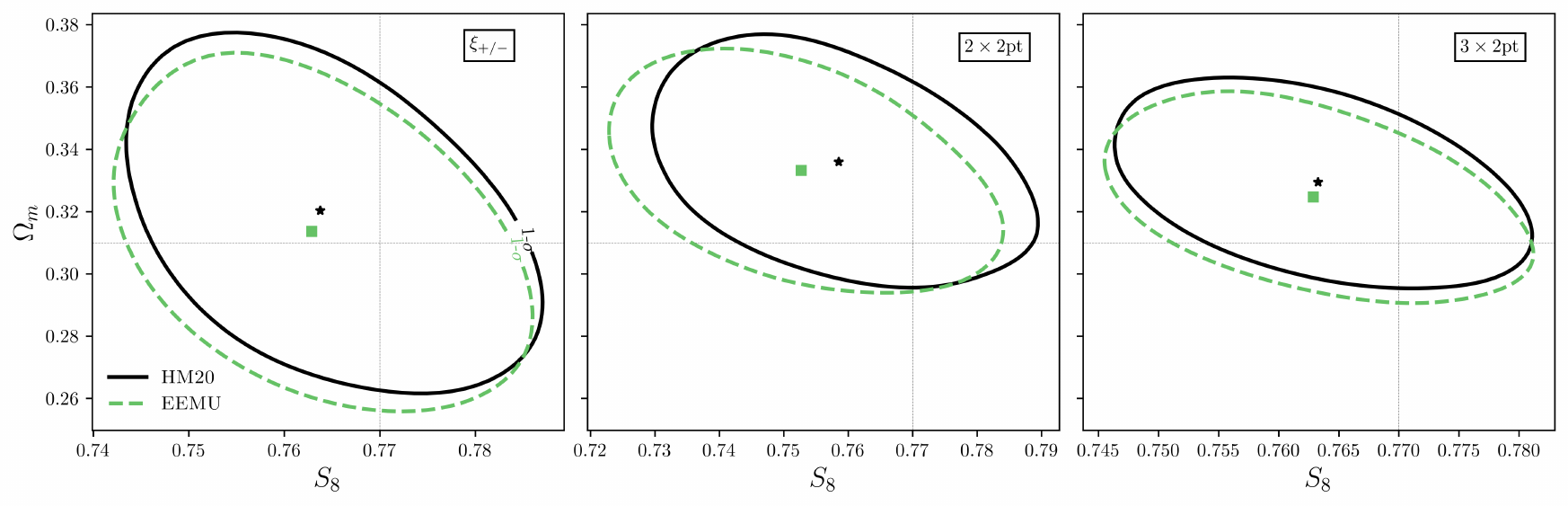}
    \caption{\label{fig:pk-robustness} Robustness against nonlinear matter power spectra modeling: $1\sigma$ countours for two choices of dark matter-only power spectrum: \textsc{HMCode 2020} (HM20, our fiducial choice) and \textsc{Euclid Emulator v2} (EEMU). Here the results are for our linear bias $\Lambda$CDM analysis. The marker denotes the mean of the posterior distributions.} 
\end{figure*}

\begin{figure*}
    \centering
    \includegraphics[width=0.9\linewidth]{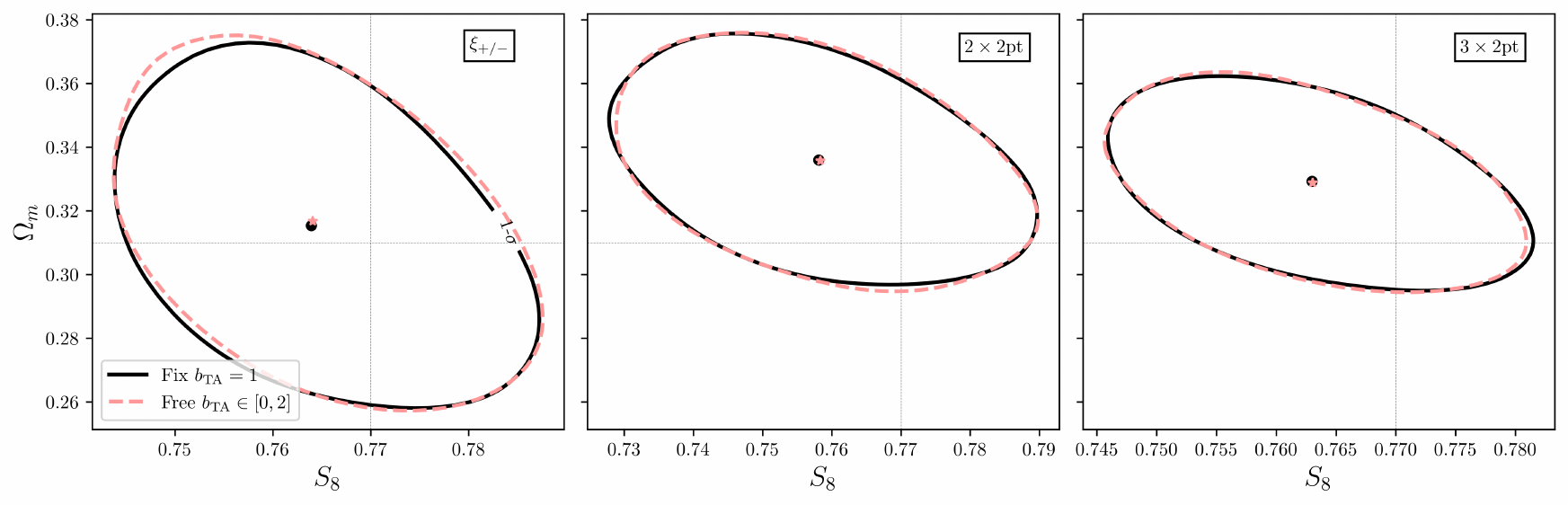}
    \caption{\label{fig:bta}Robustness against instrinsic alignment modeling choices: $1\sigma$ posterior contours from the analysis of mock signal produced with a different value of the tidal alignment bias parameter ($b_{\rm TA} = 0.2$) with respect to our fiducial model (which fixes it to be $b_{\rm TA} = 1$) and marginalizing over a broad prior range $b_{\rm TA} \in [0, 2]$. Here the results are for our linear bias $\Lambda$CDM analysis. The marker is the mean value of the posterior distribution.}
\end{figure*}

In this section, we assess the stability of our results under different robustness tests. 

The first variant we investigate concerns the \textit{calibration method of the redshift distribution}. Instead of the novel baseline approach of mode projection, we calibrate the distribution by introducing one degree of freedom per redshift bin, allowing for shifts in the mean value of both the lens and source galaxy distributions. In addition, we include one extra parameter per lens bin to permit potential stretching of the distribution. We sample over these calibration parameters assuming Gaussian priors as listed in Table \ref{tab:priors-shiftstretch}. This shift-and-stretch method was the fiducial approach adopted in previous DES analyses \cite{Abbott_2022}. A comparison of the resulting posteriors with those from the baseline method shows good agreement in both the linear and non-linear analyses, with only negligible differences in the mean value of the posterior and constraining power—at the 2–3\% level, consistent with sampling noise and thus validating the implementation of the new approach.

\begin{table}
    \centering
    \setlength{\tabcolsep}{4pt}
        \begin{tabular}{lcr}
            \hline \midrule
            Parameter & Fiducial & Prior \\ 
            \midrule
            \multicolumn{3}{l}{\textbf{Lens photo-$z$}} \\
            $\Delta z^1_l \times10^2$ & 0 & $\mathcal{N}(0, 0.7)$ \\
            $\Delta z^2_l \times10^2$ & 0 & $\mathcal{N}(0, 1.1)$ \\
            $\Delta z^3_l \times10^2$ & 0 & $\mathcal{N}(0, 0.6)$ \\
            $\Delta z^4_l \times10^2$ & 0 & $\mathcal{N}(0, 0.6)$ \\
            $\Delta z^5_l \times10^2$ & 0 & $\mathcal{N}(0, 0.7)$ \\
            $\Delta z^6_l \times10^2$ & 0 & $\mathcal{N}(0, 0.8)$ \\
            $\sigma^1_{z,l}$ & 1 & $\mathcal{N}(1, 0.062)$ \\
            $\sigma^2_{z,l}$ & 1 & $\mathcal{N}(1, 0.093)$ \\
            $\sigma^3_{z,l}$ & 1 & $\mathcal{N}(1, 0.054)$ \\
            $\sigma^4_{z,l}$ & 1 & $\mathcal{N}(1, 0.051)$ \\
            $\sigma^5_{z,l}$ & 1 & $\mathcal{N}(1, 0.067)$ \\
            $\sigma^6_{z,l}$ & 1 & $\mathcal{N}(1, 0.073)$ \\
            \midrule
            \multicolumn{3}{l}{\textbf{Source photo-$z$}} \\
            $\Delta z^1_s \times10^2$ & 0 & $\mathcal{N}(0, 1.8)$ \\
            $\Delta z^2_s \times10^2$ & 0 & $\mathcal{N}(0, 1.5)$ \\
            $\Delta z^3_s \times10^2$ & 0 & $\mathcal{N}(0, 1.1)$ \\
            $\Delta z^4_s \times10^2$ & 0 & $\mathcal{N}(0, 1.7)$ \\
            \midrule \hline
        \end{tabular}
    \caption{\label{tab:priors-shiftstretch} Fiducial values and Gaussian priors for the alternative shift and stretch redshift calibration method, where $\Delta z_{l/s}$ denotes shifts in the lens/source redshift distributions and $\sigma_{z,l}$ represents the stretch of the lens distributions.}
\end{table}

We then calibrate potential posterior shifts arising from \textit{matter power spectrum mismodeling} alone. Since our scale cuts are based on a datavector contaminated by both baryonic \textit{and} matter power spectrum effects, our goal here is to isolate the effects from matter power spectrum mismodeling specifically and confirm they remain small. The procedure is as follows: we generate two dark-matter-only signals, one using \textsc{Hm20} and the other using the \textsc{Eemu}, and analyze both with \textsc{Hm20} (our fiducial choice) in the cosmic shear, $2 \times 2$pt, and $3 \times 2$pt cases, as shown in Fig.~\ref{fig:pk-robustness}. We quantify the impact of this mismatch by computing the difference between the mean values of the posterior distributions, finding the resulting shifts to be well controlled.

We next explore the \textit{complexity of the intrinsic alignment parameterization}. In our fiducial setup, the $b_{\rm TA}$ parameter is fixed to unity in order to reduce model complexity. We have verified that using an input datavector with $b_{\rm TA}$ set to a different value or marginalizing over it has a negligible impact on the resulting posteriors. This validation is illustrated in Fig.~\ref{fig:bta}, where we compare the posteriors from cosmic shear, $2 \times 2$pt, and $3 \times 2$pt analyses of a mock signal generated with $b_{\rm TA} = 0.2$. In all cases, the posteriors exhibit no significant shifts in the mean values and retain comparable constraining power.
\section{\label{sec:conclusions}Conclusions}

In this work we have presented the analysis framework for cosmic shear, the joint analysis of galaxy clustering and galaxy–galaxy lensing ($2 \times 2$pt), and the full $3 \times 2$pt combination in both $\Lambda$CDM and $w$CDM scenarios for the DES Year~6 data set. The design and validation of the pipelines were carried out in a blinded fashion, using both noiseless synthetic data sets mimicking the expected properties of the Year~6 source and lens samples, as well as the realistic \textsc{Cardinal} mock galaxy catalog.  

The modeling pipeline for the final DES weak lensing and galaxy clustering analyses incorporates several improvements relative to previous iterations. These include the computation of the matter power spectrum with \textsc{HMCode 2020}, assuming a fixed baryonic feedback contribution of \tagn=7.7; the delivery of two parallel analyses describing galaxy bias down to linear and non-linear scales; and the optimized choice of intrinsic alignment (IA) parameterizations depending on the constraining power of each analysis. Specifically, we use both NLA and TATT-4 models for cosmic shear, while only TATT-4 is considered for $2 \times 2$pt and $3 \times 2$pt analyses. Additional methodological advances include the introduction of the novel redshift-distribution calibration method based on mode projection, as well as scale cuts to mitigate poorly modeled small-scale systematics such as baryonic feedback and galaxy bias, whose leakage could otherwise bias cosmological inferences. We design and validate scale cuts for cosmic shear analyses, as for linear and non-linear galaxy bias analyses of $2 \times 2$pt and $3 \times 2$pt in $\Lambda$CDM and $w$CDM.

We have also provided an analytical covariance for the $3 \times 2$pt data vector computed with \textsc{CosmoCov}, and tested the sensitivity of signal error estimates to modeling choices related to the matter power spectrum, IA, and survey geometry. The statistical inference framework is based on a Bayesian formalism for cosmological parameter estimation. In the case of weak lensing and galaxy clustering, the main constraining power is captured by the combination of the matter density and the amplitude of matter fluctuations, $S_8 \equiv \sigma_8 \left( \Omega_{\rm m}/0.3 \right)^{\alpha}$, with $\alpha=1/2$.  

As results, we have validated the robustness of the scale cuts designed for each analysis, applying them to both the \textsc{Cardinal} mock catalog and DES Year~6-like synthetic 2PCFs. During this process, we highlighted the challenges posed by an increasingly complex parameter space, including projection and volume effects in the posteriors of parameters that remain only mildly constrained.

After validating the analysis framework, we stress-test and study the robustness of the results against alternative modeling choices and fixing calibration parameters. We also summarize the properties of the DES signal after applying scales cuts, amounting up to a total of 660 (717) data points for the $3 \times 2$pt linear (non-linear) galaxy bias analysis. Similarly, we forecast the constraining power in the best-constrained plane, $\Omega_{\rm m}-\sigma_8$. We find similar constraining power from the linear and non-linear galaxy bias analyses in $\Lambda$CDM. In $w$CDM, the constraining power is worse for the non-linear galaxy bias compared to the linear result due to the increased dimensionality of the parameter space. Finally, we compare with respect a DES Year 3 scenario in which we remove the two highest redshift bins of the lens sample, finding an increase of around $60 \%$ in the constraining power even under that conservative scenario.

The results, methodology, and challenges presented here establish a foundation for ongoing and upcoming photometric galaxy surveys pursuing weak-lensing and clustering analyses. We highlight key requirements that must be met for the analysis of future data sets: (i) carrying out analyses with minimally complex parameterizations, guided by prior information from complementary studies, and (ii) achieving technical advances that accelerate likelihood evaluations and parameter inference pipelines. Regarding the first point, we stress the need for continued work on major theoretical systematics—IA, galaxy bias, and baryonic feedback—so that prior information can be reliably integrated into future $3 \times 2$pt analyses. Such progress will be essential to fully exploit the scientific potential of these studies and to deepen our understanding of dark-sector physics and possible new phenomena. Regarding the second point, the demanding MCMC campaigns required for this type of work represent a major challenge, emphasizing the need for technical developments that will make future $3 \times 2$pt analyses both feasible and successful. This will require faster analysis pipelines powered by matter power spectrum emulators and more efficient MCMC sampling algorithms, among other technological improvements, to meet the demands of next-generation surveys such as the \textit{Vera C. Rubin Observatory}, the \textit{Euclid mission}, and the \textit{Nancy Grace Roman Space Telescope}.
\section*{Author Contributions}
All authors contributed to this paper and/or carried out infrastructure work that made this analysis possible. DSC, AF, and JB served as leads of the analysis. DSC and AF developed and maintained the code to derive and validate the scale cuts in this paper, based on the Year 3 Extensions framework created by AF and JM. JB contributed to the non-linear galaxy bias modeling and intrinsic alignment parameterization. AA contributed to baryonic feedback, matter power spectrum, and intrinsic alignment modeling. JP contributed to the point-mass marginalization modeling, calibration projection effects and scale cuts definition. SS validated cosmic shear scale cuts and the design of the intrinsic alignment parameterization, and provided tools to calculate distances between posterior distributions. AP and JC validated the \textsc{Nautilus} sampler for our setup. GG, WdA, and BY contributed galaxy redshift distributions and calibration. MY and MRB contributed the galaxy shape catalog. CHT provided the \textsc{Cardinal} mock galaxy catalog. PC, APou, MS, AF, and FAO contributed to matter power spectrum emulator. JZ contributed technical support and development of \textsc{CosmoSis}.
DSC, AF, JCN, MY, PR, AA, and JP contributed running MCMC analyses. FAO, SD, EK, and PR contributed expertise on the covariance matrix. DSC, AF, JB, SS, AA, JP, JM, AP, JCN, and WdA contributed to the writing of the manuscript. DSC, SS, JM, and JCN contributed to the production of the figures. JMF, EK, and RR served as the internal review committee and provided valuable feedback on the writing and presentation of this paper. MAT, CC, MC, and MRB contributed to the development of this work as coordinators of the Year 6 analysis. Overall guidance and feedback was provided by JB, SS, AA, FAO, JM, AP, JP, NW, and NM. The remaining authors have made contributions to this paper that include, but are not limited to, the construction of DECam and other aspects of collecting the data: data processing and calibration; developing broadly used methods, codes, and simulations; running the pipelines and validation tests; and promoting the science analysis.
\section*{Acknowledgments}

Funding for the DES Projects has been provided by the U.S. Department of Energy, the U.S. National Science Foundation, the Ministry of Science and Education of Spain, 
the Science and Technology Facilities Council of the United Kingdom, the Higher Education Funding Council for England, the National Center for Supercomputing 
Applications at the University of Illinois at Urbana-Champaign, the Kavli Institute of Cosmological Physics at the University of Chicago, 
the Center for Cosmology and Astro-Particle Physics at the Ohio State University,
the Mitchell Institute for Fundamental Physics and Astronomy at Texas A\&M University, Financiadora de Estudos e Projetos, 
Funda{\c c}{\~a}o Carlos Chagas Filho de Amparo {\`a} Pesquisa do Estado do Rio de Janeiro, Conselho Nacional de Desenvolvimento Cient{\'i}fico e Tecnol{\'o}gico and 
the Minist{\'e}rio da Ci{\^e}ncia, Tecnologia e Inova{\c c}{\~a}o, the Deutsche Forschungsgemeinschaft and the Collaborating Institutions in the Dark Energy Survey. 

The Collaborating Institutions are Argonne National Laboratory, the University of California at Santa Cruz, the University of Cambridge, Centro de Investigaciones Energ{\'e}ticas, 
Medioambientales y Tecnol{\'o}gicas-Madrid, the University of Chicago, University College London, the DES-Brazil Consortium, the University of Edinburgh, 
the Eidgen{\"o}ssische Technische Hochschule (ETH) Z{\"u}rich, 
Fermi National Accelerator Laboratory, the University of Illinois at Urbana-Champaign, the Institut de Ci{\`e}ncies de l'Espai (IEEC/CSIC), 
the Institut de F{\'i}sica d'Altes Energies, Lawrence Berkeley National Laboratory, the Ludwig-Maximilians Universit{\"a}t M{\"u}nchen and the associated Excellence Cluster Universe, 
the University of Michigan, NSF NOIRLab, the University of Nottingham, The Ohio State University, the University of Pennsylvania, the University of Portsmouth, 
SLAC National Accelerator Laboratory, Stanford University, the University of Sussex, Texas A\&M University, and the OzDES Membership Consortium.

Based in part on observations at NSF Cerro Tololo Inter-American Observatory at NSF NOIRLab (NOIRLab Prop. ID 2012B-0001; PI: J. Frieman), which is managed by the Association of Universities for Research in Astronomy (AURA) under a cooperative agreement with the National Science Foundation.

The DES data management system is supported by the National Science Foundation under Grant Numbers AST-1138766 and AST-1536171.
Data access is enabled by Jetstream2 and OSN at Indiana University through allocation PHY240006: Dark Energy Survey from the Advanced Cyberinfrastructure Coordination Ecosystem: Services and Support (ACCESS) program, which is supported by U.S. National Science Foundation grants 2138259, 2138286, 2138307, 2137603, and 2138296.
The DES participants from Spanish institutions are partially supported by MICINN under grants PID2021-123012, PID2021-128989 PID2022-141079, SEV-2016-0588, CEX2020-001058-M and CEX2020-001007-S, some of which include ERDF funds from the European Union. IFAE is partially funded by the CERCA program of the Generalitat de Catalunya.

We  acknowledge support from the Brazilian Instituto Nacional de Ci\^encia
e Tecnologia (INCT) do e-Universo (CNPq grant 465376/2014-2).

This document was prepared by the DES Collaboration using the resources of the Fermi National Accelerator Laboratory (Fermilab), a U.S. Department of Energy, Office of Science, Office of High Energy Physics HEP User Facility. Fermilab is managed by Fermi Forward Discovery Group, LLC, acting under Contract No. 89243024CSC000002.

DSC thanks the support provided by the Ministry of Sciences and Education of Spain through the research grant PRE2019-088032 and the Swiss National Science Foundation (SNSF) under grant number 10002981.
JB is partially supported in this work by NSF awards AST2206563 and AST2442796 and DOE grant DE-SC0024787.

Part of this research was carried out at the Jet Propulsion Laboratory, California Institute of Technology, under a contract with the National Aeronautics and Space Administration (80NM0018D0004).

This research used resources of the National Energy Research Scientific Computing Center, a DOE Office of Science User Facility supported by the Office of Science of the
U.S. Department of Energy under Contract No. DE-AC02-
05CH11231. Some of the computing for this project was performed on the Sherlock cluster. We would like to thank Stanford University and the Stanford Research Computing Center for providing computational resources and support that contributed to these research results. We acknowledge the use of Spanish Supercomputing Network (RES) resources provided by the Barcelona Supercomputing Center (BSC) in MareNostrum 5 under allocations AECT-2024-3-0020, 2025-1-0045 and 2025-2-0046.



\appendix
\section{Matter power spectrum emulators}
\label{app:pk-emu}

A typical cosmological inference for 3x2pt in the present study requires of the order of a few to tens of thousands of CPU-hours, depending on the choice of sampler (\textsc{Nautilus} and \textsc{Polychord}) and modeling choice (linear and non-linear galaxy bias). In order to curb the computing time in our scale cuts procedure, which required running around one hundred and fifty chains, we made use of an emulator of the \textsc{HMCode 2020} matter power spectrum from \url{https://github.com/felipeaoli/HMcode2020Emu} based on \textsc{Cosmopower}~\cite{SpurioMancini2022}, specifically trained within the prior ranges ~\cite{Tsedrik2024} of the present analysis considering one massive neutrino. We used this emulated power spectrum in place of \textsc{Camb} for the intermediate scale cuts exploration that we operated, thus improving the runtime of the cosmological inference by a factor of around 3. However, we have shown the emulator did not give reliable results on smaller scales, especially when considering a non-linear galaxy bias model. For the final determination of scale cuts, we therefore decided to use our fiducial full pipeline for all probes and both for the linear and non-linear galaxy bias models.    

\section{Matter power spectrum reconstruction}
\label{app:pk-reconstruction}

The baseline modeling approach described in previous sections marginalizes over cosmological parameters alongside nuisance and astrophysical parameters. A complementary approach one can take is to instead fix cosmological parameters and allow the shape of the matter power spectrum to deviate freely \cite{Preston_2024}. 

In doing this, whilst the cosmology we fix to will be $\Lambda$CDM, the power spectrum is free to deviate from this shape. We marginalize instead over $N$ parameters $C_{i}$ evenly spaced in $\log(k)$, defined such that 

\begin{equation}
    \frac{P_{\rm{m}}(k,z)}{P^{\rm{DMO}}_{\rm{m}}(k,z)} = C(k_{i})  \hspace{1mm} , \hspace{5mm} i=[1,N] 
\end{equation}

where ${P^{\rm{DMO}}_{\rm{m}}(k,z)}$ is the dark matter only matter power spectrum at fixed cosmology from the fiducial $3\times2$pt analysis, $P_{\rm{m}}(k,z)$ is the measured matter power spectrum, and $C(k_{i})$ are bins measuring the amplitude of the matter power spectrum relative to the fixed spectrum. In this analysis method we marginalize over $N$ amplitude parameters. 

The binning scheme chosen reflects a choice between maximizing shape information without compromising constraining power. For this analysis, systematic parameters are marginalized over as described in Table~\ref{tab:priors}, with scale cuts implemented as described in Section~\ref{sec:cuts}.

Although not directly used in this work, the matter power spectrum reconstruction method is being applied in forthcoming cosmic shear (\textit{in prep.}) and $3 \times 2$pt (\textit{in prep.}) analyses to probe the suppression of power on small scales.

\section{MCMC Samplers}
\label{app:mcmc_samplers}

Accurate posteriors require samplers that can sufficiently explore the distribution, but what tools and settings are sufficient depend on the inference problem, requiring analysis-specific checks to confirm that posteriors can be trusted \cite*{Lemos_Weaverdyck_2022}. We perform consistency checks using the validated \textsc{Polychord} sampler \cite{Handley_2015} and \textsc{Nautilus}, an importance sampling code based on neural networks for efficient phase-space exploration. The configuration used is detailed in section \ref{sec:likelihood}. On the other hand, we test two different configurations of \textsc{Polychord}: the \textit{low-resolution}, which has been optimized for speed, making it suitable for development and preliminary checks, using: \texttt{nlive} = 250, \texttt{num\_repeats} = 30, and \texttt{tolerance} = 0.01, and the \textit{high-resolution}, for high-accuracy results, including: \texttt{nlive} = 675, \texttt{num\_repeats} = 120, and \texttt{tolerance} = 0.01.

\begin{figure}[t]
    \includegraphics[width=0.8\linewidth]{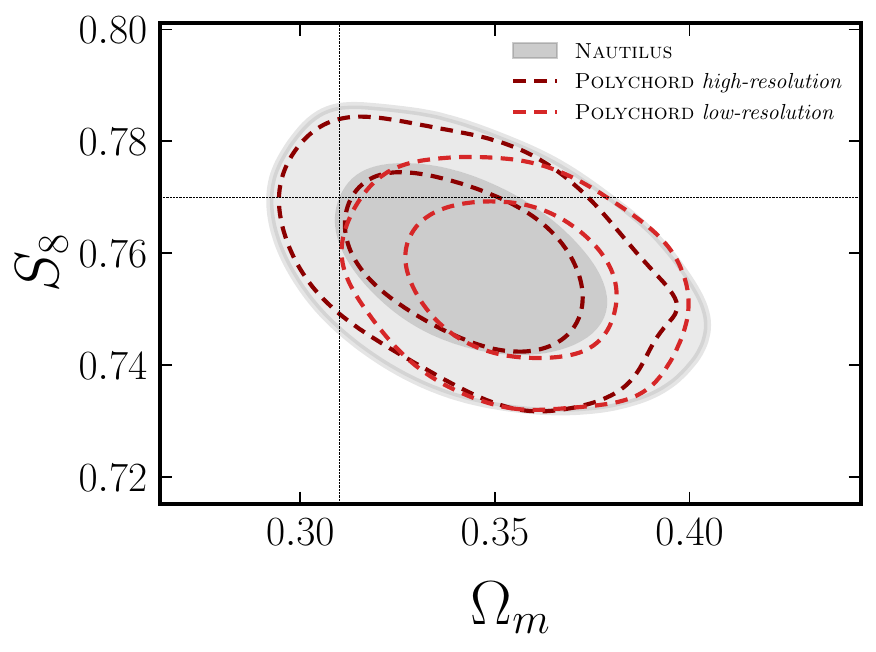}
    \caption{\label{fig:nautilus-test}
    Sampler comparison in the S$_8$-$\Omega_{\rm m}$ plane for a $3 \times 2$pt $\Lambda$CDM non-linear bias chain. The \textsc{Nautilus} and the \textit{high-resolution} \textsc{Polychord} results show reasonable agreement and are consistent with the fiducial cosmology within 2$\sigma$. The \textit{low-resolution} \textsc{Polychord} differs from the fiducial cosmology at more than the 2$\sigma$ level.
    }
\end{figure}

\textsc{Nautilus} produces contours consistent with those of \textsc{Polychord} but in significantly less time. For a $3 \times 2$pt run, \textsc{Nautilus} requires approximately three and four times fewer core hours than Polychord with, respectively, the \textit{low-} and  \textit{high-resolution} settings. The weighted mean shifts between both samplers in the S$_8$-$\Omega_{\rm m}$ plane are smaller than 0.03 sigma in both cases. However, we find $\mathcal{O}$(0.2$\sigma$) inconsistencies in the \textit{low-resolution} case for non-linear bias chains, as displayed in Fig.~\ref{fig:nautilus-test}. Additional metrics are also relevant to this comparison, such as the number of effective samples. Our \textsc{Nautilus} chains contain about 80000 samples, which is two and four times more than in \textit{high-} and \textit{low-resolution} Polychord, ensuring that the posterior is not under-sampled. Furthermore, the global evidence is nearly identical across three cases, but uncertainty on its estimate is an order of magnitude smaller with Nautilus. Finally, in our runs, \textsc{Nautilus} requires about four times less likelihood evaluations than the \textit{high-resolution} \textsc{Polychord} to reach convergence. These results lead us to adopt Nautilus as the default sampler to speed-up the Year 6 cosmological analysis.

\section{Parameterizing the amplitude of density fluctuations}
\label{app:parametrizing-amplitude}

Among the different ways to parametrize the amplitude of matter density fluctuations---including $\ln(10^{10} A_s)$, $\sigma_8$, and $S_8$---we choose $A_s \times 10^9$ for both physical and computational reasons.

\begin{table}
    \centering
    \resizebox{\columnwidth}{!}{%
    \begin{tabular}{lccc}
        \midrule \midrule
        Parameterization & Prior & $\text{FoM}_{\Omega_{\rm m}, \sigma_8}$ & $\Delta \text{FoM}_{\Omega_{\rm m}, \sigma_8}$ [\%] \\
        \midrule
        $A_s$ & $\mathcal{U}[0.5, 5.0] \times 10^{-9}$ & 688.53 & - \\
        $\ln(10^{10} A_s)$ & $\mathcal{U}[1.5, 6.0]$ & 677.56 & -1.59 \\
        $S_8$ & $\mathcal{U}[0.1, 1.3]$ & 693.21 & 0.68 \\
        $\sigma_8$ & $\mathcal{U}[0.6, 0.9]$ & 824.82 & 19.79 \\
        $A_s \times 10^9$ & $\mathcal{U}[0.5, 5.0]$ & 674.40 & -2.05 \\
        \midrule
    \end{tabular}
    } 
    \caption{\label{tab:amplitude-param} Figure-of-merit in the $\Omega_{\rm m}$–$\sigma_8$ plane obtained from cosmic shear analyses, for various parameterizations of the amplitude of density fluctuations. Each approach is sampled using the flat priors listed in the table. The rightmost column reports the relative difference in constraining power, $\Delta \text{FoM}_{\Omega_{\rm m}, \sigma_8}$ (in percent), using the sampling of $A_s$ as the reference.}
\end{table}

From a computational perspective, directly sampling $A_s$ without rescaling introduces numerical instabilities due to the mismatch in scale between $A_s$ (on the order of $10^{-9}$) and other parameters (typically $\mathcal{O}(1)$). This motivates the use of a scaled version such as $A_s \times 10^9$ or $\ln(10^{10} A_s)$. Between these two computationally viable options, we choose $A_s \times 10^9$. We avoid sampling $S_8$ directly to prevent placing prior assumptions on the specific parameter combination that our data constrain most tightly. To assess the practical impact of these choices, we compute the FoM for the $\Omega_{\rm m}$--$\sigma_8$ plane from cosmic shear chains using our baseline modeling (Table~\ref{tab:amplitude-param}). The relative differences in FoM across parameterizations are negligible ($\lesssim 2\%$) for all choices except $\sigma_8$, where the larger deviation (20\%) likely reflects the difficulty of specifying an appropriate prior range for this derived quantity without inadvertently introducing prior-driven bias. This confirms that our choice of $A_s \times 10^9$ yields stable and unbiased results while remaining computationally tractable.

\section{Intrinsic Alignment}
\label{app:ia}

\begin{figure}[t]
    \centering
    \includegraphics[width=\columnwidth]{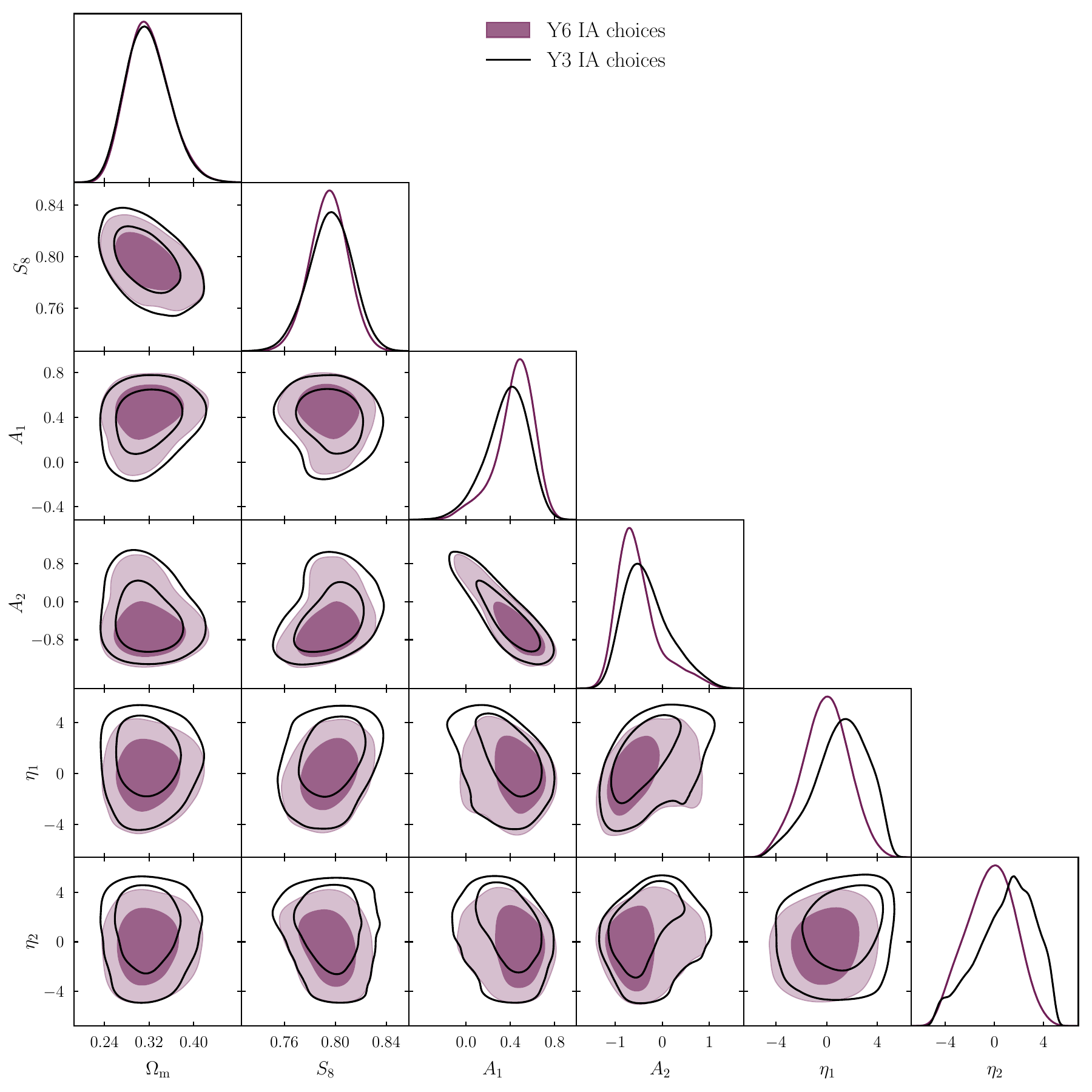}
    \caption{\label{fig:ia_priors}The combined impact of IA modeling choices on simulated cosmic shear data. The purple contours were run with our fiducial IA priors and pivot redshift. In black we show an analysis of the same simulated data using Year 3 priors (flat in the range [-5,5] for $A_1$,$A_2$,$\eta_1$,$\eta_2$) and pivot redshift ($z_{\rm piv}=0.62$ instead of $z_{\rm piv}=0.3$). As seen, the Year 6 priors have almost no impact on cosmological constraining power, but help regulate the pile-up of power at high $\eta_{1/2}$.}
\end{figure}

In Year 6 we have included a number of minor updates to the intrinsic alignment model relative to previous DES analyses. Specifically (a) the pivot redshift $z_{\rm piv}$ has been reduced from 0.62 to 0.3; (b) the flat priors on the IA amplitudes $A_1$ and $A_2$ have been narrowed from [-5,5] to the ranges shown in Table \ref{tab:priors}; (c) the flat priors on the redshift evolution parameters $\eta_1$ and $\eta_2$ have been replaced with wide Gaussian priors $\mathcal{N}(0,3)$. We will briefly discuss each of these in turn below. 

The change in pivot redshift is intended to better reflect where our data constrain IAs -- the IA sensitivity is dominated by the lowest source bin, which peaks below $z=0.5$. Bringing the pivot closer to this range helps to avoid projection effects, which can cause a build-up of posterior weight at the upper edge of the priors on $\eta_1$ and $\eta_2$ (see, for example, \cite{deskids} Fig. 3). \journal{The fiducial amplitude values in Table~\ref{tab:priors}, $A_1 = 0.13$ and $A_2 = -0.2$, are obtained by translating the DES Year 3 IA constraints ($A_1 = 0.12$, $A_2 = -0.3$ at $z_{\rm piv} = 0.62$) to the updated pivot redshift $z_{\rm piv} = 0.3$, preserving the physical IA amplitude at all 
redshifts.}

The change in the amplitude priors reflects improving constraining power of lensing data. Previous lensing results have constrained $A_1$ and $A_2$ to well within $A_1,A_2=\pm3$. Although the exact values will be sample dependent, shifts out to 3 or more would be highly unexpected. The lower bound on $A_1$ at -1 reflects the fact that we have some knowledge of the physics driving such alignments. Linear alignment from pressure-supported red galaxies is expected to produce positive $A_1$. Any negative value coming from interaction with other forms of alignment (e.g. tidal torquing in blue galaxies) will be small, given null constraints on IAs in such galaxies in the literature \cite{Samuroff_2023,navarro2025}. However, cutting off the negative $A_1$ tail, which correlates with $S_8$ and so can be a source of projection effects, in principle should help to keep such effects under control. 

Finally, the Gaussian prior on $\eta_{1,2}$ is designed to slowly downweight extreme (and physically implausible) values without excluding them. Again, the main motivation here is to minimize volume effects from the IA model without restricting it to the point of causing possible parameter biases.

We show the combined impact of these changes in Fig.~\ref{fig:ia_priors}. As seen here, in a simulated cosmic shear analysis, the changes reduce the concentration at high $\eta_1$, but otherwise do not impact cosmological constraining power.

\section{\texorpdfstring{Incorporating Source $n(z)$ Modes–Shear Bias Correlations}{Incorporating Source n(z) Modes–Shear Bias Correlations}}
\label{app:corrpriors}

\begin{figure*}
    \centering
    \includegraphics[width=0.8\linewidth]{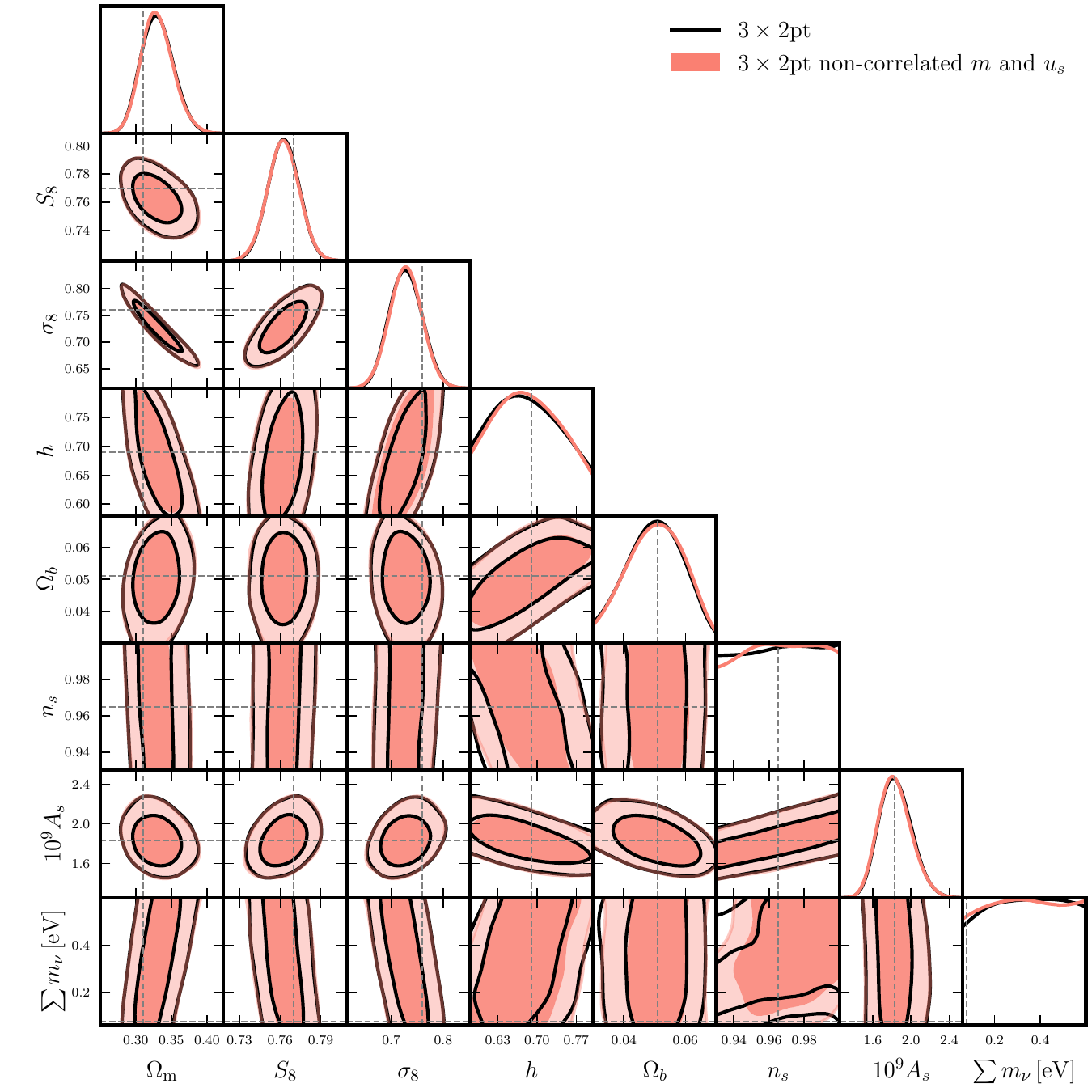}
    \caption{\label{fig:source-corrp}
    Posterior distributions of the cosmology parameters derived from the fiducial $3 \times 2$pt analysis. The solid black contours include the correlation between the multiplicative shear-bias parameters $m$ and the source redshift-distribution modes $u_s$, whereas the salmon contours show results obtained when this correlation is ignored.
    }
\end{figure*}

Building on the scale-cut validation described in Section~\ref{subsec:cuts-results}, we extend our treatment of source-galaxy calibration by explicitly linking uncertainties in shape-measurement biases to those in the source redshift distributions. Concretely, we introduce correlations between the multiplicative shear-bias parameters $m$ and the source redshift-distribution modes $u_s$, as detailed in Section~\ref{subsec:parameters-and-priors}.  

Rather than repeating the full scale-cut validation, we assess the impact of this additional modeling ingredient by applying it to the fiducial $3 \times 2$pt analysis. As illustrated in Fig.~\ref{fig:source-corrp}, incorporating the $m$–$u_s$ correlation produces negligible shifts in the inferred cosmological posteriors and only minor changes in the SNR. Our main conclusions based on the original scale-cut validation, which neglected these correlations, therefore remain unchanged.

\section{\label{app:larger-scales}Exploiting larger angular scales}
We investigated alternative binning of the 2PCFs, extending beyond the fiducial maximum angular scale of $250$ arcmin. and 20 angular bins. In particular, we explored the impact of including six additional bins up to $1000$ arcmin. to assess whether complementary constraining power resides at these larger scales.  

To this end, we performed a cosmic shear analysis of a noiseless mock signal that included the extended set of angular bins, and compared the resulting constraining power to that obtained with the fiducial binning scheme. As in the main analysis, we applied small-scale cuts to remove data points where baryonic feedback processes are poorly modeled. We evaluated the impact of the additional large-scale information using two metrics: the SNR and the relative differences in the marginalized parameter uncertainties.  

The results can be summarized as follows. Including the extended angular scales produces a clear increase in SNR, from $50.5$ in the fiducial setup to $68.3$ with the additional bins. However, this increase does not translate into improved parameter constraints: the analyses with and without the extra bins yield statistically consistent error bars. The relative differences in parameter uncertainties are at the level of $\sim 9\%$ worse constraints on the matter abundance and $\sim 5\%$ tighter constraints on $S_8$. Constraints on IA parameters are also similar. These shifts are compatible with statistical fluctuations due to sampling variance.  

Based on these results, we conclude that the additional large-scale bins do not provide meaningful improvements and therefore are not included in our fiducial analysis. We interpret the negligible gain in constraining power as evidence that the information at very large angular scales is largely redundant with that already captured at intermediate and small scales.

\section{\label{app:mapsearches}MAP estimation}
The high dimensionality of our parameter space makes estimating the maximum a posteriori (MAP) parameter for a given analysis is a non-trivial process.  Throughout this work, we estimate MAP parameters via a procedure in which we use the $N$ highest posterior samples from a chain  as input guesses to optimizer searches, and then select highest posterior output of those $N$ searches. 

To assess the need for this procedure and its performance, we performed a series of tests on noisy synthetic data realizations using a fast approximate version of the $3 \times 2$pt pipeline. These synthetic data were generated by adding a Gaussian noise realization based on the data covariance to a model prediction at our fiducial cosmology. This approximate pipeline, initially developed for tests of posterior predictive distribution methods of internal consistency (to be described in an upcoming publication), makes it  feasible to repeat the simulated analyses for multiple noise realizations. It assumes $\Lambda$CDM and linear bias, uses linear-only power spectrum model from the Eisenstein and Hu approximation of Ref.~\cite{Eisenstein:1997jh}, and instead of using mode projection parametrizes photometric uncertainties through shift and stretch nuisance parameters as described in Section~\ref{sec:robustness-tests}. Also, because these simulated analyses were run prior to the finalization of fiducial scale cuts they use slightly different set of angular scales than the main analysis. Despite these differences, the similar number of model parameters and observables, as well as the fact that the dependence of the posterior on parameters is roughly similar, means that simulated analyses using this approximate pipeline should still serve as a useful laboratory for testing our MAP estimation procedure. 

With this approximate pipeline we ran {\sc Nautilus} chains for ten noise realizations, and conducted $N=20$ optimizer searches per posterior estimate. In these simulated analyses, we find that the maximum posterior estimated by this two-step search procedure is fairly reliable, but that there is significant noise in the associated parameter values. In other words, the parameter values reported for the MAP will reliably produce a theory prediction that has a good fit to the data, but are not a good predictor of the true underlying parameters. 

\begin{figure}[t]
\includegraphics[width=0.95\columnwidth]{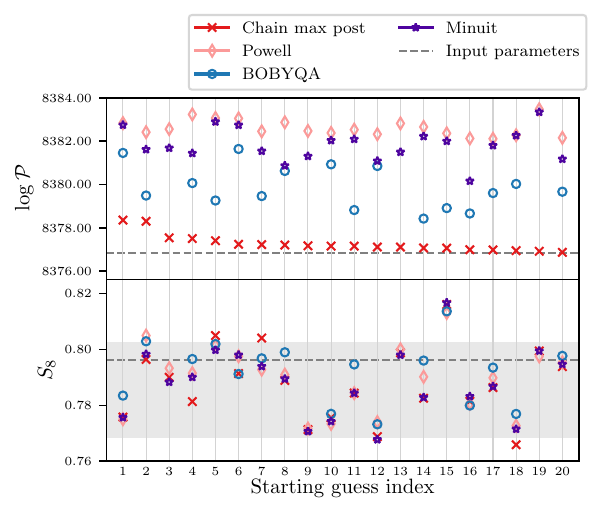}  
    \caption{\label{fig:map_optimization}Results of MAP estimation searches for synthetic $3\times 2$pt data generated by adding a Gaussian noise realization to a model prediction at our fiducial cosmology. Red X markers show the 20 highest posterior estimates from the {\sc Nautilus} chain, which were used as starting guesses for a set of optimizer searches for the maximum posterior. Pink, purple, and blue markers show the results of those searches with the Powell, {\sc minuit } and {\sc bobyqa} algorithms, respectively. Dashed gray lines show the posterior and $S_8$ values at the input parameters used to generate the simulated data, and the light gray shaded region shows the $1\sigma$ confidence region of the marginalized posterior. }
\end{figure}

To illustrate this in more detail, Fig.~\ref{fig:map_optimization} shows the results of optimizer searches on estimates of the MAP values for the posterior and $S_8$ for one simulated noise realization. In that figure, the red ``X'' markers report values from chain samples, while the other markers show the outputs of posterior maximization searches from various optimizer algorithms. Our fiducial MAP estimation searches are performed with the Powell algorithm~\cite{powell_1964} as implemented in {\sc SciPy.}~\cite{2020SciPy-NMeth},  which is part of the {\sc CosmoSIS}~\cite{Zuntz_2015} ``maxlike'' sampler. We found similar performance with {\sc minuit}~\cite{minuit}, while  {\sc bobyqa}~\cite{pybobyqa1, pybobyqa2, pybobyqa3} found somewhat less optimal estimates of the maximum posterior. In these tests, all three algorithms were run with the tolerance setting 0.05. Powell and {\sc bobyqa} were run with a maximum number of iterations set to 5000, while {\sc minuit} was run for a maximum of 1000 iterations. 

The top panel, which shows the posterior values, clearly demonstrates that the optimizer step non-negligibly improves the estimate of the maximum posterior. We see that, for this noise realization, the highest posterior found came from an optimizer search starting at the 19th highest posterior sample from the chain. The other noise realizations studied show a similar lack of correlation between which optimizer search results in the highest posterior estimate and the index of the starting guess, which motivates us to retain this relatively high number of searches ($N=20$) as part of our MAP estimation procedure. Considering MAP parameter estimates, we note that the optimizer searches do tend to move $S_8$ towards the simulation input value, but are still fairly broadly scattered around it. This behavior is seen for other parameters as well, we are finding a number of points with significant scatter in parameter space which are likely local maxima with comparably high posteriors. Based on these results we caution that such estimates of MAP parameters have significant scatter that can depend on the specific choices made in the optimization procedure. 

Interestingly, we note that all of these searches found samples with a posterior higher than that evaluated at the fiducial parameter values used to generate the simulated data, shown with gray dashed lines. This highlights the fact that  noise in the MAP parameter estimation around the input parameters is not simply due to the optimizer failing to find a global best-fit. It occurs because there are a number points in our parameter space that by chance fit a given noise realization better than the true underlying cosmology. This impact of noise on MAP parameter estimation likely explains why our results are in contrast to those found for a noiseless simulated analysis in Ref.~\cite{Joachimi:2020abi}, which demonstrated for KiDS 1000 methodology that a similar MAP optimization procedure could accurately recover the input parameters.    

\bibliographystyle{apsrev4-2}
\bibliography{biblio}

\end{document}